\documentclass[prb,floatfix,twocolumn,showpacs,amsmath,amssymb]{revtex4}
\usepackage{graphicx}
\usepackage{dcolumn}
\usepackage{bm}

\begin{document}

\title{Quantum Hall conductance and 
de Haas van Alphen oscillation
 in a tight-binding model with electron and hole pockets 
for (TMTSF)$_2$NO$_3$
}

\author
{Keita Kishigi}
\affiliation{Faculty of Education, Kumamoto University, Kurokami 2-40-1, 
Kumamoto, 860-8555, Japan}

\author{Yasumasa Hasegawa}
\affiliation{Department of Material Science, 
Graduate School of Material Science, 
University of Hyogo, Hyogo, 678-1297, Japan}

\date{\today}

\begin{abstract}

Quantized Hall conductance and de Haas van Alphen (dHvA) oscillation are studied theoretically
in the tight-binding model for (TMTSF)$_2$NO$_3$, in which there are small pockets of electron and hole 
due to the periodic potentials of anion ordering in the $a$-direction. 
The magnetic field is treated by hoppings as complex numbers
due to the phase caused by the vector potential, i.e. Peierls substitution.
In realistic values of parameters and the magnetic field, the energy as a function of a magnetic field (Hofstadter butterfly diagram) is obtained. 
It is shown that energy levels are broadened and the gaps are closed or almost closed periodically as a function of the inverse magnetic field, which are not seen in a semi-classical theory of the magnetic breakdown. Hall conductance is quantized with an integer obtained by Diophantine equation when the chemical potential lies in an energy gap. When electrons or holes are doped in this system, Hall conductance is quantized in some regions of a magnetic field but it is not quantized in other regions of a magnetic field due to the broadening of the Landau levels. The amplitude of the dHvA oscillation at zero temperature decreases as the magnetic field increases, while it is constant in the semi-classical Lifshitz Kosevich formula. 
\end{abstract}

\date{\today}

\maketitle

\section{Introduction}

\begin{figure}[bt]
\begin{flushleft} \hspace{0.5cm}(a) \end{flushleft}\vspace{-0.0cm}
\includegraphics[width=0.3\textwidth]{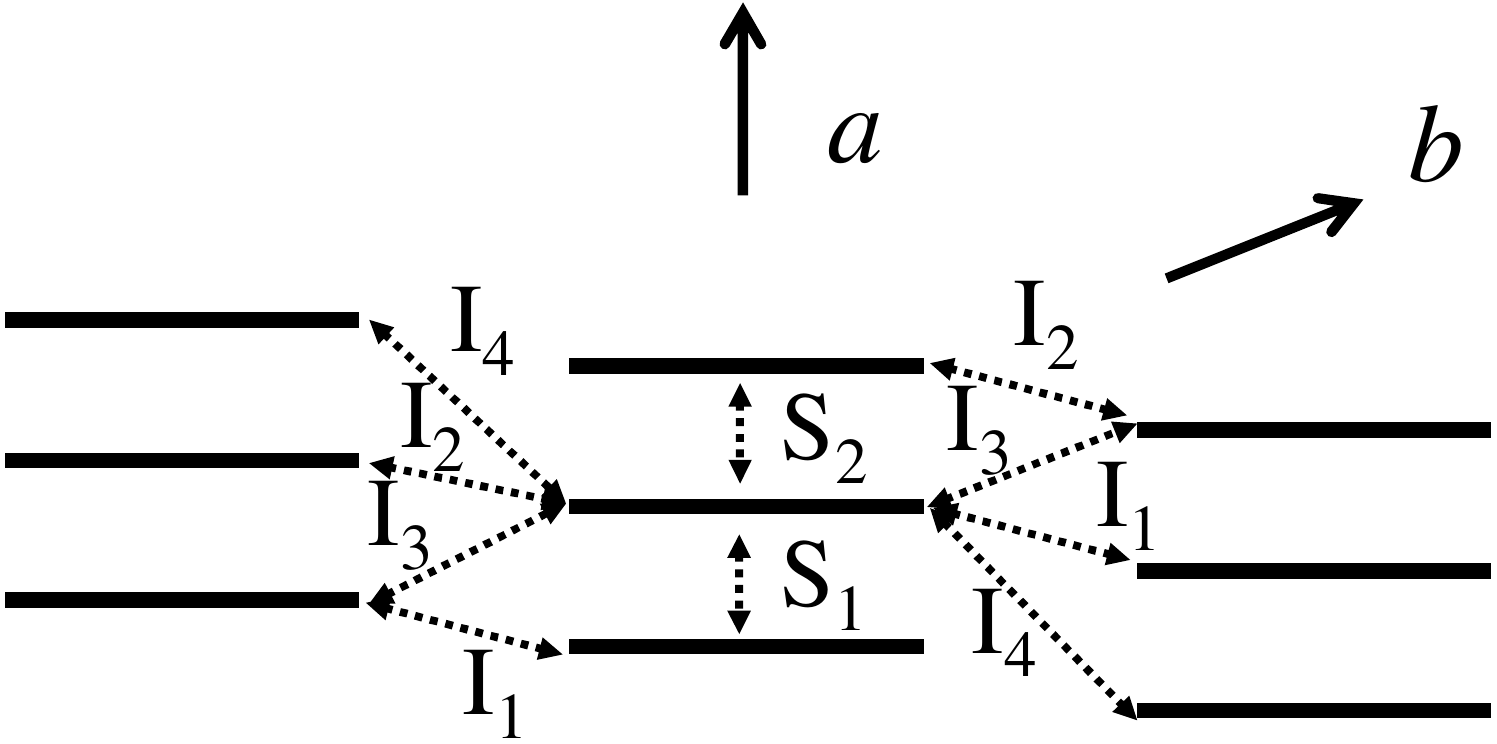}
\begin{flushleft} \hspace{0.5cm}(b) \end{flushleft}\vspace{0.1cm}
\includegraphics[width=0.37\textwidth]{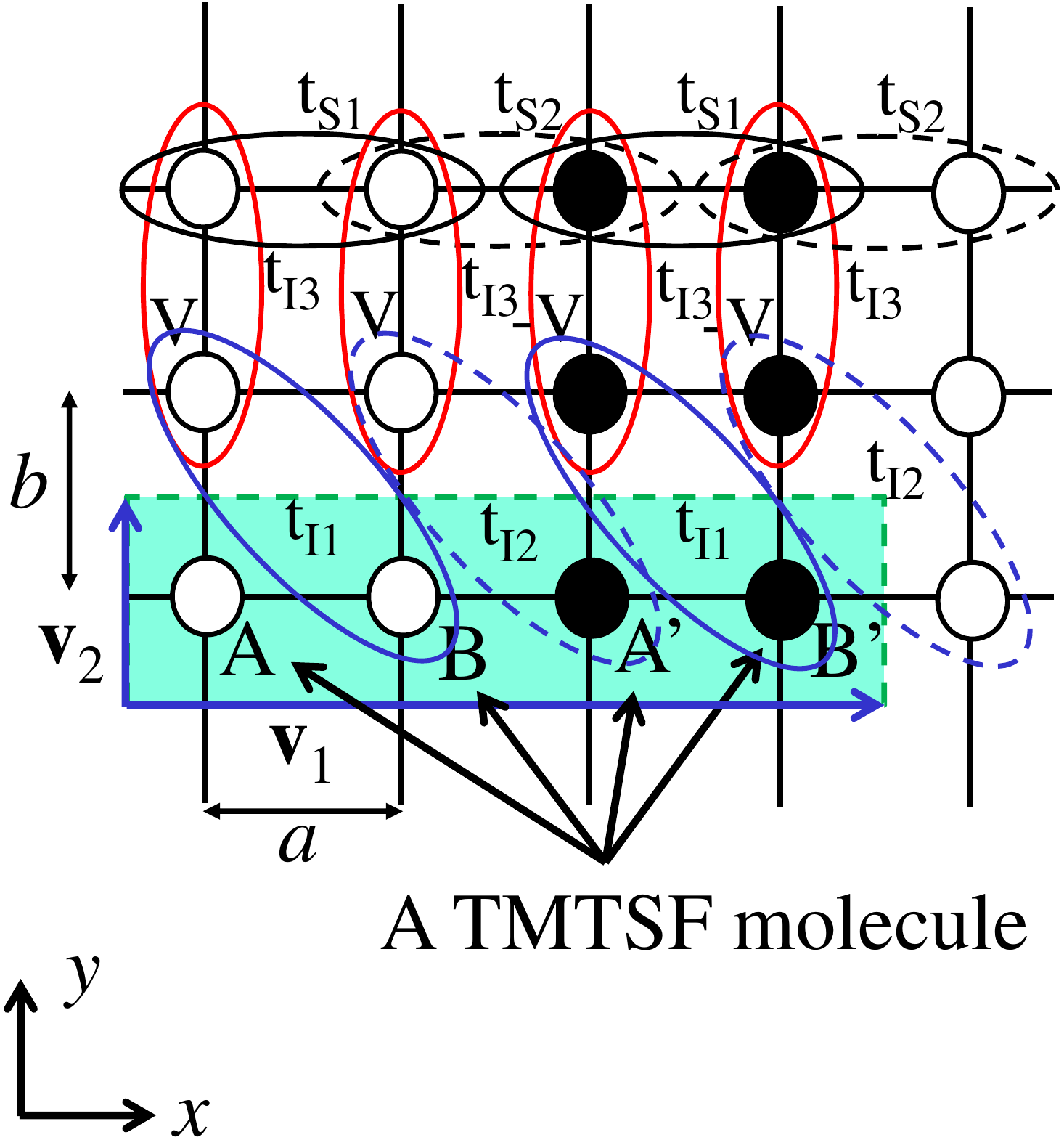}
\begin{flushleft} \hspace{0.5cm}(c) \end{flushleft}\vspace{0.3cm}
\includegraphics[width=0.39\textwidth]{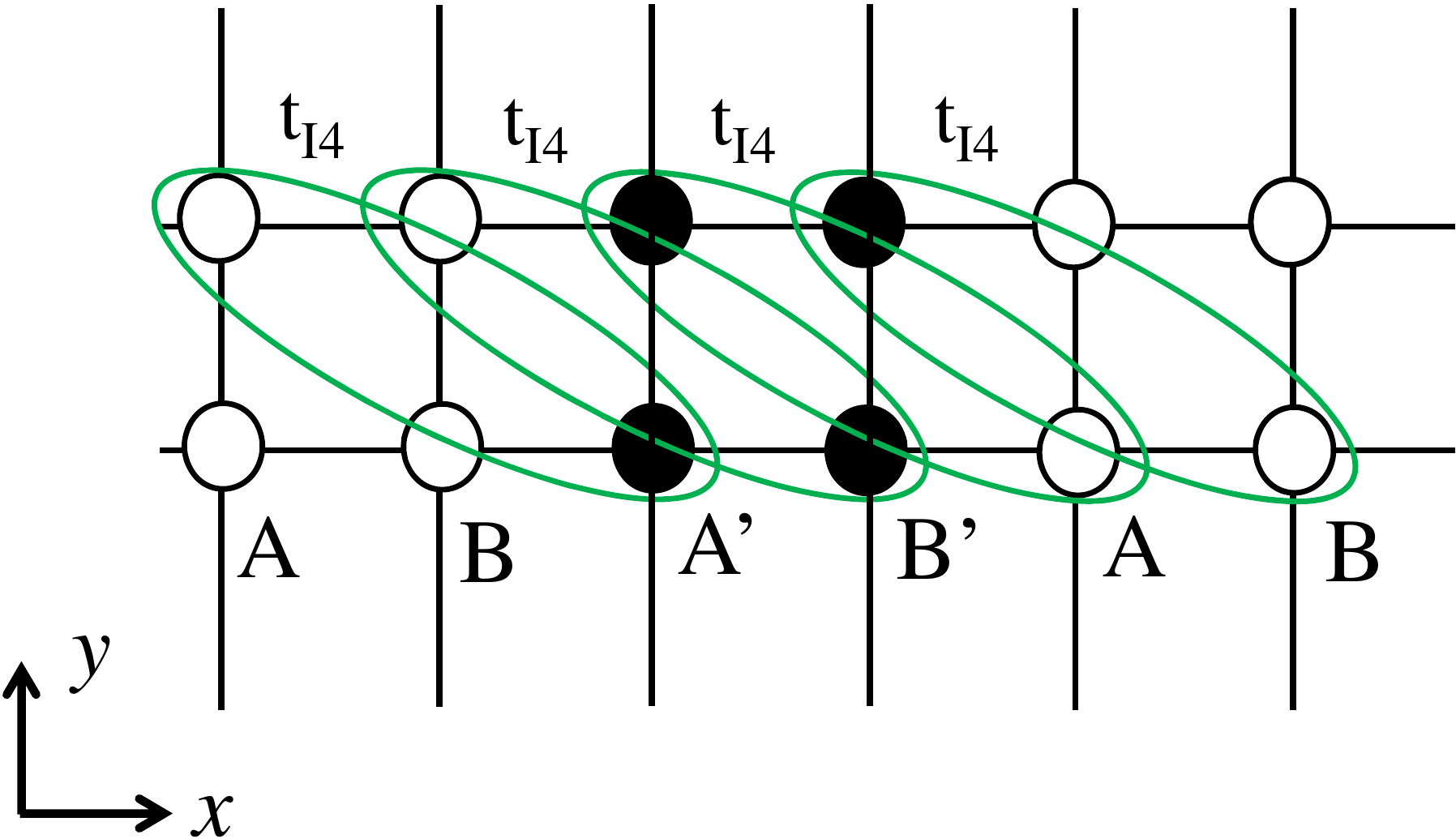}
\caption{(Color online) (a) Schematic side view of (TMTSF)$_2$X. Solid lines are for TMTSF molecules and dotted lines are transfer integrals\cite{review}.
 (b) 
The simplified tight-binding model for (TMTSF)$_2$X in the 
rectangular lattice, where $2a$ and $b$ are the lattice constants (note that
there are two sites (A and B) in the unit cell when $V=0$.
The definition of $a$ is a half of that used in Ref. \cite{review}.).
The unit cell is shown as the light green rectangle. 
The transfer integrals ($t_{\mathrm{S1}}$,  $t_{\mathrm{S2}}$, $t_{\mathrm{I1}}$, $t_{\mathrm{I2}}$, $t_{\mathrm{I3}}$) are shown as ovals. The effect of the ordering of the orientation of the anion NO$_3$ is taken as the on-site potentials $\pm V$. 
(c) Transfer integrals of $t_{\textrm{I4}}$.
}
\label{Figure1}
\end{figure}

Organic conductors, (TMTSF)$_2$X, where TMTSF is
tetra-methyl-tetra-selena-fulvalence and
X is anion (X=NO$_3$, PF$_6$, ClO$_4$ {\it etc.})\cite{review,grant}, have the structure of stacked planer molecules,
TMTSF, in the $a$-direction as shown in Fig. \ref{Figure1} (a). 
We can neglect the hoppings perpendicular to $a$-$b$ plane, because they are very small\cite{review}. The energy band structure is well described\cite{review} by six hopping integrals ($t_{\mathrm{S1}}$,  $t_{\mathrm{S2}}$, $t_{\mathrm{I1}}$, $t_{\mathrm{I2}}$, $t_{\mathrm{I3}}$ and $t_{\mathrm{I4}}$) which are shown in Fig. \ref{Figure1} (a). 
Since the absolute values of the hoppings in the chain along the 
$a$-direction are about ten times larger than those between chains, the Fermi surface consists of quasi-one dimensional sheets as shown in Fig. \ref{Figure2} (a). 

The unit cell of (TMTSF)$_2$NO$_3$ is doubled along the $a$-direction due to the ordering of the orientation of the anion NO$_3$ below $T_{\mathrm{AO}}\simeq 45$~K\cite{pouget,fisdw_no3,kang_2009}. 
The Brillouin zone is halved and there appear small electron and hole pockets, as seen in Fig. \ref{Figure2} (b). 
When the magnetic field $(H$) is applied perpendicular to the $a$-$b$ plane, the energy of electrons is quantized. In this case, the de Haas van Alphen (dHvA) effect\cite{shoenberg} is expected. 
Fortin and Audourad\cite{fortin,fortin_2009} adopt the phenomenological 
network model\cite{Pippard62,Falicov66} of a semi-classical theory for the magnetic breakdown and a semi-classical quantization of energies\cite{onsager}. 
In two-dimensional systems, the oscillation of the chemical potential as a function of a magnetic field cannot be neglected in general\cite{shoenberg,nakano,alex2001,champel,KH}, whereas it is safely neglected in dHvA effect in three-dimensional systems
as in the Lifshitz-Kosevich (LK) formula.
\cite{shoenberg,LK,alex2001,champel,CM,KH,Igor2004PRL,Igor2011,Sharapov}
Fortin and Audourad\cite{fortin,fortin_2009} have shown that the oscillation of the chemical potential is very small and the LK formula explains
the field and temperature dependences of the amplitudes of the dHvA oscillation, if the effective masses of electron and hole are nearly the same.

In a tight-binding model, the energy under a magnetic field can be obtained without a phenomenological parameter for the probability amplitude of the tunneling, which is used in the semi-classical theory of the magnetic breakdown. The quantized Landau levels of the two-dimensional free electrons 
are described by delta functions. When the periodic potentials exist or the tight-binding model is used\cite{Harper,Harper2}, the energy levels are broadened. These energy levels as a function of the magnetic field 
are known as the Hofstadter butterfly diagram\cite{Hof,HLRW,HHKM}.
The study of the dHvA oscillation has been done in the tight-binding model\cite{machida,kishigi,sandu,Han2000,GV2007} in the systems where quasi-one dimensional Fermi surface and two-dimensional Fermi surface coexist. This Fermi surface is suitable to study the magnetic breakdown in the dHvA oscillation and is realized, for example, in (BEDT-TTF)$_2$Cu(NCS)$_2$\cite{review}. 
Fortin and Ziman\cite{fortin1998} have calculated the dHvA oscillation in the similar system by using the network model\cite{Pippard62,Falicov66}. 
In both studies of tight-binding model and the semi-classical network model, combination frequency, $\beta-\alpha$, has been obtained due to the chemical potential oscillation as a function of the magnetic field. 

The dHvA oscillation in the tight-binding model\cite{KM1996,KM1997} for (TMTSF)$_2$NO$_3$ has been studied theoretically. The model studied previously was, however, much simplified one and the exaggerated parameters were taken (half-filled band on the rectangular lattice with $t_b/t_a=0.6$ and $t_b'/t_a=0.2$, where $t_a$ and $t_b$ are the nearest-neighbor hoppings in $a$ and $b$ directions, respectively, and $t_b'$ is the next-nearest-neighbor hopping in $b$-direction). On the other hand, the quantum Hall effect in (TMTSF)$_2$NO$_3$ have never been studied in the actual parameters in the tight-binding model, as far as we know. 
The integer quantum Hall effects in two-dimensional electron systems are understood as topological phenomena. 
The quantized value of the Hall conductance is obtained as a first Chern number
or the solution of the Diophantine equation\cite{TKNN,Kohmoto_1985,Kohmoto_1989}. 


In this paper we adopt the tight-binding model with the realistic parameters for (TMTSF)$_2$NO$_3$ in the magnetic field treated quantum-mechanically. In experimentally accessible magnetic field ($\sim 6$~T),
we obtain an interesting structure of the energy as a function of the magnetic field (Hofstadter butterfly diagram), quantum Hall conductance, and dHvA oscillation. 
We show the difference between the results in quantum mechanical theory and those in semi-classical theory.


\section{Spin density wave in (TMTSF)$_2$NO$_3$}

\begin{figure}[bt]
\begin{flushleft} \hspace{0.5cm}(a) \end{flushleft}\vspace{-0.5cm}
\includegraphics[width=0.4\textwidth]{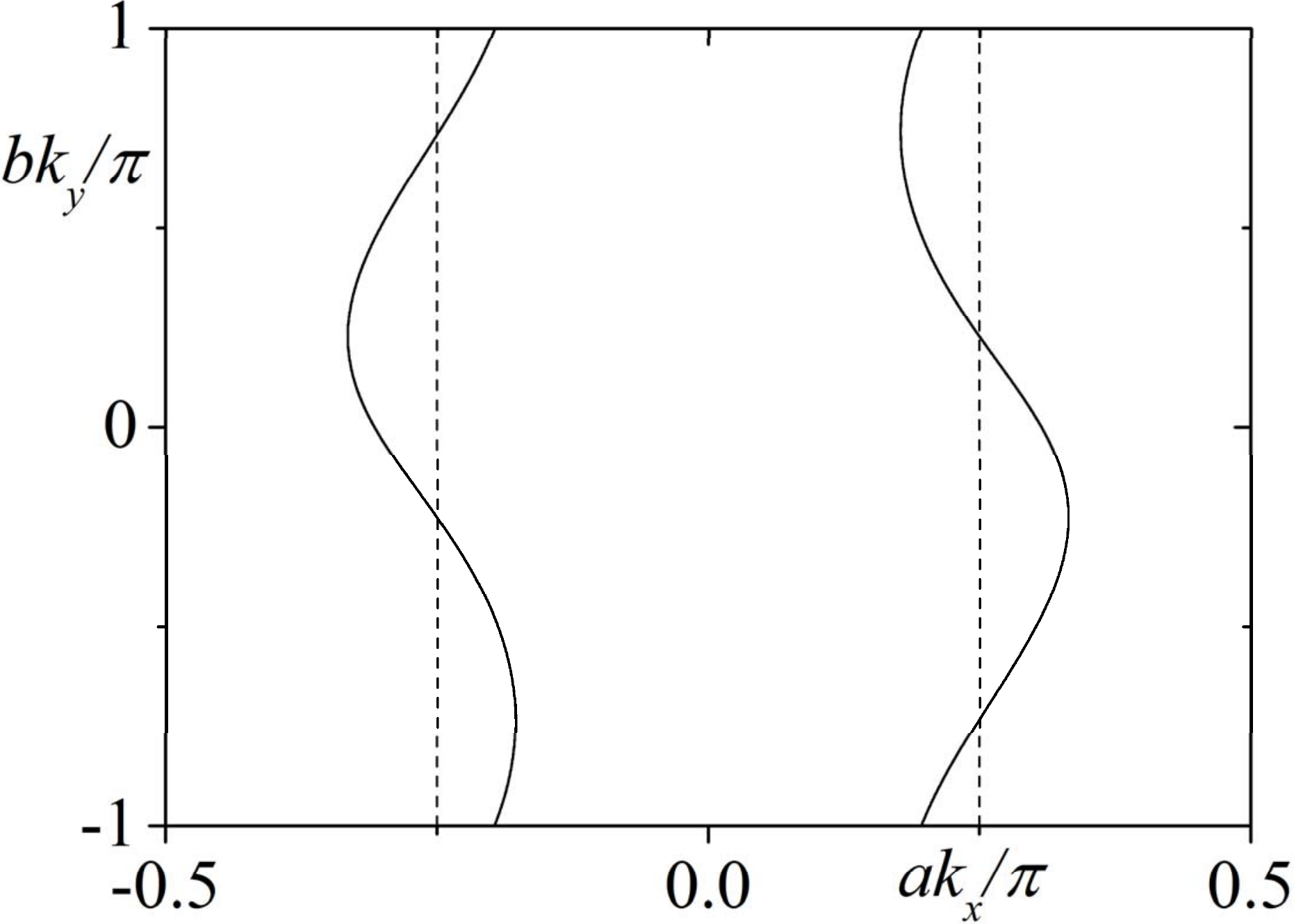} \\
\begin{flushleft} \hspace{0.5cm}(b) \end{flushleft}\vspace{-0.5cm}
\includegraphics[width=0.4\textwidth]{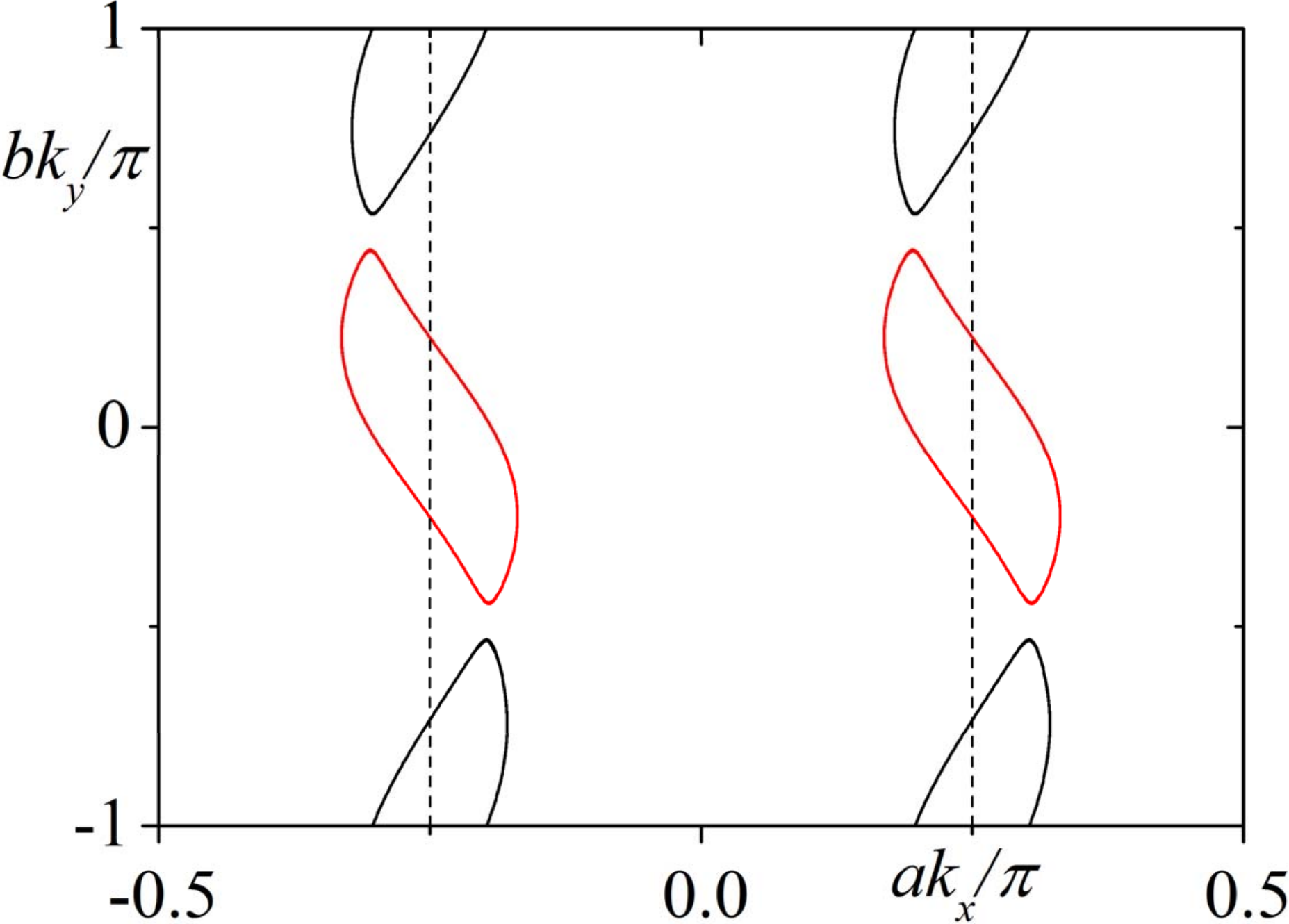}
\caption{(Color online)
(a) Fermi surface at 3/4-filling for $V=0$ and the transfer integrals are
$t_{\rm S1}=274.4$, $t_{\rm S2}=250.5$, $t_{\rm I1}=-29.1$, $t_{\rm I2}=-42.7$, $t_{\rm I3}=56.6$, and
$t_{\rm I4}=-6.3$ in the unit of meV. 
(b) Fermi surface at 3/4-filling for $V=12.38$ meV. The Brillouin zone is halved ($-1/4 < ak_x /\pi \leq 1/4$). Black  
and red curves are electron and hole pockets, respectively. 
}
\label{Figure2}
\end{figure}

The shape and the dimensionality of the Fermi surface in (TMTSF)$_2$NO$_3$ are controversial at high pressure and strong magnetic field\cite{fisdw_no3,kang_2009}. 
At the ambient pressure the spin density wave (SDW)\cite{tomic_1989,kang_1990,tomic_1991,Le} is stabilized in (TMTSF)$_2$NO$_3$ below $T_{\rm SDW}\simeq 9 \sim$ 12 K. 
The wave vector of the SDW has been observed in NMR experiments\cite{hiraki,satsukawa} to be $(q_x, q_y) \simeq (0, 0.25 (2\pi/b))$. 
That vector is indicated by an arrow in Fig. \ref{Figure2} (b), which is a good nesting vector. 
By applying pressure the nesting of the Fermi surface becomes less perfect 
and SDW is suppressed. Indeed, the metallic state in the absence of SDW is 
reported above 8.5 kbar in the magnetoresistance experiment by Vignolles {\it et al.}\cite{fisdw_no3}. 
The orientational order of NO$_3$ occurs even at high pressure. 
They have shown the 
difference between the frequency of the Shubnikov-de Haas 
(SdH) oscillation at low pressure and that at high pressure (above 8.5 kbar). 
They suggested that there exist two-dimensional pockets even above 8.5 kbar. 
However, Kang and Chung\cite{kang_2009} have observed that the angular-dependent magnetresistance oscillations in (TMTSF)$_2$NO$_3$ at 14 T and the pressures (6.0, 7.0 and 7.8 kbar) are similar to those in (TMTSF)$_2$ClO$_4$. They suggested that 
(TMTSF)$_2$NO$_3$ in the metallic state at high pressure has a quasi-one 
dimensional Fermi surface even in the presence of the anion ordering. 

The theoretical study of the angular-dependent magnetoresistance, however, has been done only semi-classically in quasi-one dimensional systems\cite{Danner-Chaikin,osada1996,yoshino1997,lee1998,lebed2003} and quasi-two dimensional systems\cite{Yamaji}. The SdH oscillation has not been studied quantum-mechanically in (TMTSF)$_2$NO$_3$, either. As we will show below, we study the tight-binding model for (TMTSF)$_2$NO$_3$  under the magnetic field at $T < T_{\rm AO}$ without SDW order in quantum mechanically and we obtain the results unexpected in the semi-classical theory. Therefore, in order to identify the shape and the dimensionality of the Fermi surface in (TMTSF)$_2$NO$_3$ at high pressure and magnetic field, we have to compare the experiments with the theory treated not in semi-classical theory but in quantum mechanics. Thus, our study will be a first step to understand the shape and the dimensionality of the Fermi surface in
(TMTSF)$_2$NO$_3$, where there are electron and hole pockets in the absence of magnetic field. The studies in the presence of the SDW order or under high pressure will be needed in future. 

The field-induced spin density wave (FISDW) has been also observed in (TMTSF)$_2$NO$_3$ at strong magnetic field ($\sim20$ T) and at the high pressure ($\sim8.5$ kbar) \cite{fisdw_no3,kang_2009}. The FISDW is caused by electron-electron interactions in similar quasi-one-dimensional organic conductors such as (TMTSF)$_2$PF$_6$ and (TMTSF)$_2$ClO$_4$\cite{GL1984,GL1995}. Since the instabilities of the FISDW are expected to be strong in (TMTSF)$_2$NO$_3$, the study including electron-electron interactions will be needed as a future problem. 

\section{electron and hole pockets at $H=0$}

Since the direction of the anion is random above $T_{\mathrm{AO}}$, we can neglect the effects of the anion potential. Then, the tight-binding model for (TMTSF)$_2X$ is described by six hopping integrals, $t_{\rm S1}$, $t_{\rm S2}$, $t_{\rm I1}$, $t_{\rm I2}$, $t_{\rm I3}$ and $t_{\rm I4}$ which are shown in Fig.~\ref{Figure1}\cite{review}.
Although the real lattice is monoclinic, the energy as a function of wave number is topologically the same as that 
in the rectangular lattice as shown in Fig.~\ref{Figure1}(b) and (c). 
The energy as a function of the uniform magnetic field (Hofstadter butterfly diagram) is also the same. (The similar situation has been known in the triangular lattice and the honeycomb lattice.
 For example, the tight-binding electrons on the triangular lattice have the same energy versus magnetic field as those on the square lattice with diagonal hoppings
 along one direction.\cite{HLRW,HHKM})
Since $t_{\mathrm{S1}} \neq t_{\mathrm{S2}}$ and $t_{\mathrm{I1}} \neq t_{\mathrm{I2}}$, there are two nonequivalent sites A and B in the unit cell. 
There are two bands in this case. 
Electrons are 3/4 filled for the bands made of highest occupied molecular orbits (HOMO) of TMTSF, 
since one electron is removed from two TMTSF molecules.
Then the lower band is completely filled and the upper
band is half-filled. By diagonalizing Eq. (\ref{J3_v0}) ($2\times2$ matrix) in Appendix~\ref{appendixA}, we plot the Fermi surface in Fig.~\ref{Figure2}(a), in which we take the parameters reported by Alemany, Pouget and Canadell\cite{Pere}; 
$t_{\rm S1}=274.4$, 
$t_{\rm S2}=250.5$, $t_{\rm I1}=-29.1$, $t_{\rm I2}=-42.7$, $t_{\rm I3}=56.6$, 
$t_{\rm I4}=-6.3$ in the unit of meV.

\begin{figure}[bt]
\begin{flushleft} \hspace{0.5cm}(a) \end{flushleft}\vspace{-0.5cm}
\includegraphics[width=0.3\textwidth]{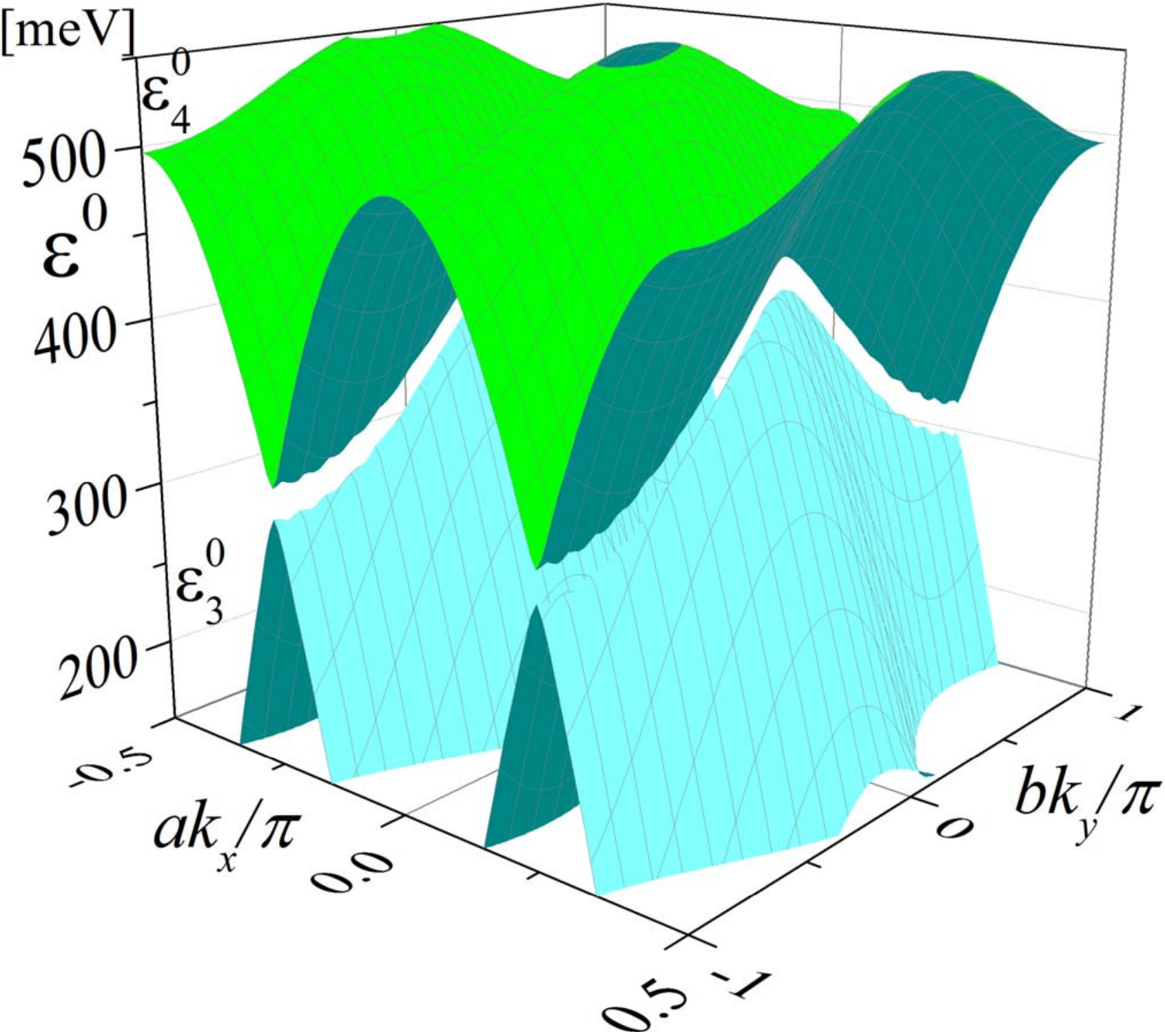}
\begin{flushleft} \hspace{0.5cm}(b) \end{flushleft}\vspace{-0.5cm}
\includegraphics[width=0.4\textwidth]{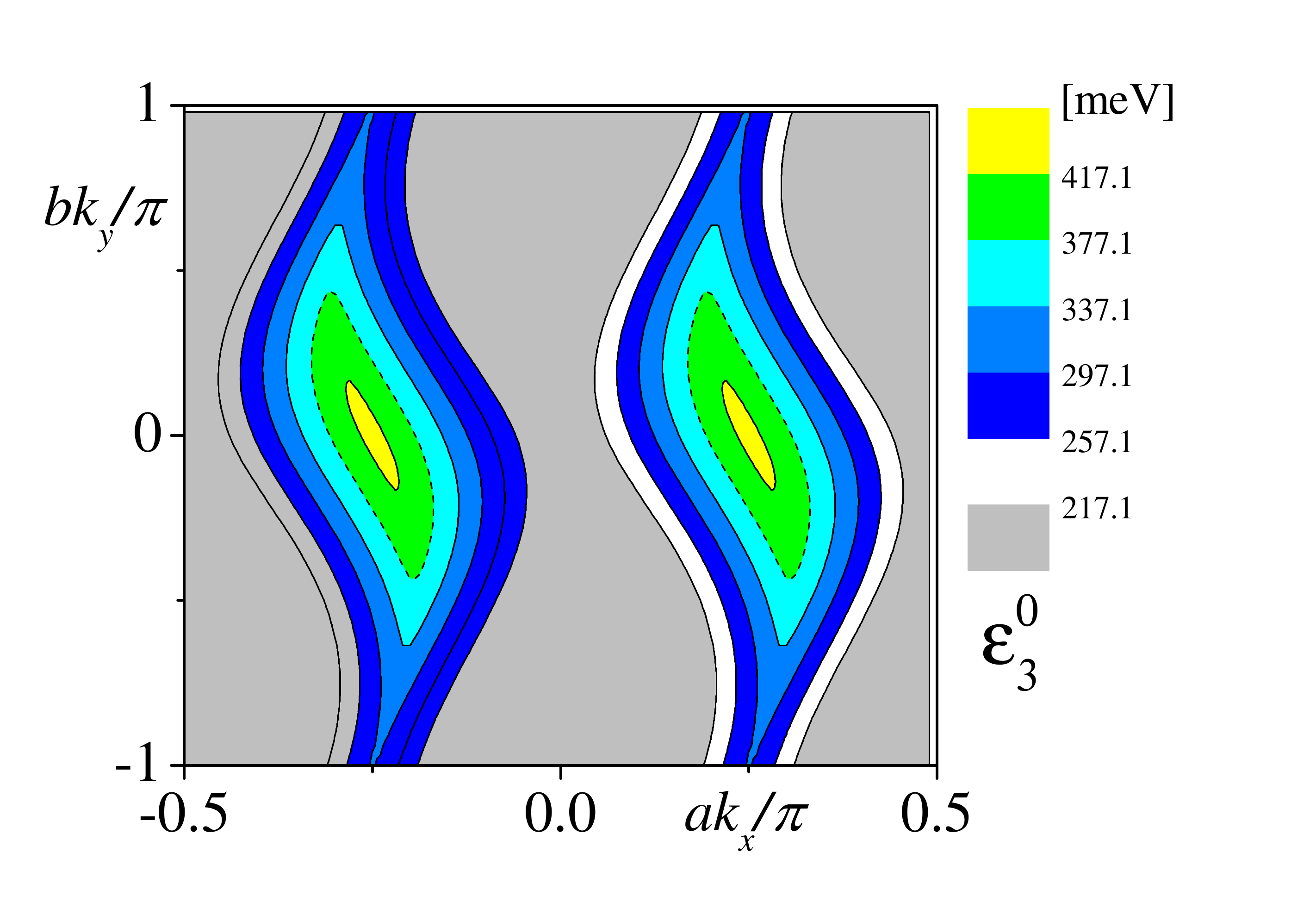}
\begin{flushleft} \hspace{0.5cm}(c) \end{flushleft}\vspace{-0.5cm}
\includegraphics[width=0.4\textwidth]{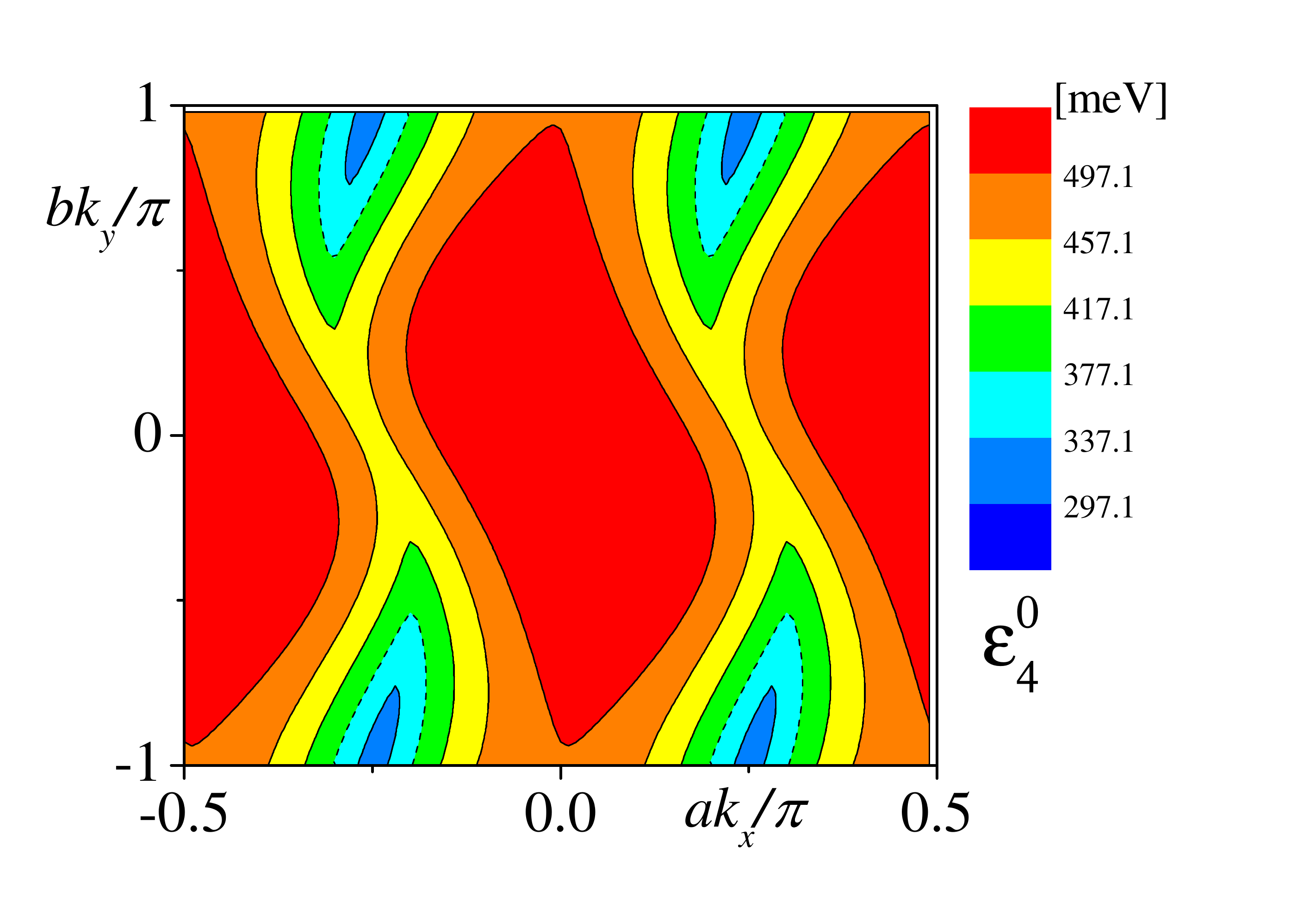}
\caption{(Color online) (a) The third and fourth energy bands near the Fermi energy 
($\varepsilon^0_{\rm F}\simeq 377.1$ meV) at 3/4-filling with
the same parameters as those in Fig.~\ref{Figure2} (b) ($V=12.38$ meV). 
In this case the energy gap at $(ak_x/\pi, bk_y/\pi)=(1/4, 1)$ is $2\Delta\simeq 17.80$~meV, 
the top energy of the third band is $\varepsilon_{\rm 3t}^0\simeq 425.3$~meV, 
and the bottom energy of the forth band is $\varepsilon_{\rm 4b}^0\simeq 317.8$ ~meV.
Contour plots of (b) the third energy band and (c) the fourth energy band. 
Dotted lines are for the Fermi surface at  3/4-filling. 
}
\label{Figure3}
\end{figure}

\begin{figure}[bt]
\begin{flushleft} \hspace{0.5cm}(a) \end{flushleft}\vspace{-0.5cm}
\includegraphics[width=0.3\textwidth]{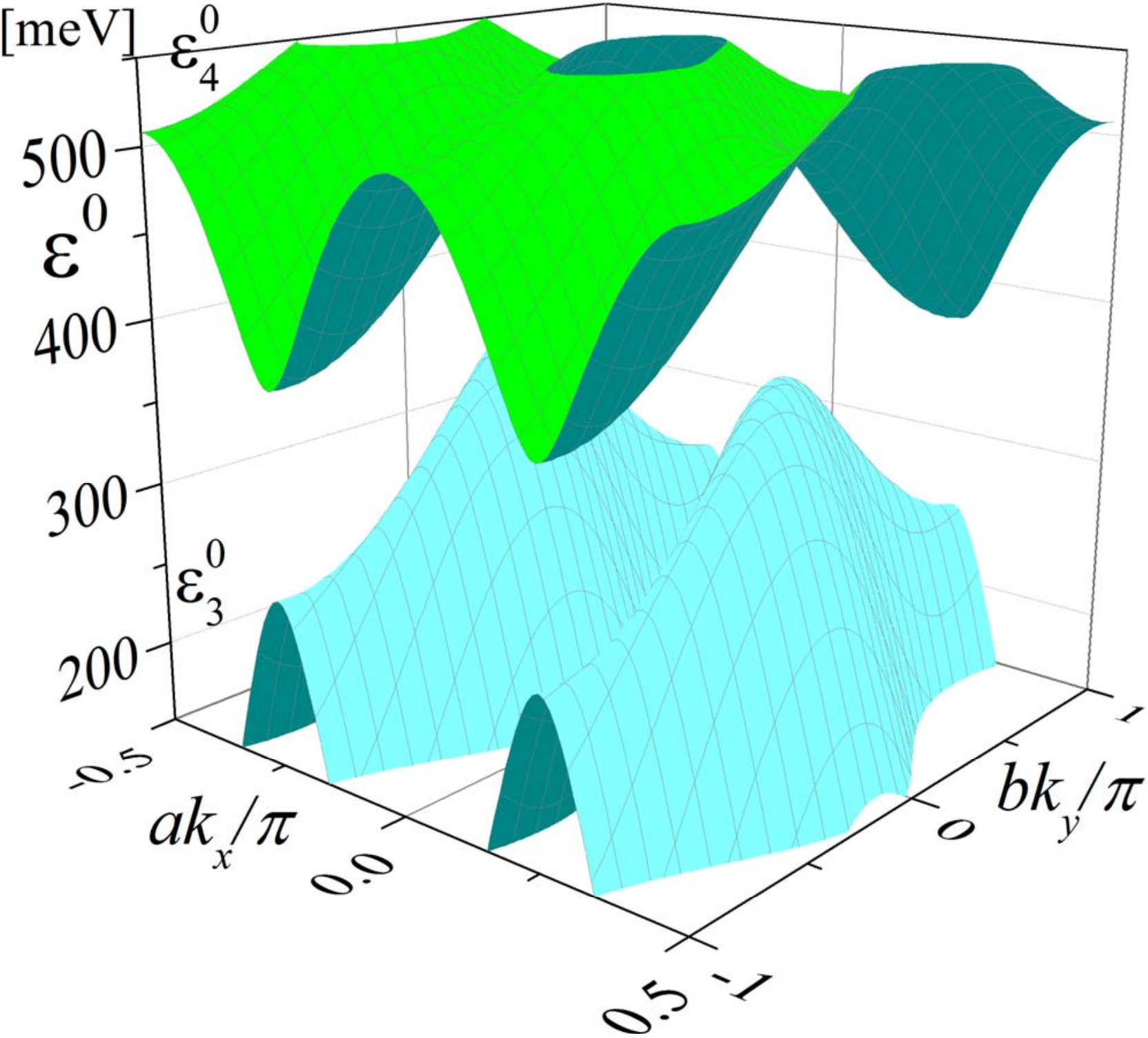}
\begin{flushleft} \hspace{0.5cm}(b) \end{flushleft}\vspace{-0.5cm}
\includegraphics[width=0.4\textwidth]{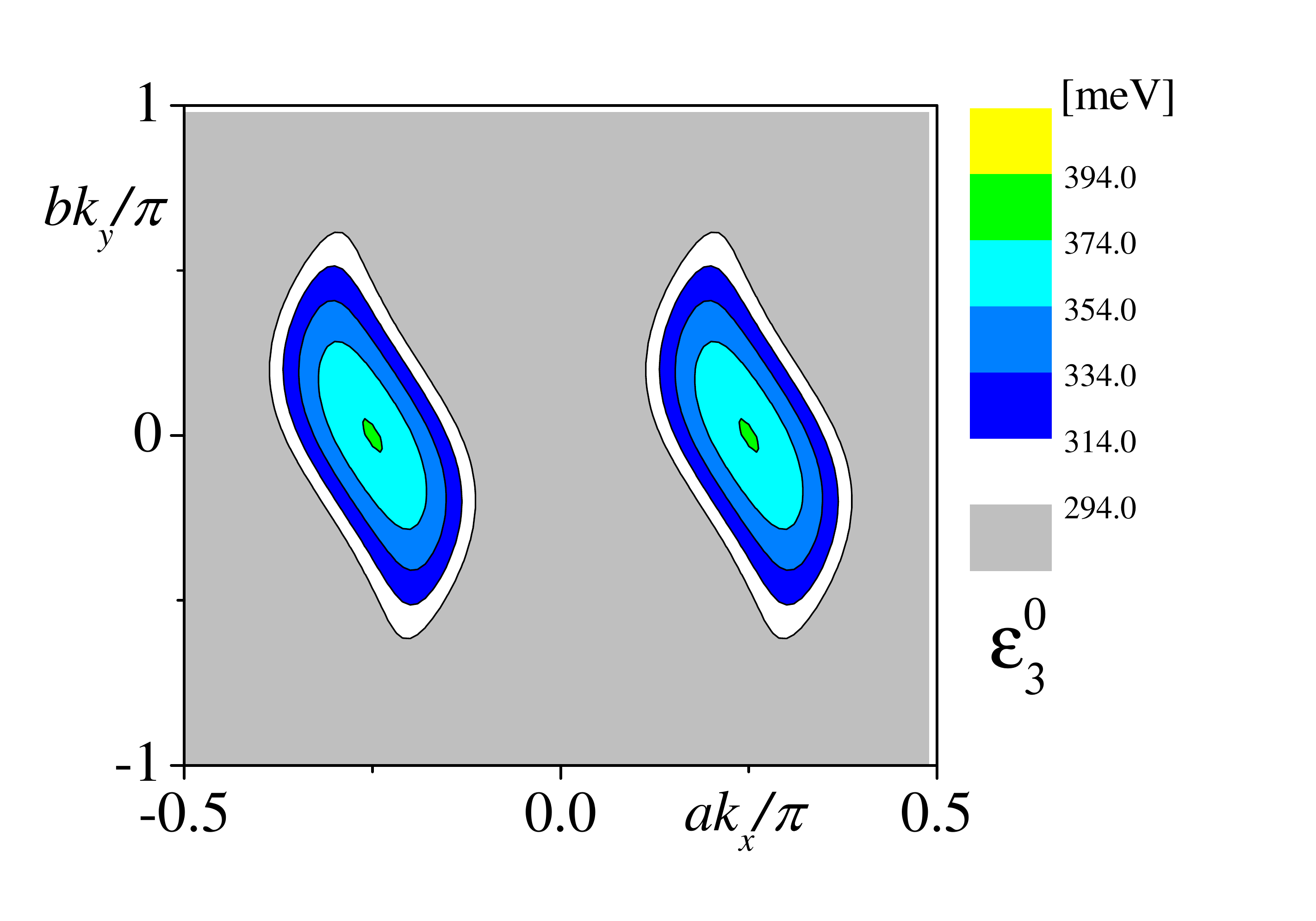}
\begin{flushleft} \hspace{0.5cm}(c) \end{flushleft}\vspace{-0.5cm}
\includegraphics[width=0.4\textwidth]{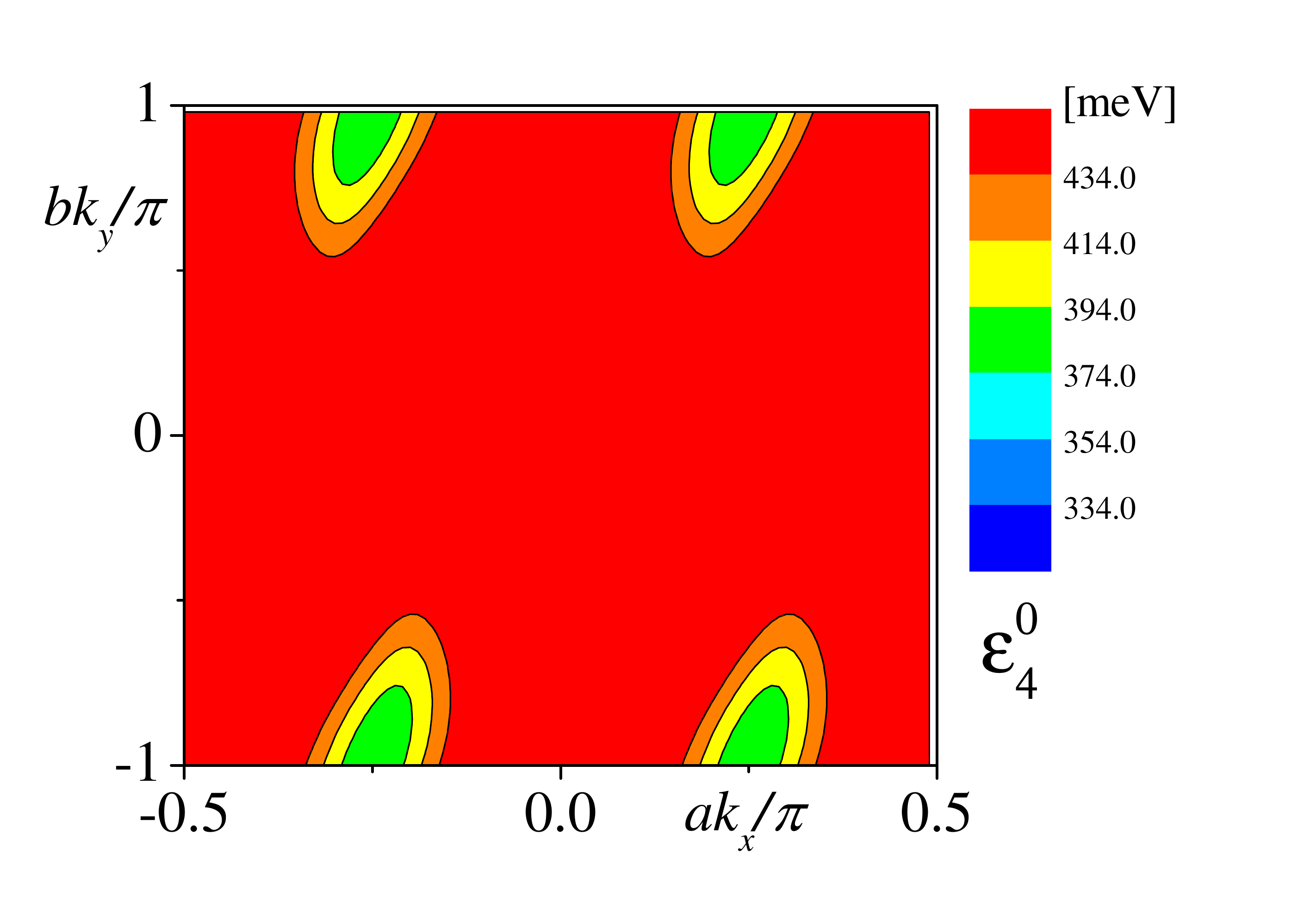}
\caption{(Color online) (a) The third and fourth energy bands near the Fermi energy 
($\varepsilon^0_{\rm F}\simeq 374.7$ meV) at 3/4-filling for $V=86.50$ meV. 
Other parameters are the same as those in Fig.~\ref{Figure2}. 
Contour plots of (b) the third energy band and (c) the fourth energy band. 
}
\label{Figure4}
\end{figure}

The effect of the ordering of the anion NO$_3$ is taken as the 
on-site potential $V$ and $-V$ as shown in Fig.~\ref{Figure1}(b). 
In this case, there are four sites (A, B, A$^{\prime}$ and B$^{\prime}$) in the 
unit cell which are indicated by a light green rectangle in 
Fig. ~\ref{Figure1}(b) and becomes twice larger than that without the anion ordering. 
The Brillouin zone is halved along $k_x$-direction. 
The energy is obtained by the diagonalization of Eq. (\ref{J3}) ($4\times 4$ matrix) in Appendix~\ref{appendixA}. 
The minimum gap made 
at $(ak_x/\pi, bk_y/\pi)=(1/4, 1)$ between a third band and a fourth band is obtained to be about 17.80 meV when we set $V=12.38$ meV. 
Alemany, Pouget and Canadell\cite{Pere} have obtained the minimum gap between the third band and the fourth band to be 17.8 meV. 
Therefore, we take $\pm V= \pm 12.38$~meV as the on-site potential 
 of (TMTSF)$_2$NO$_3$ at $T<T_{\textrm{AO}}$. We show the Fermi surface in Fig.~\ref{Figure2}(b) in the extended zone scheme, where 
there exist electron and hole pockets with the same area. When $V$ becomes large, the areas of electron and hole pockets become small.
In Figs.~\ref{Figure3} and \ref{Figure4} we show the 3D plots and the contour plots of the third band and the fourth band 
as a function of the wave number $\mathbf{k}$ for $V=12.38$ meV and $86.50$ meV, respectively.
When $V \geq 86.50$ meV, the areas of electron and hole pockets are zero. 




\begin{figure}[bt]
\begin{flushleft} \hspace{0.0cm}(a) \end{flushleft}\vspace{-0.5cm}
\includegraphics[width=0.35\textwidth]{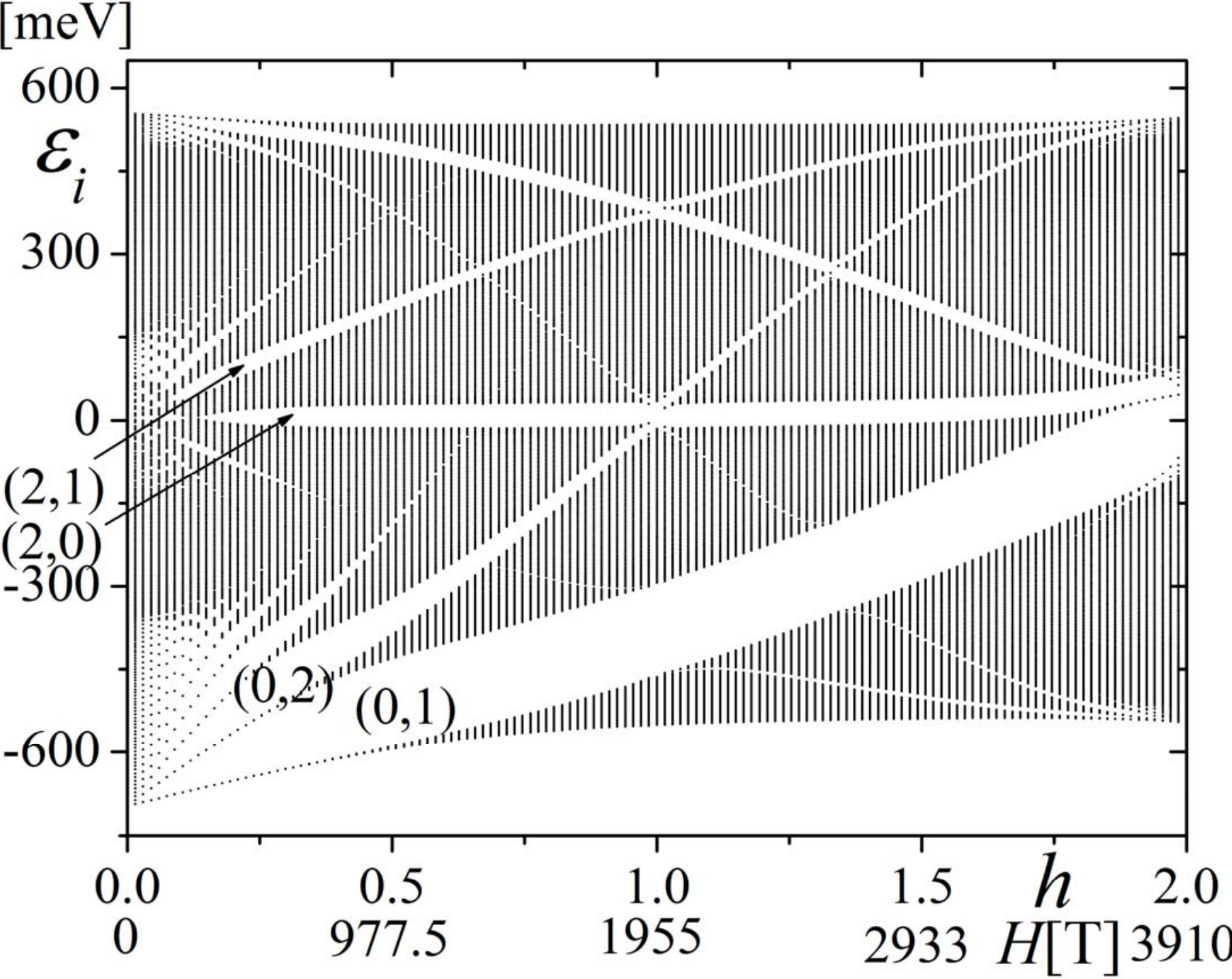}\vspace{-0.0cm}
\begin{flushleft} \hspace{0.0cm}(b) \end{flushleft}\vspace{-0.5cm}
\includegraphics[width=0.35\textwidth]{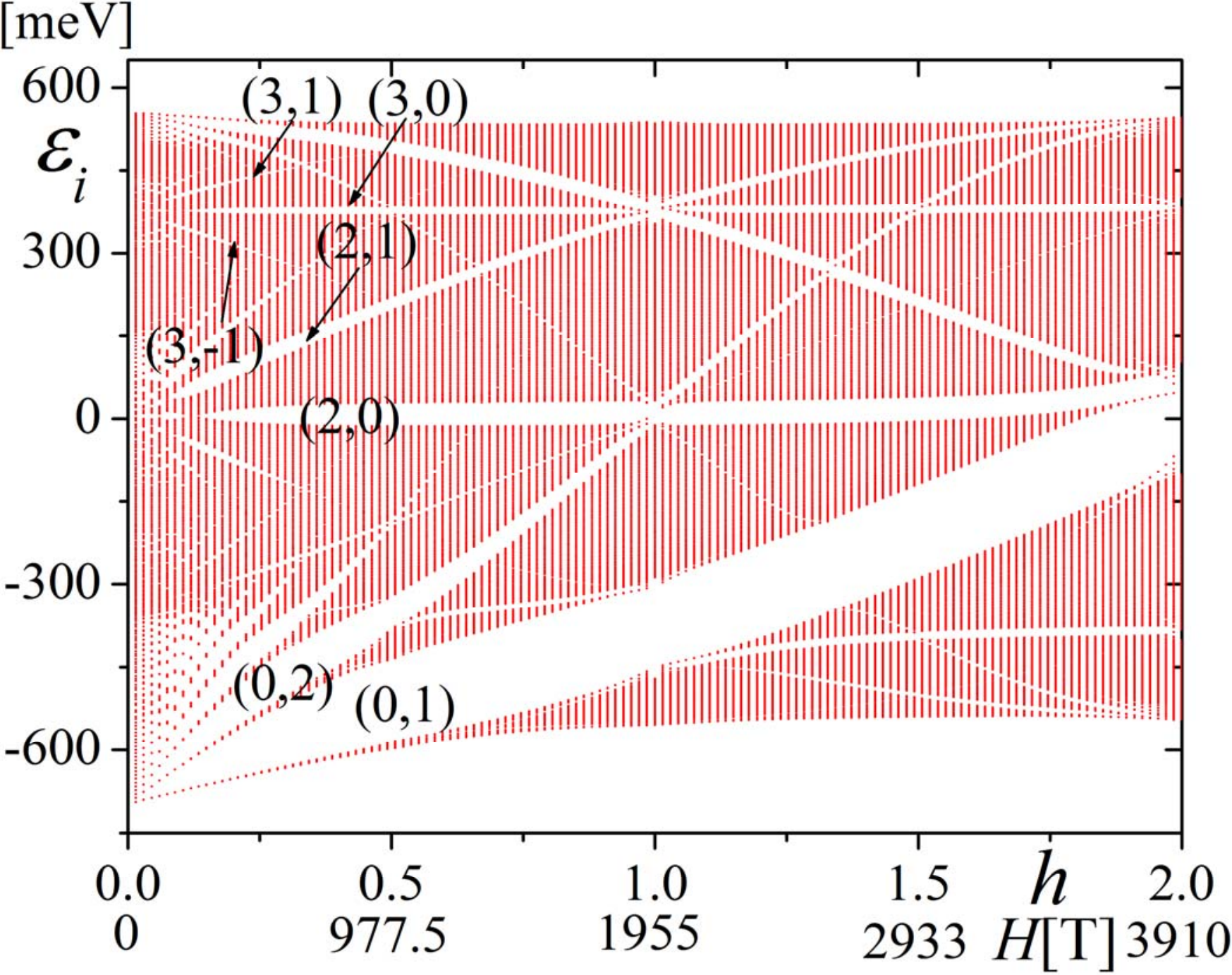}\vspace{-0.0cm}
\begin{flushleft} \hspace{0.0cm}(c)\end{flushleft}\vspace{-0.5cm}
\includegraphics[width=0.35\textwidth]{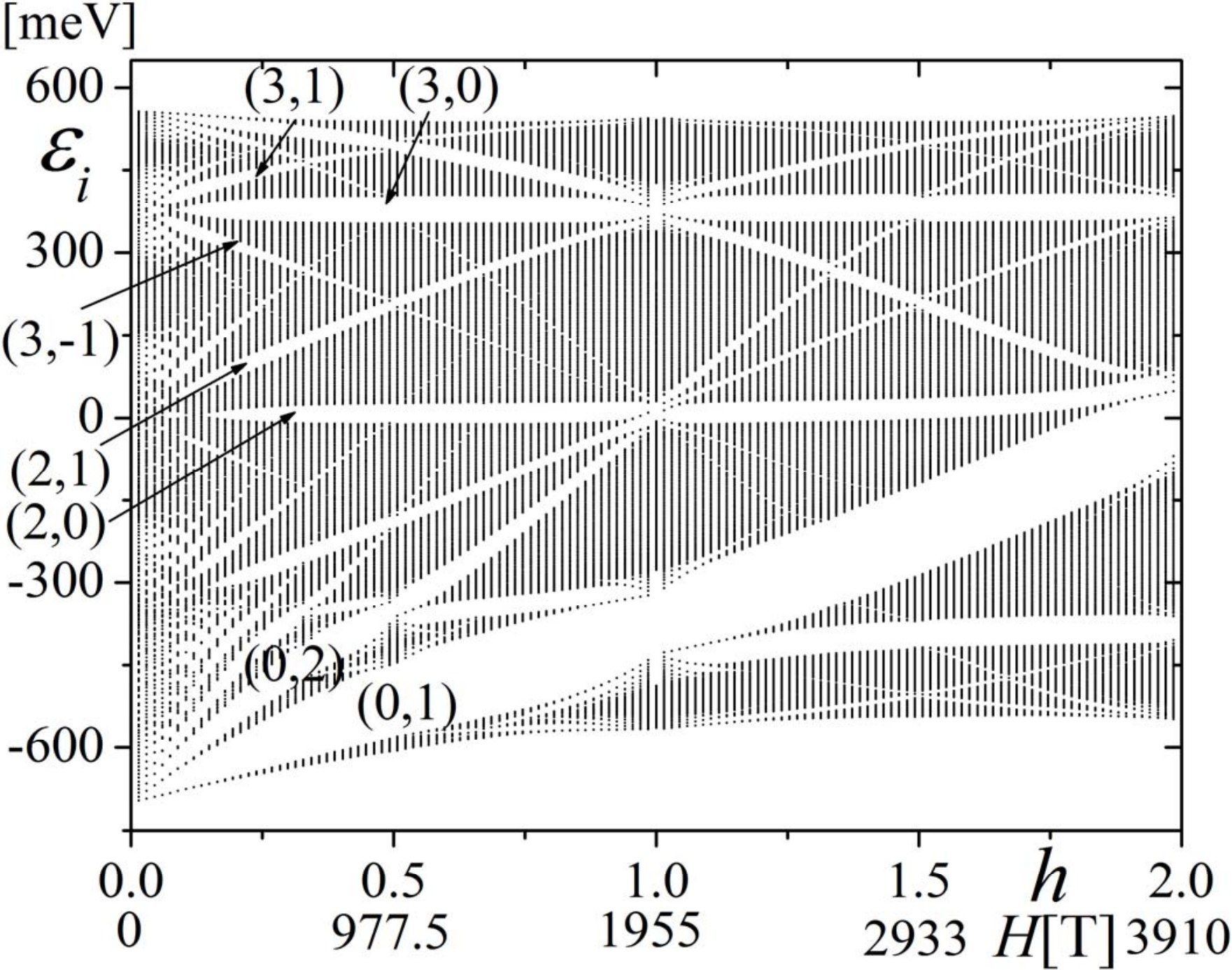}\vspace{-0.0cm}
\caption{(Color online) 
Energy as a function of $h$. (a) Parameters are the same as those in Fig.~\ref{Figure2}~(a) ($V=0$).
(b)  Parameters are the same as those in Fig.~\ref{Figure2}~(b) and Fig.~\ref{Figure3} ($V=12.38$~meV).
 (c) $V=37.14$~meV and other parameters are the same as those in Fig.~\ref{Figure2}. 
 We take $h=p/q$ with 
$q=67$  and $p=1, 2, 3, \cdots, 2q$.  
Wave numbers are taken as $(k_x, k_y)=(0, 0)$, $(\pi/(2 aq), 0)$ , and $(2\pi/(2aq), 0)$. 
Other parameters are the same as those in Fig.~\ref{Figure2}. 
Quantum numbers $(s_r, t_r)$ for some gaps are shown.
}
\label{Figure5}
\end{figure}

%
\begin{figure}[bt]
\begin{flushleft} \hspace{0.5cm}
\end{flushleft}\vspace{-0.5cm}
\includegraphics[width=0.35\textwidth]{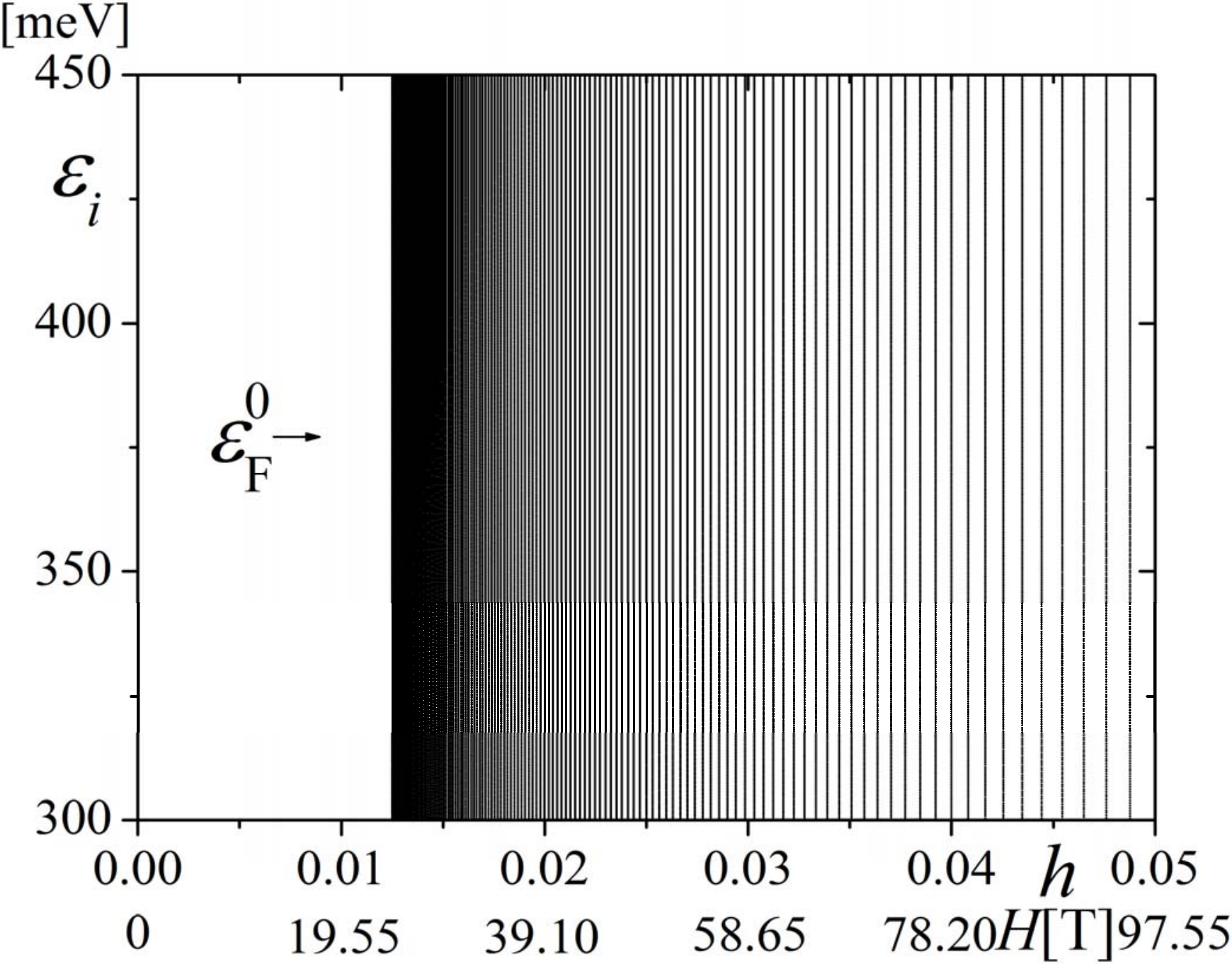}\vspace{-0.0cm}
\caption{(Color online) 
Energy as a function of $h$ for $V=0$. (A close up figure of Fig.~\ref{Figure5} (a) 
at $h \approx 0$ and  $\epsilon_i \approx \epsilon_{\rm F}^0 \simeq 377.0$~meV which is the Fermi energy
at $h=0$ for $3/4$ filled case). 
We take $h=1/q$ with $20 \leq q \leq 80$ and $h=2/(2m+1)$ with $20 \leq m \leq 79$, where the wave number $(k_x, k_y) = (n_x \pi/(30a),0)$ with $0 \leq n_x \leq 30$. 
The energy gaps cannot be seen in this scale.
}
\label{Figure6}
\end{figure} 
%
\section{Quantum Hall effect and Landau quantization 
}

The energy of tight-binding electrons in the uniform magnetic field is obtained by 
taking the phase factor in the hoppings as shown in Appendix \ref{refappB}.
The energy can be calculated only when the magnetic flux ($\Phi$) through the area of the unit cell $(4ab$) is a rational number $p/q$ in the unit of the flux quantum ($\phi_0$), where $p$ and $q$ are mutually prime numbers. Thus, 
we define $h$ as
\begin{equation}
 h = \frac{\Phi}{\phi_0} = \frac{4 a b H}{\phi_0},
\end{equation}
and we take $h$ as a rational number, 
\begin{equation}
 h=\frac{p}{q}. 
\end{equation}
The value of the flux quantum is $\phi_0=2\pi\hbar c /e\simeq 4.14\times 10^{-15}$ Tm$^2$, where $2\pi\hbar$, $c$ and $e$ are the Planck constant, the speed of light and the absolute value of electron charge, respectively. 
Since $a\simeq 7.02$~{\AA}   and $b\simeq 7.54$~{\AA}   in (TMTSF)$_2$NO$_3$\cite{barrans}, 
$h=1$ corresponds to about $H=1955$~T. 


In the presence of a weak periodic potential\cite{Harper,Harper2}, each Landau level is broadened (which is called Harper broadening) and separates into 
$p$ bands when the magnetic flux $\Phi$ through the unit cell is $\Phi=(p/q)\phi_0$. 
On the other hand, the electron energy becomes $q$ bands when the magnetic flux $\Phi$ is applied to the tight-binding electrons with one site and one orbit in the unit cell.

When the chemical potential is in the $r$th gap from the bottom 
in the tight-binding model, the quantized Hall conductance is given as
\begin{equation}
 \sigma_{xy} = \frac{e^2}{h} t_r,
\end{equation}
where the integer $t_r$ is given by the Diophantine equation,\cite{TKNN,Kohmoto_1985,Kohmoto_1989}
\begin{equation}
 r = q s_r + p t_r.
 \label{sr}
\end{equation}

Two cases (the weak potential case and the tight-binding electrons) are reconciled,
when $p/q \ll 1$ and electron filling is small in the tight-binding model on the rectangular (or triangular, honeycomb etc.)  lattice; 
every set of $p$ bands from the bottom of the energy
is considered as a broadened Landau level, i.e. each Landau level is separated into $p$ bands. 
The energy gaps above the $(p t_r)$th band from the bottom are larger than other  energy gaps. 
When chemical potential is in the $(p t_r)$th gap from the bottom, $s_r=0$.   
The Hall conductance given by $t_r$ is understood as the result of the $t_r$ Landau levels, each of which is broadened and separates into $p$ sub-bands.
The smaller $(p-1)$ gaps are considered as the gaps between the $p$ bands within the $t_r$th Landau level, as in the weak potential case.
In this way, the trivial value of quantum Hall effect ($s_r$=0) can be understood in the Landau levels for free electrons,
\begin{equation}
 \varepsilon_n^{\mathrm{parabola}} \propto ( n+\gamma^{\mathrm{parabola}}) h,
\label{eqLandauQ}
\end{equation}
where $n$ is zero or positive integer and the phase is $\gamma^{\mathrm{parabola}}=1/2$.
The Landau quantization of the energy levels (Eq.~(\ref{eqLandauQ})) is
obtained in the approximation that the energy dispersion near the bottom of the band at $h=0$ is 
treated as that of free electrons, i.e. parabolic ($\varepsilon_{\mathbf{k}}^{(h=0)} \propto (k_x^2 + k_y^2)$). 
The Landau levels are obtained by the condition that the area of the Fermi surface at $h=0$ is quantized\cite{onsager} to be proportional to  $( n+\gamma) h$. 
We call this quantization as the semi-classical Landau quantization. 

In order to observe the 
non-trivial values of Hall conductance ($s_r \neq 0$) in the tight-binding electrons on rectangular and triangular lattices, 
very strong magnetic field (the flux through the unit cell should be the same order as the flux quantum) is required. 
In the honeycomb lattice, which has two sites in the unit cell and there are $2q$ bands, the gaps labeled by $s_r=1$ are also large at small magnetic field near half-filling. \cite{HK2006} 
The quantum Hall effect with $s_r=1$ is observed
in graphene when electrons or holes are doped\cite{Novo2005}. 
The quantum Hall effect in graphene with $s_r=1$ can be also understood semi-classically, if we approximate the energy dispersion near the 
massless Dirac points ($ \pm \mathbf{k}^0 = \pm (k_x^0, k_y^0 )$) at $h=0$ as 
\begin{equation}
 \varepsilon_{\mathbf{k}}^{(h=0)} \propto \pm \sqrt{(k_x - k_x^0) ^2 + (k_y - k_y^0)^2}
\end{equation}
and adopt the semi-classical quantization\cite{onsager} of the area of the Fermi surface;
\begin{equation}
 \varepsilon_n^{\mathrm{Dirac}} \propto \mathrm{sign}(n) \sqrt{(\lvert  n \rvert +\gamma^{\mathrm{Dirac}} )h} ,
\end{equation}
where $n$ is integer and $\gamma^{\mathrm{Dirac}}=0$. 
In the semi-classical treatment of Landau quantization, the broadening of the Landau levels and 
a rich structure of the Hofstadter butterfly diagram do not appear. In a real system of graphene, a very strong magnetic field is 
necessary to observe the quantum Hall effect for $s_r \neq 1$. 
However, when the area of the unit cell is large, the rich structure of the Hofstadter butterfly diagram can be observed experimentally at the accessible magnetic field. Indeed, the moire pattern in twisted bilayer graphene or graphene on the hexagonal boron nitride (h-BN) substrates\cite{Dean}, 
graphene anti-dot lattice\cite{Pedersen}, cold atoms in optical lattice\cite{aidel,miyake}, etc. 
 are shown to have a Hofstadter butterfly diagram with various values of $s_r$ and $t_r$.

\begin{figure}[bt]
\begin{flushleft} \hspace{0.0cm}(a) \end{flushleft}\vspace{-0.5cm}
\hspace{0.2cm}\includegraphics[width=0.35\textwidth]{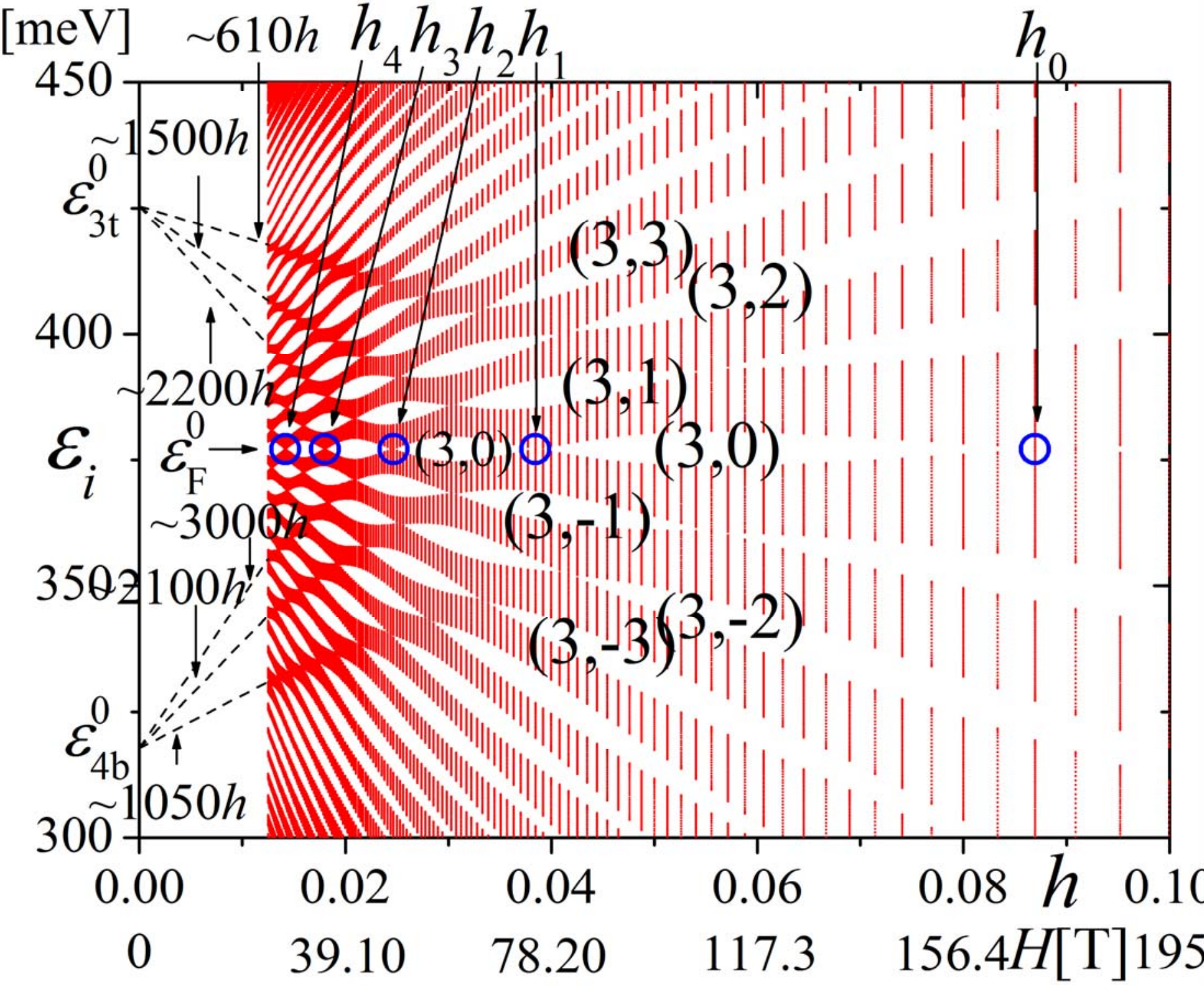}\vspace{-0.0cm}
\begin{flushleft} \hspace{0.0cm}(b) \end{flushleft}\vspace{-0.5cm}
\hspace{0.2cm}\includegraphics[width=0.35\textwidth]{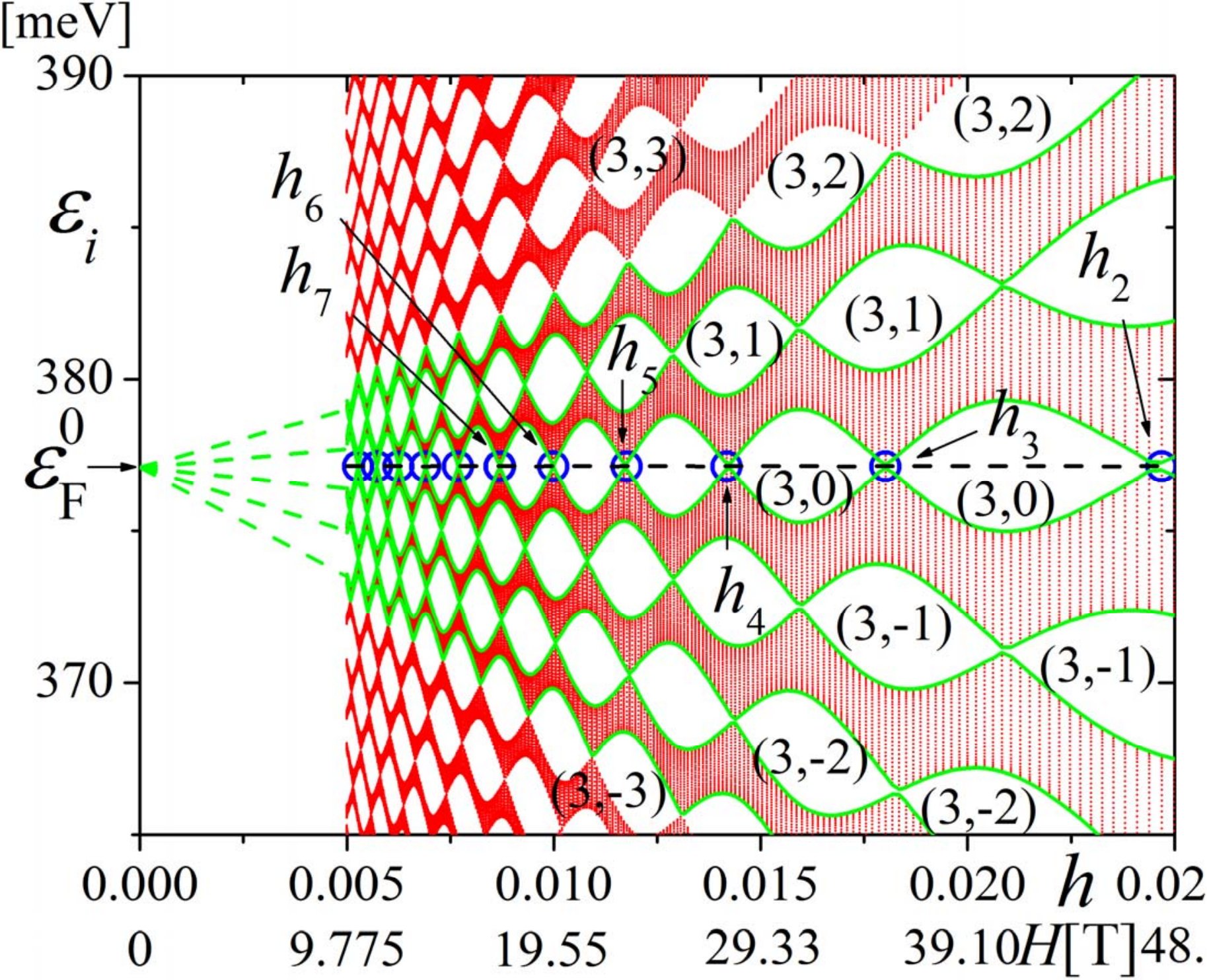}\vspace{-0.0cm}
\begin{flushleft} \hspace{0.0cm}(c) \end{flushleft}\vspace{-0.5cm}
\hspace{0.5cm}\includegraphics[width=0.45\textwidth]{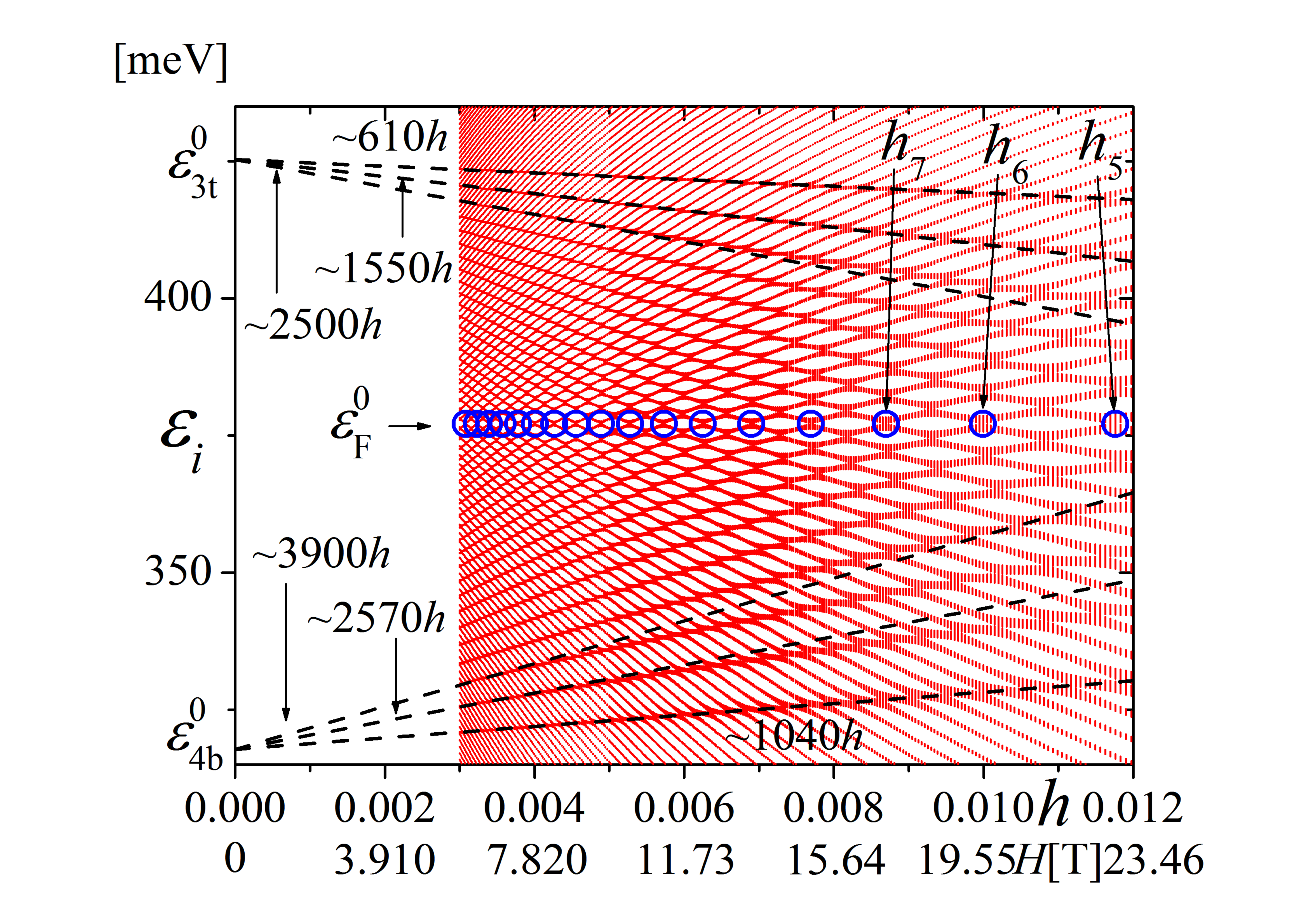}\vspace{0.2cm}
\caption{(Color online) 
(a) Energy as a function of $h$ with the same parameters as those in Fig.~\ref{Figure2} (b), Fig.~\ref{Figure3}, and Fig.~\ref{Figure5}~(b)
($V=12.38$~meV), where $\varepsilon^0_{\rm F}\simeq 377.1$ meV. 
We take $h=1/q$ with $10 \leq q \leq 80$ and $h=2/(2m+1)$ with $10 \leq m \leq 79$.
 (b) An enlarged figure of (a). We take $h=1/q$ with $40 \leq q \leq 200$ and $h=2/(2m+1)$ with $40 \leq m \leq 199$. A dotted black line is the chemical potential as a function of $h$. 
(c) A figure for smaller $h$. The parameters are the same as those of (a) and (b). We take $h=1/q$ with $84 \leq q \leq 333$ and $h=2/(2m+1)$ with  $84 \leq m \leq 332$. 
We take the wave number $(k_x, k_y) = (n_x \pi/(18a),0)$ with $0 \leq n_x \leq 18$ for $q>200$, $(k_x, k_y) = (n_x \pi/(30a),0)$ with $0 \leq n_x \leq 30$ for $80<q\leq 200$ 
and $(k_x, k_y) = (n_x \pi/(61a),0)$ with $0 \leq n_x \leq 61$ for $q\leq 80$. 
}
\label{Figure7}
\end{figure} 
\begin{figure}[bt]
\begin{flushleft} \hspace{0.5cm}(a) \end{flushleft}\vspace{-0.5cm}
\includegraphics[width=0.5\textwidth]{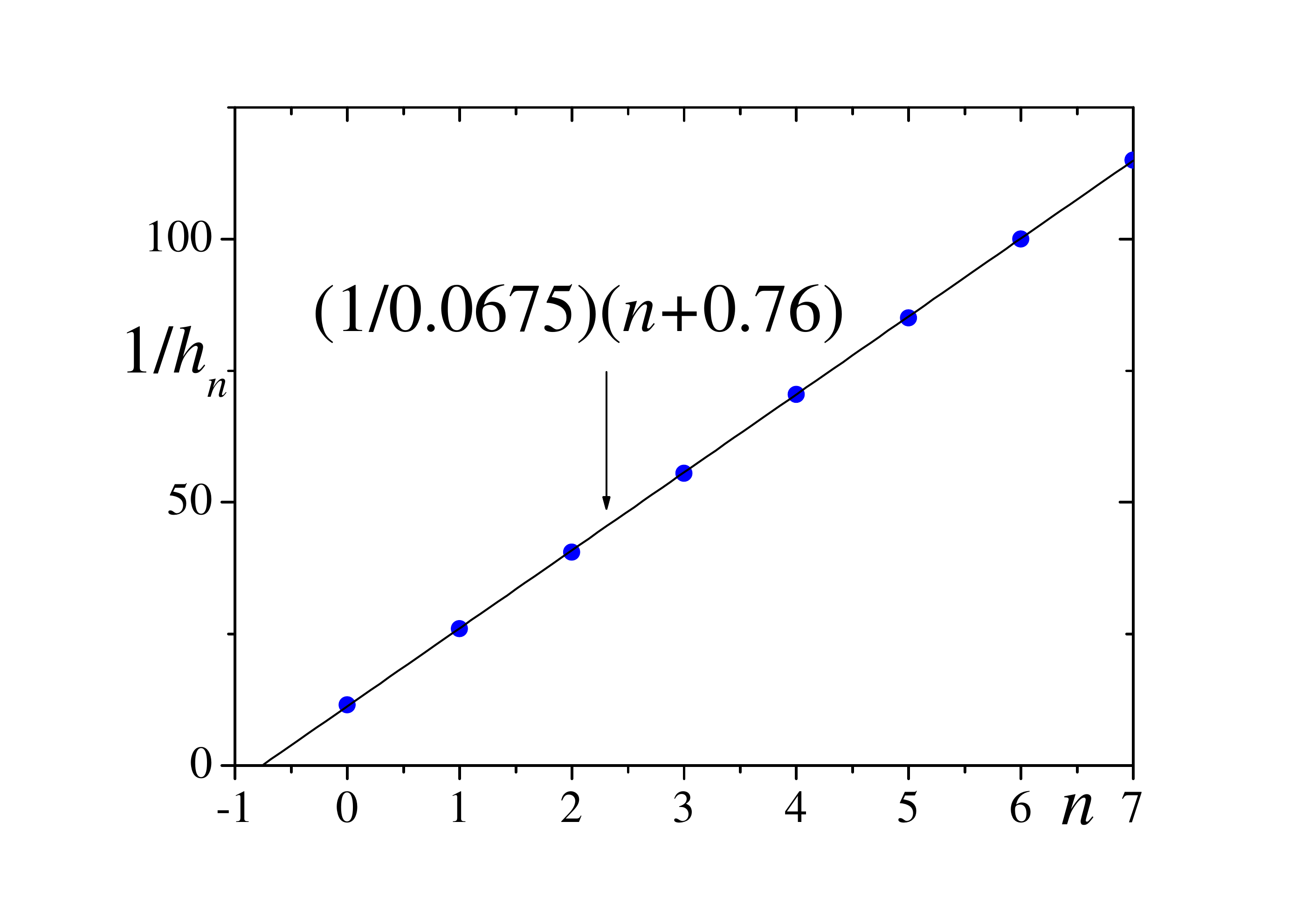}\vspace{-0.5cm} \\
\begin{flushleft} \hspace{0.5cm}(b) \end{flushleft}\vspace{-0.5cm}
\includegraphics[width=0.5\textwidth]{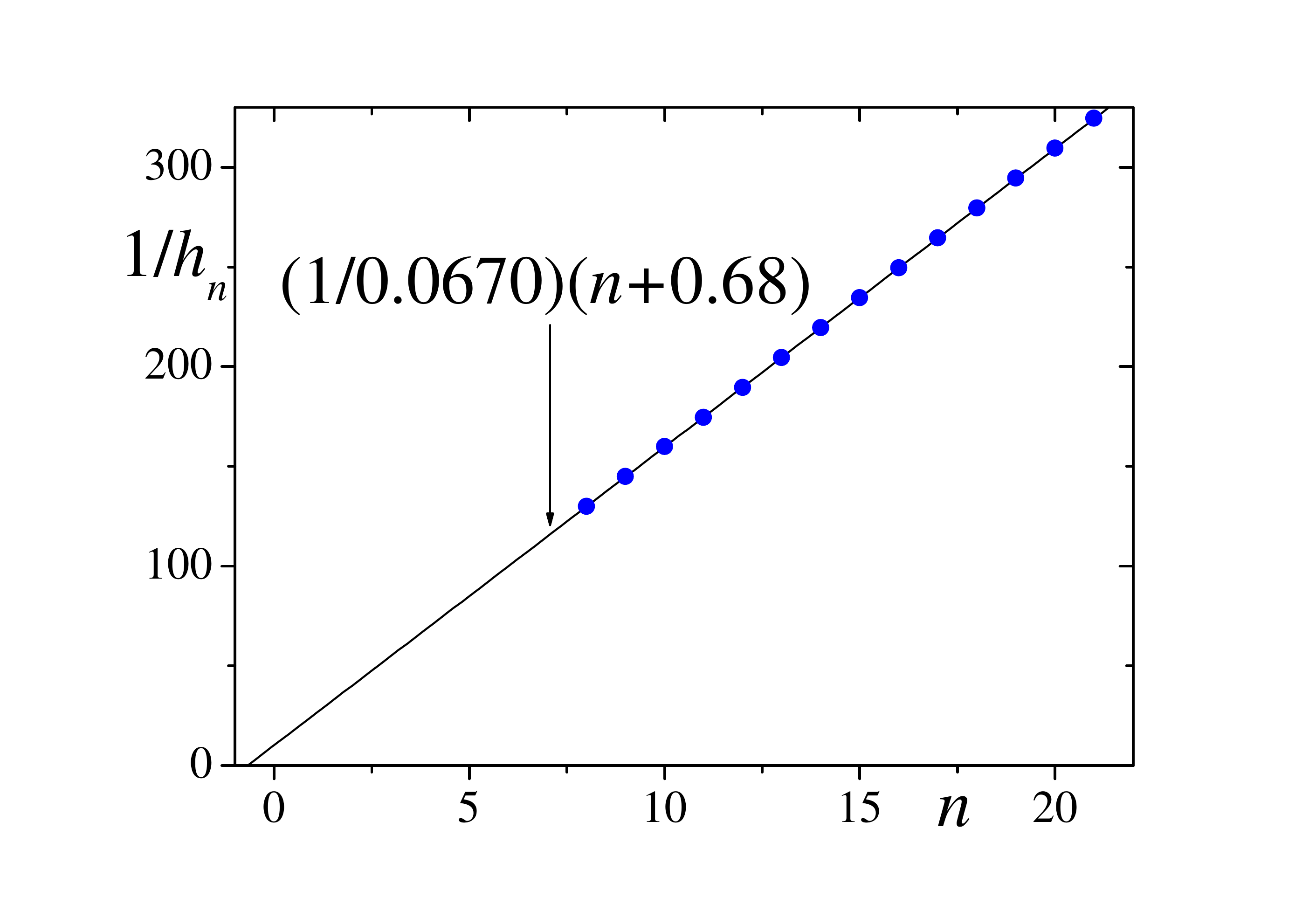}\vspace{-0.5cm}
\caption{(Color online) 
$1/h_n$ as a function of $n$ for (a) $0 \leq n \leq 7$ and (b) $8 \leq n \leq 21$.  
At the magnetic fields $h_n$ ($n=0,1,2, \cdots$),  the energy gap with the index $(3,0)$ is closed, as shown in Fig.~\ref{Figure7}.
}
\label{Figure8}
\end{figure}
\begin{figure}[bt]
\begin{flushleft} \hspace{0.5cm}(a)
\end{flushleft}\vspace{-0.5cm}
\includegraphics[width=0.41\textwidth]{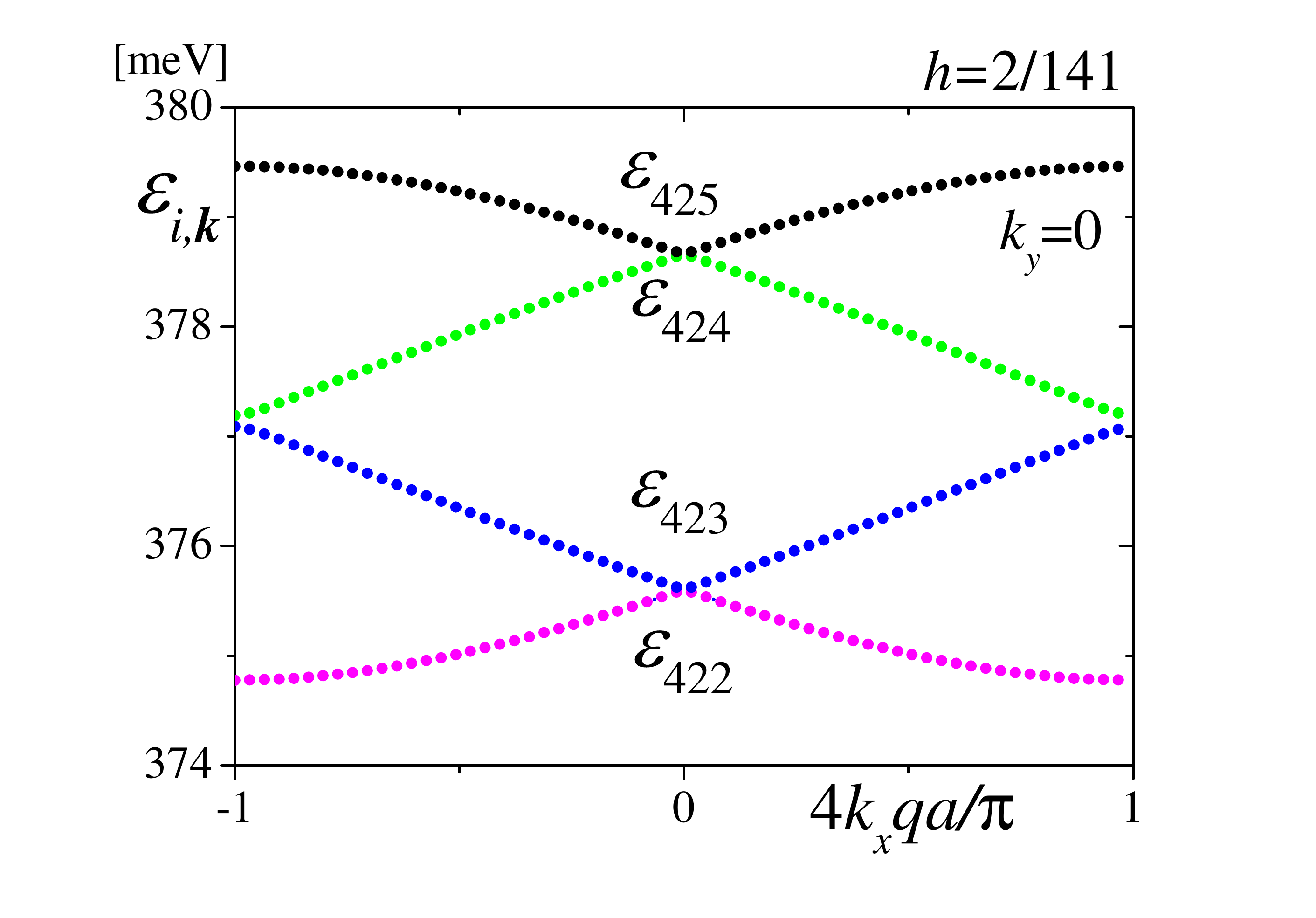}\vspace{-0.5cm}\\
\begin{flushleft} \hspace{0.5cm}(b) \end{flushleft}\vspace{-0.5cm}
\includegraphics[width=0.39\textwidth]{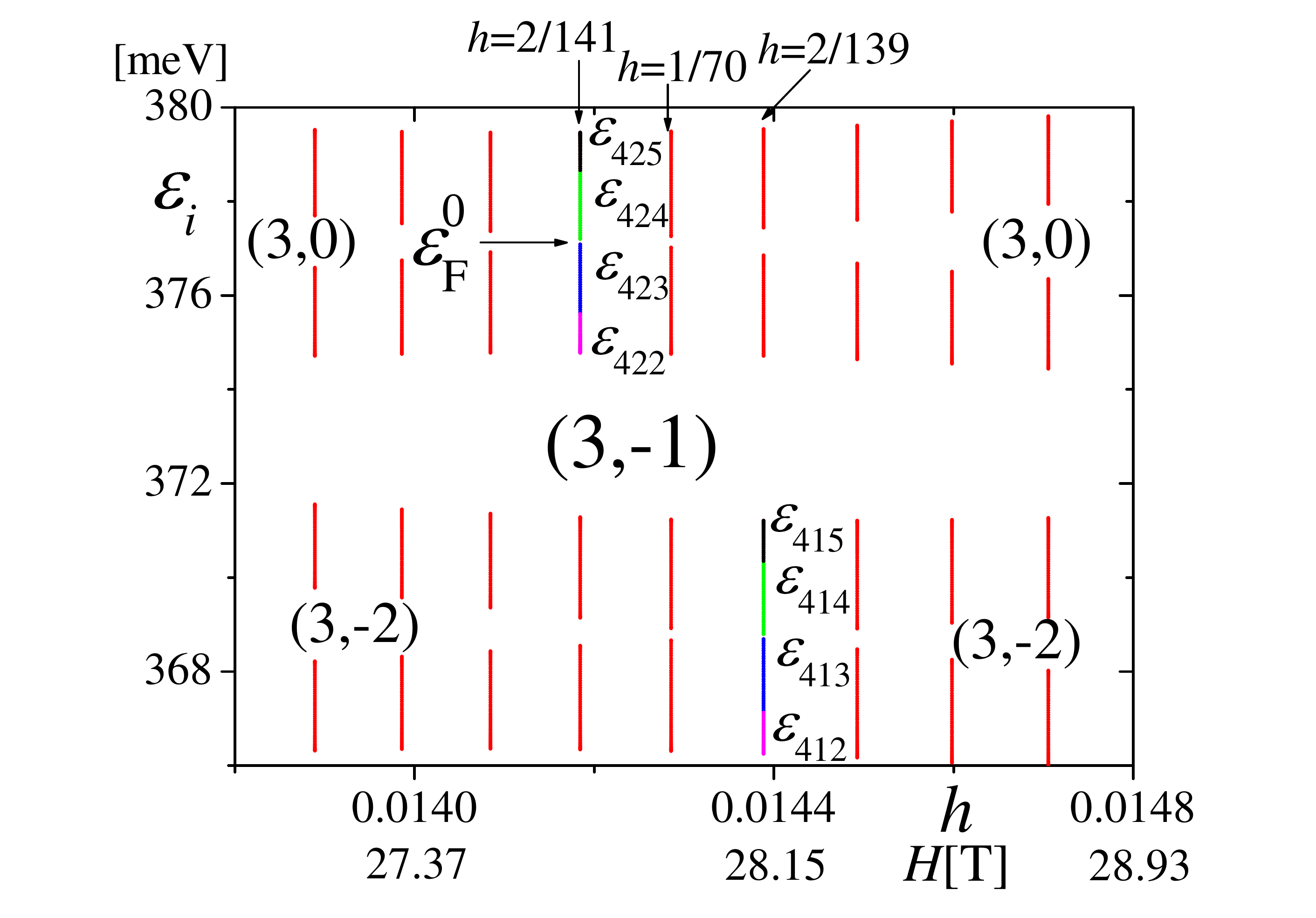}\vspace{-0.5cm}\\
\begin{flushleft} \hspace{0.5cm}(c)  \end{flushleft}\vspace{-0.5cm} 
\includegraphics[width=0.31\textwidth]{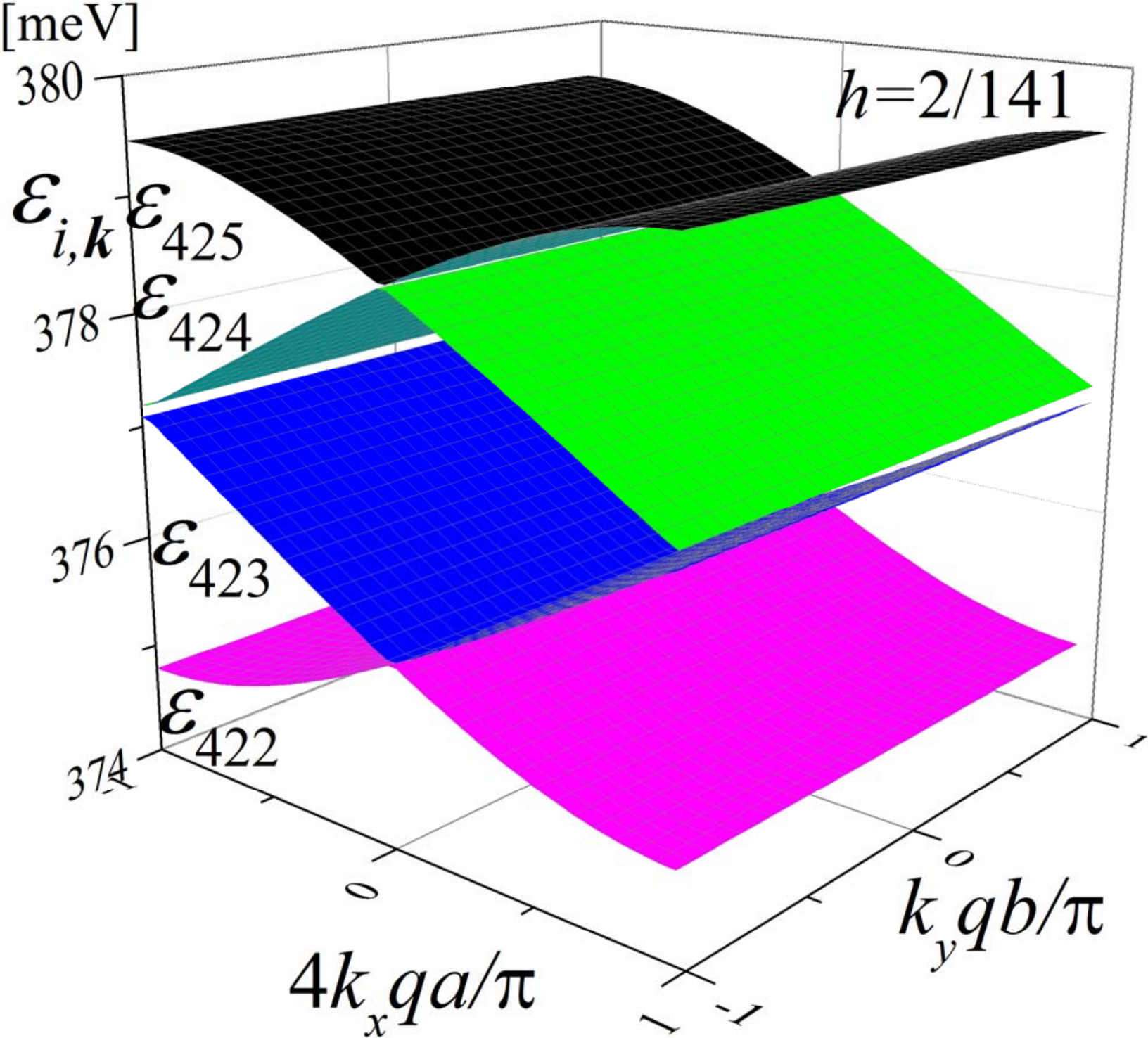}\vspace{-0.5cm} \\
\begin{flushleft} \hspace{0.5cm}(d) \end{flushleft}\vspace{-0.5cm}
\includegraphics[width=0.39\textwidth]{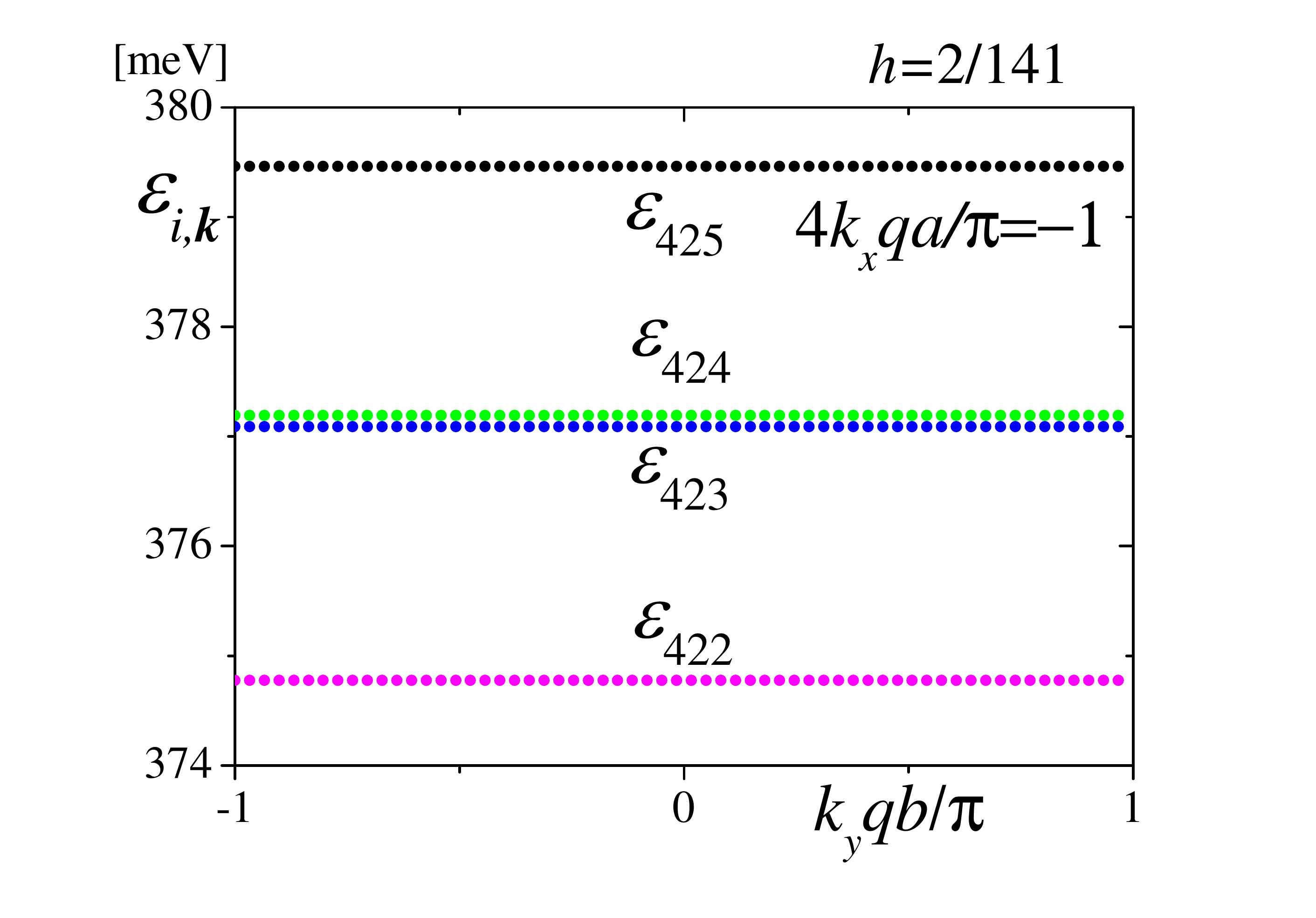}
\vspace{-0.5cm}
\caption{(Color online) 
(a) A close up figure of Fig.~\ref{Figure7} near $h_4$ ($V=12.38$ meV). 
We take $h = 1/72$, $2/143$, $1/71$, $2/141$, $1/70$, $2/139$, $1/69$, $2/137$, and $1/68$.
(b) Energies as a function of $k_x$ at $h=2/141$ and $k_y=0$. 
(c) 3D plot of the energies at $h=2/141$.
(d) There is a small gap between $\varepsilon_{423}$ and $\varepsilon_{424}$ at $k_x= \pm \pi/(4qa)$. 
The gap is almost independent of $k_y$. 
}
\label{Figure9}
\end{figure}

\section{Energy in the magnetic field 
}

\begin{figure}[bt]
\begin{flushleft} \hspace{0.0cm}(a) \end{flushleft}\vspace{-0.5cm}
\hspace{0.2cm}\includegraphics[width=0.35\textwidth]{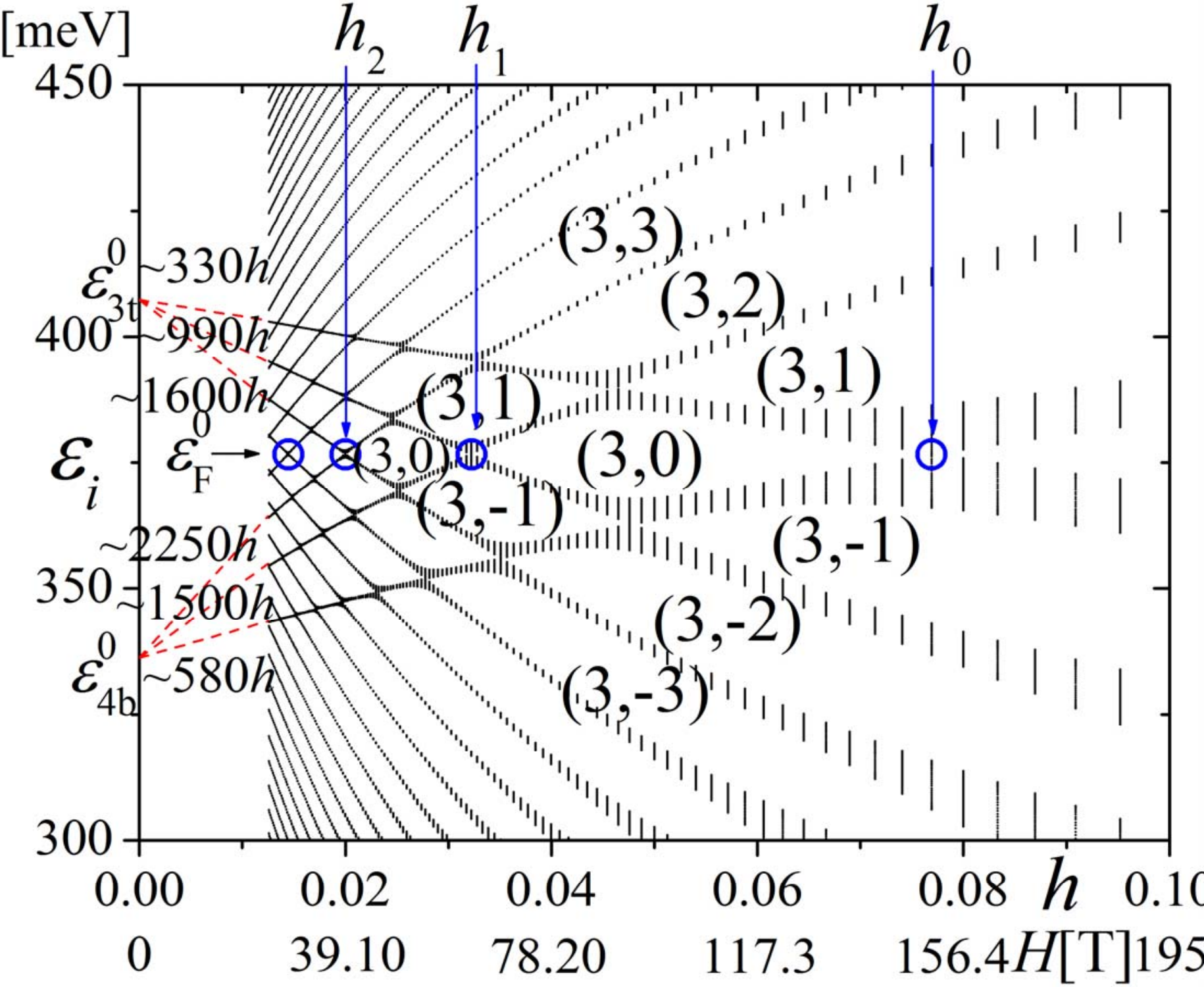}\vspace{-0.0cm}
\begin{flushleft} \hspace{0.0cm}(b) \end{flushleft}\vspace{-0.5cm}
\hspace{0.2cm}\includegraphics[width=0.35\textwidth]{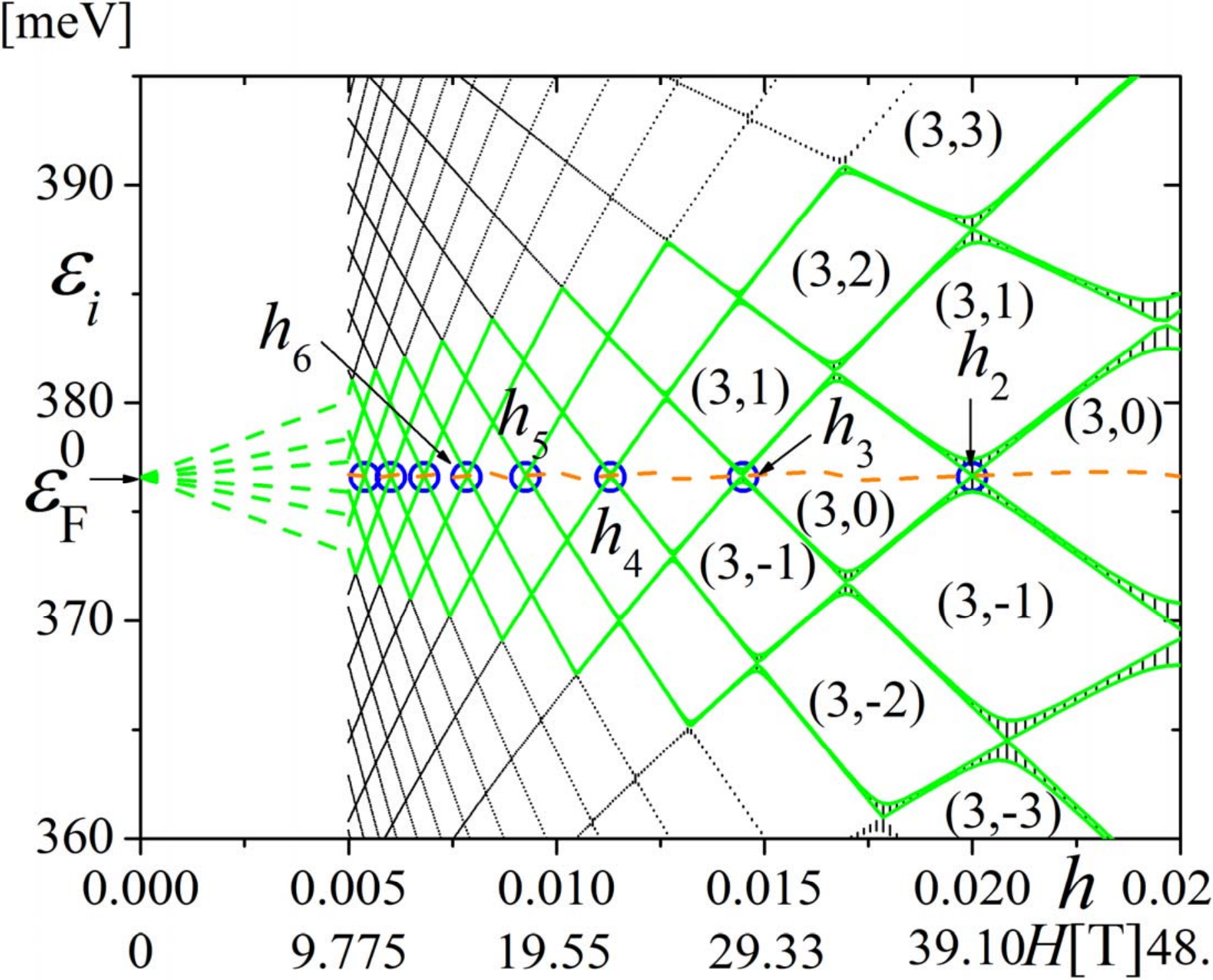}\vspace{-0.0cm}
\begin{flushleft} \hspace{0.0cm}(c) \end{flushleft}\vspace{-0.5cm}
\hspace{0.5cm}\includegraphics[width=0.45\textwidth]{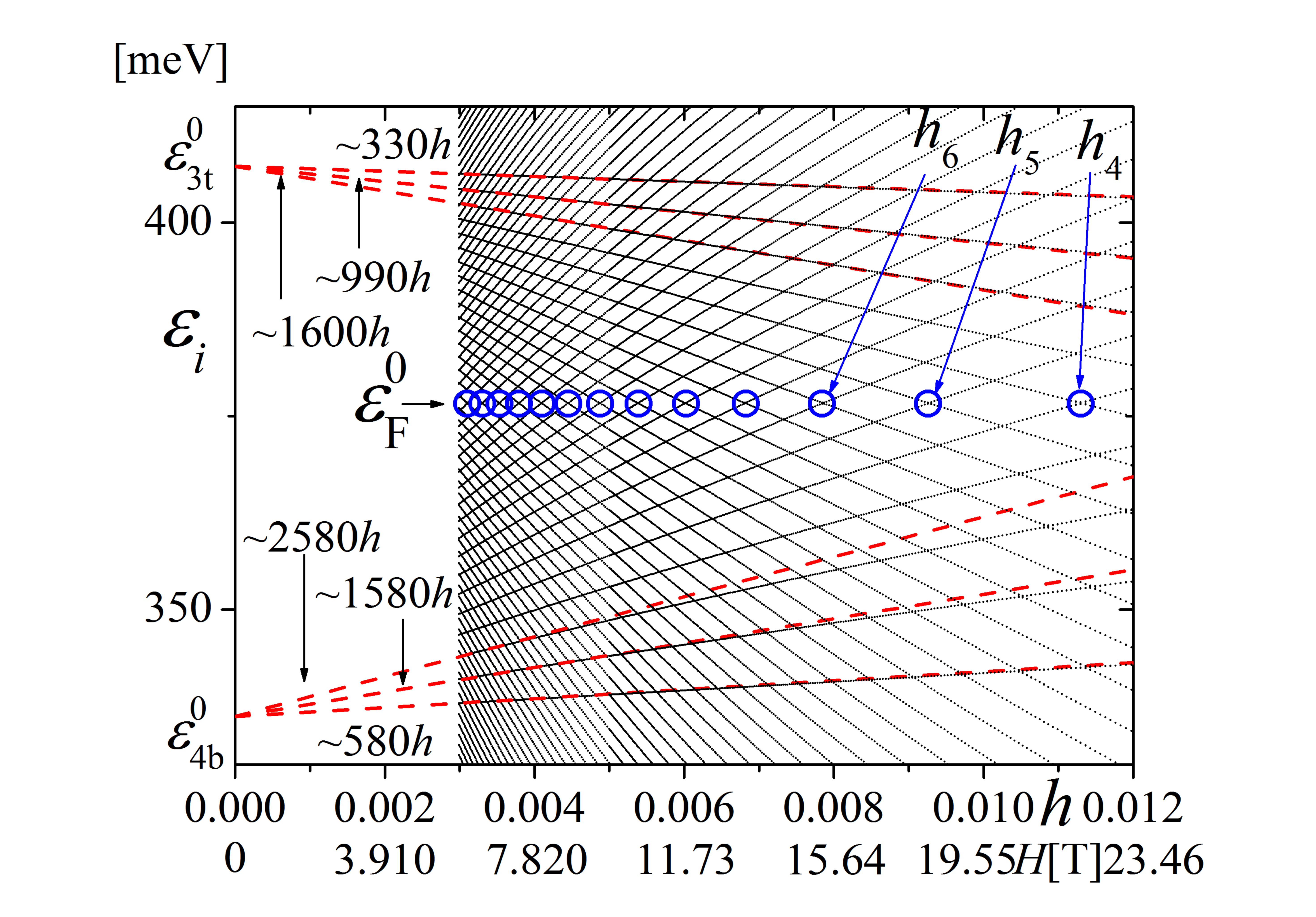}\vspace{-0.0cm}
\caption{(Color online) 
(a) Energy as a function of $h$ for $V=37.14$ meV. 
Other parameters are the same as those in Fig.~\ref{Figure2} (b). 
The direct band gap at $h=0$ is $2\Delta\simeq 53.32$~meV, 
the Fermi energy for the $3/4$ filled case is $\varepsilon^0_{\rm F}\simeq 376.6$ meV, 
the top energy of the third band is $\varepsilon_{\rm 3t}^0\simeq 407.3$~meV and 
the bottom energy of the forth band is $\varepsilon_{\rm 4b}^0\simeq 336.2$~meV. 
(b) An enlarged figure of (a). A dotted orange line is the chemical potential as a function of $h$. 
(c) A figure for smaller $h$. 
The parameters are the same as those of (a) and (b). 
In all figures, the values of $q$ and the wave number $(k_x, k_y)$ are the same as those of Figs. \ref{Figure7}. 
Small blue circles indicate the magnetic fields $h_0, h_1, h_2, \cdots$ at which the gaps indexed by $(3,0)$ are closed or almost closed. 
}
\label{Figure10}
\end{figure}

\begin{figure}[bt]
\begin{flushleft} \hspace{0.5cm}(a) \end{flushleft}\vspace{-0.5cm}
\includegraphics[width=0.5\textwidth]{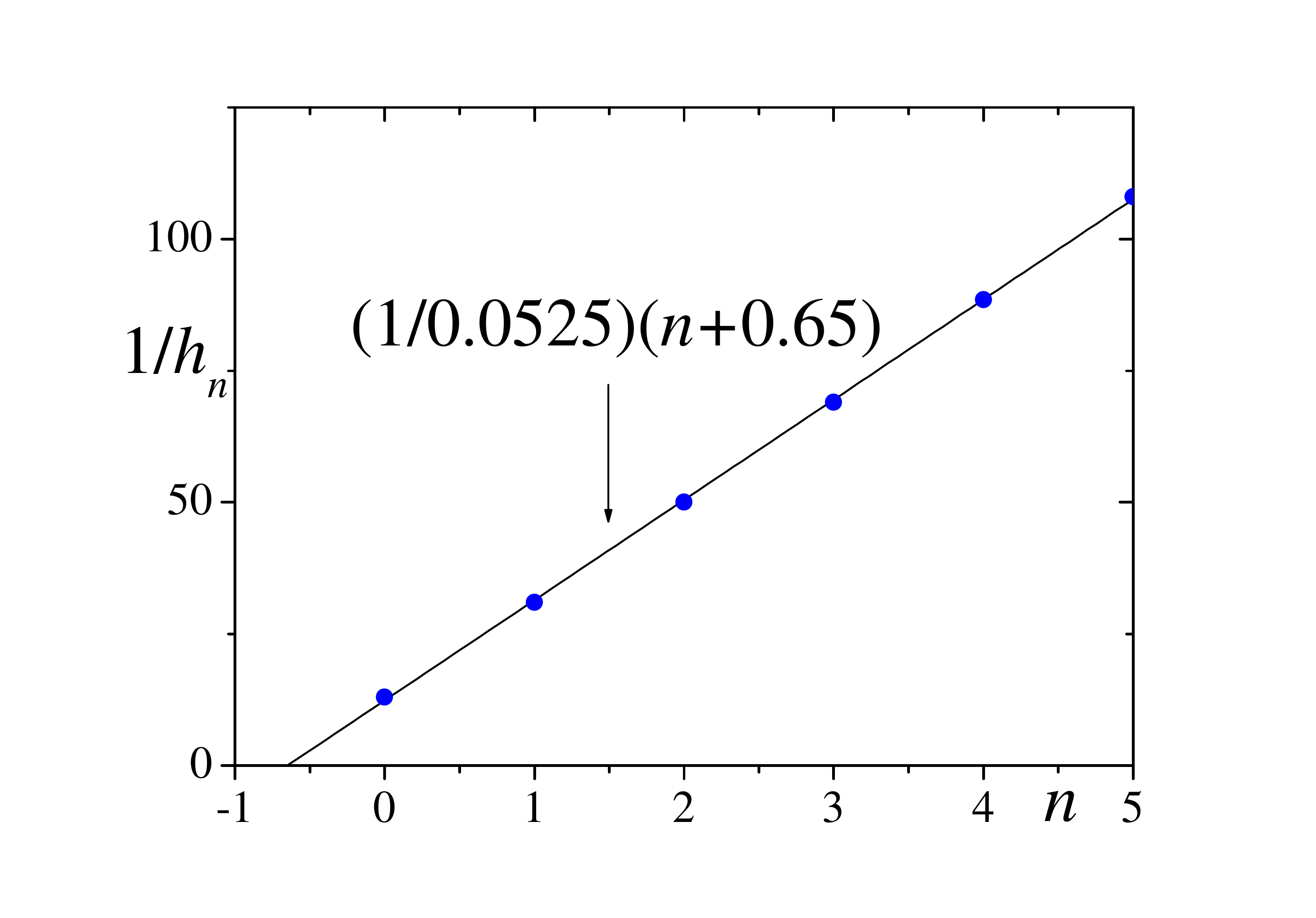}\vspace{-0.5cm} \\
\begin{flushleft} \hspace{0.5cm}(b) \end{flushleft}\vspace{-0.5cm}
\includegraphics[width=0.5\textwidth]{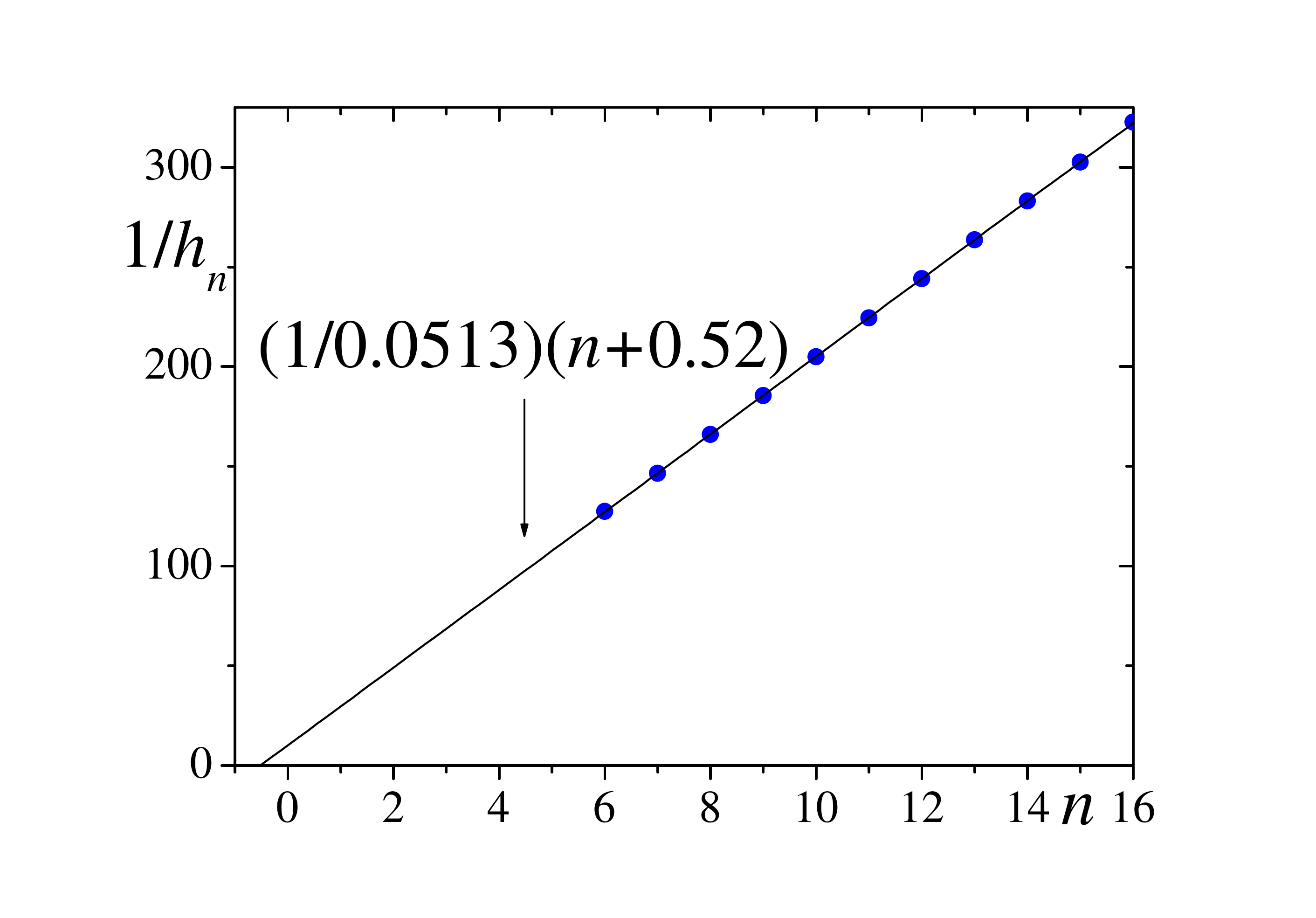}\vspace{-0.5cm}
\caption{(Color online) 
Similar plot as Fig.~\ref{Figure8} for $V=37.14$~meV.
}
\label{Figure11}
\end{figure}

By numerically diagonalizing the matrix of Eq. (\ref{j5}), 
we plot the energy ${\varepsilon}_{i, \mathbf{k}}$ as a function of $h$ in Fig.~\ref{Figure5}(a) 
for (TMTSF)$_2$NO$_3$ at $T>T_{\mathrm{AO}}$, where the Fermi surface consists of two warped lines as shown in Fig. \ref{Figure2}(a). 
There are $2q$ bands at $h=p/q$ and each bands are doubly degenerate. Since the Fermi surface is not closed, the Landau quantization is not expected 
to occur near the Fermi energy in the semi-classical treatment\cite{onsager}. 
Even in this case 
there should be very small gaps in the tight-binding electrons in principle, but
there are no visible gaps in the energy spectrum near 3/4-filling, as shown in Fig.~\ref{Figure6}. 
It is consistent with the semi-classical picture that the Landau quantization 
occurs only when the Fermi surface is closed. At $T<T_{\rm AO}$, where the orientation of the anion orders, $V$ is finite and electron and hole pockets appear at $h=0$ as shown in  Fig.~\ref{Figure2} (b).
The Hofstadter butterfly diagrams for $V=12.38$ meV and the three times larger value ($V=37.14$ meV) are shown in Figs.~\ref{Figure5} (b) and (c), respectively. 
The gaps are labeled by $(s_r, t_r)$ 
(Eq.~(\ref{sr})) in Figs.~\ref{Figure5} (a), (b) and (c). 
The overall structures for $V\neq 0$, especially for smaller filling ($(s_r, t_r) = (0,1), (0,2),$ {\it etc}.) are  similar as that for $V=0$ (Fig.~\ref{Figure5} (a)). 
We plot the Hofstadter butterfly diagram near the Fermi energy for $3/4$ filled case in Fig.~\ref{Figure7} ($V=12.38$~meV), Fig. 
\ref{Figure10} ($V=37.14$~meV), and Fig. \ref{Figure11} ($V=86.50$~meV, no pockets).

In Fig.~\ref{Figure7} 
the energy is not quantized as delta functions for finite value of $h$, but we can see the broadened Landau levels starting from $\varepsilon_i=\varepsilon_{\mathrm{4b}}^0$ and $\varepsilon_{\mathrm{3t}}^0$ at $h=0$,
where $\varepsilon_{\mathrm{4b}}^0$ and $\varepsilon_{\mathrm{3t}}^0$ are the bottom energy of the fourth band and the top energy of the third band, respectively
(see Fig.~\ref{Figure3}). Note that the broadening of the Landau levels is not seen in the semi-classical theory\cite{fortin} of the magnetic breakdown.

If we approximate electron and hole pockets in eigenvalues of Eq. (\ref{J3}) as the anisotropic parabolic bands, it is expected 
that the Landau levels are semi-classically given by
\begin{equation}
\varepsilon_n^{\mathrm{electron\ pocket}} \simeq \varepsilon_{\mathrm{4b}}^0 + \frac{1}{C_{ep}} \left( n+\gamma \right) h, \label{e_4b}
\end{equation}   
and
\begin{equation}
 \varepsilon_n^{\mathrm{hole\ pocket}} \simeq \varepsilon_{\mathrm{3t}}^0 - \frac{1}{C_{hp}} \left( n+\gamma \right) h,\label{e_3t}
\end{equation}   
where $\gamma =1/2$, $C_{ep}$ and $C_{hp}$ 
are constants depending on the curvature of the anisotropic parabolic bands, respectively, and $n=0, 1, 2, \cdots$. 
If this is the case, the ratio of the slope of the Landau levels as a function of the magnetic field should be
\begin{equation}
 (0+\frac{1}{2}) : (1+\frac{1}{2}) : (2+\frac{1}{2}) : \cdots = 1 : 3 : 5 : \cdots.
 \label{eqslope}
\end{equation}
We obtain, however, that  the ratio of the slope is 
approximately $1050:2100:3000:\cdots \approx 1:2:3:\cdots$, and $610: 1500: 2200: \cdots \approx 2:5:7:\cdots $
from the dotted lines in Fig.~\ref{Figure7}(a). When we approximate the slopes in the region of weaker magnetic field as shown 
in Fig.~\ref{Figure7}~(c), we obtain the ratio of the slope as
$1040:2570:3900: \cdots \approx 2:5:8:\cdots$ and $610:1550:2500:\cdots \approx 2:5:8:\cdots$.
These results are inconsistent with the expected values of Eq.~(\ref{eqslope}). 
As seen in
 Figs.~\ref{Figure7}~(a) and (c) the fittings with the dotted lines are not good. 
Therefore, the semi-classical quantization of Landau levels for free electron and free hole pockets is not a quantitatively acceptable approximation, even when we neglect the broadening of Landau levels. 


Another interesting point in Fig.~\ref{Figure7} is that there are many gaps with the same index $(s_r,t_r)$ near 3/4-filling ($s_r=3$). Gaps with the same index $(3, t_r)$ are closed or almost closed at points as a function of $h$. 
If the Landau levels (dotted lines in Figs.~\ref{Figure7}(a) and (c)), 
which were thought to be the quantized levels of electrons and holes in the electron and hole pockets,
 were broadened in the tight-binding model, bands would be overlapped in finite ranges of $h$ instead of closed at points. We can see 
the bands between the gaps $(3,t_r)$ and $(3,t_r+1)$ as if they start from the Fermi energy $\varepsilon_{\rm F}^0$ at $h=0$, which are indicated by green lines in Fig.~\ref{Figure7} (b).

We draw blue circles in Fig.~\ref{Figure7} at $\varepsilon_i=\varepsilon_{\rm F}^0$ and $h=h_0, h_1, h_2, h_3, \cdots$, at which
the energy gap labeled by $(3,0)$ is almost closed.  
We plot $1/h_n$ as a function of $n$ in Fig.~\ref{Figure8}. 
We can fit $1/h_n$ as
\begin{align}
 \frac{1}{h_n} &=\frac{1}{0.0675} (n+0.76) \hspace{1cm} (0 \leq n \leq 7), \label{17}\\
 \frac{1}{h_n} &=\frac{1}{0.0670} (n+0.68) \hspace{1cm} (8 \leq n \leq 21), \label{18}
\end{align}
in Fig.~\ref{Figure8}. In Eqs. (\ref{17}) and (\ref{18}), the region of $1/h_n$ are $11.5 \leq 1/h_n \leq 115$ and $130 \leq 1/h_n \leq 324.5$, which correspond to 170 $\geq H \geq 17$ T and 15.04 T $\geq H \geq 6.025$ T, respectively. 

When we set $\varepsilon_n^{\mathrm{electron\ pocket}}=\varepsilon_{\rm F}^0$ and $\varepsilon_n^{\mathrm{hole\ pocket}}=\varepsilon_{\rm F}^0$ in Eqs. (\ref{e_4b}) and (\ref{e_3t}), we get 
\begin{eqnarray}
\frac{1}{h_n}&\simeq&\frac{1}{
C_{ep}(\varepsilon_{\rm F}^0-\varepsilon_{\mathrm{4b}}^0)}(n+\frac{1}{2}), \label{e_4b_2} \\
\frac{1}{h_n}&\simeq&\frac{1}{
C_{hp}(\varepsilon_{\mathrm{3t}}^0-\varepsilon_{\rm F}^0)}(n+\frac{1}{2}), \label{e_3t_2} 
\end{eqnarray}
where $C_{ep}(\varepsilon_{\rm F}^0-\varepsilon_{\mathrm{4b}}^0)$ and 
$C_{hp}(\varepsilon_{\mathrm{3t}}^0-\varepsilon_{\rm F}^0)$ are given by the areas of the electron pocket and the hole pocket at $h=0$ per the area of the Brillouin zone, respectively. 
The areas of electron pocket (black curves) and hole pocket (red curves) are the same and 0.0676 times the area of the Brillouin zone, as seen in Fig.~\ref{Figure2}(b). Thus, 
the obtained values from the linear fitting (0.0675 and 0.0670) in Eqs. (\ref{17}) and (\ref{18}) are in good agreement with the semi-classical Landau quantization for parabolic bands, although $\gamma$ deviates from $1/2$.

To analyze the closing of the gap in detail, 
we plot the close up figures near $h_4=2/141$ in Fig.~\ref{Figure9}. 
There are $564 (=4 \times q)$ bands when $h=p/q=2/141$.
When the band is $3/4$ filled, the chemical potential is between the $423$th and $424$th bands, i.e. $r=423$. 
The gap is almost closed at $k_x=\pm \pi/(4qa)$ but there is a small gap, which depends on $k_y$ very slightly, as shown in Figs.~\ref{Figure9} (c) and (d).

Next, we study the energy for a larger $V=37.14$~meV (Fig. \ref{Figure10}). 
The width of the band at $V=37.14$~meV is smaller than that at $V=12.38$~meV, and it is smaller at smaller $h$. In this case the areas of the electron pocket and the hole pocket are smaller than those for $V=12.38$~meV. 
 The ratio of the slopes of the ``Landau levels'' starting from the bottom of the fourth band and the top 
of the third band (red dotted lines in Figs.~\ref{Figure10} (a) and (c)) becomes closer to that of free electrons, $1 : 3 : 5: \cdots$. This can be understood
as follows. When $V$ becomes large, the electron pocket and hole pocket are separated in the Brillouin zone and the areas of electron and hole pockets at $h=0$ and 3/4-filling 
become small. 
Then we can safely adopt the approximation that electrons and holes in the pockets are treated as free electrons and free holes.  The semi-classical picture of the magnetic breakdown between pockets may cause small effects.
We plot the inverse of the magnetic fields $h_n$, at which the gaps indexed by $(3,0)$ are closed or almost closed, 
as a function of $n$ in Fig.~\ref{Figure11}. 
This $1/h_n$ is fitted by the straight line with the larger slope 
than that of $V=12.38$~meV (Fig.~\ref{Figure8}), which 
corresponds to the smaller areas of the electron and hole pockets. The phase factor $\gamma$ obtained from the intersection with the $n$-axis is near the free electron value, $1/2$.

\begin{figure}[bt]
\begin{flushleft} \hspace{0.5cm}(a) \end{flushleft}\vspace{-0.5cm}
\includegraphics[width=0.47\textwidth]{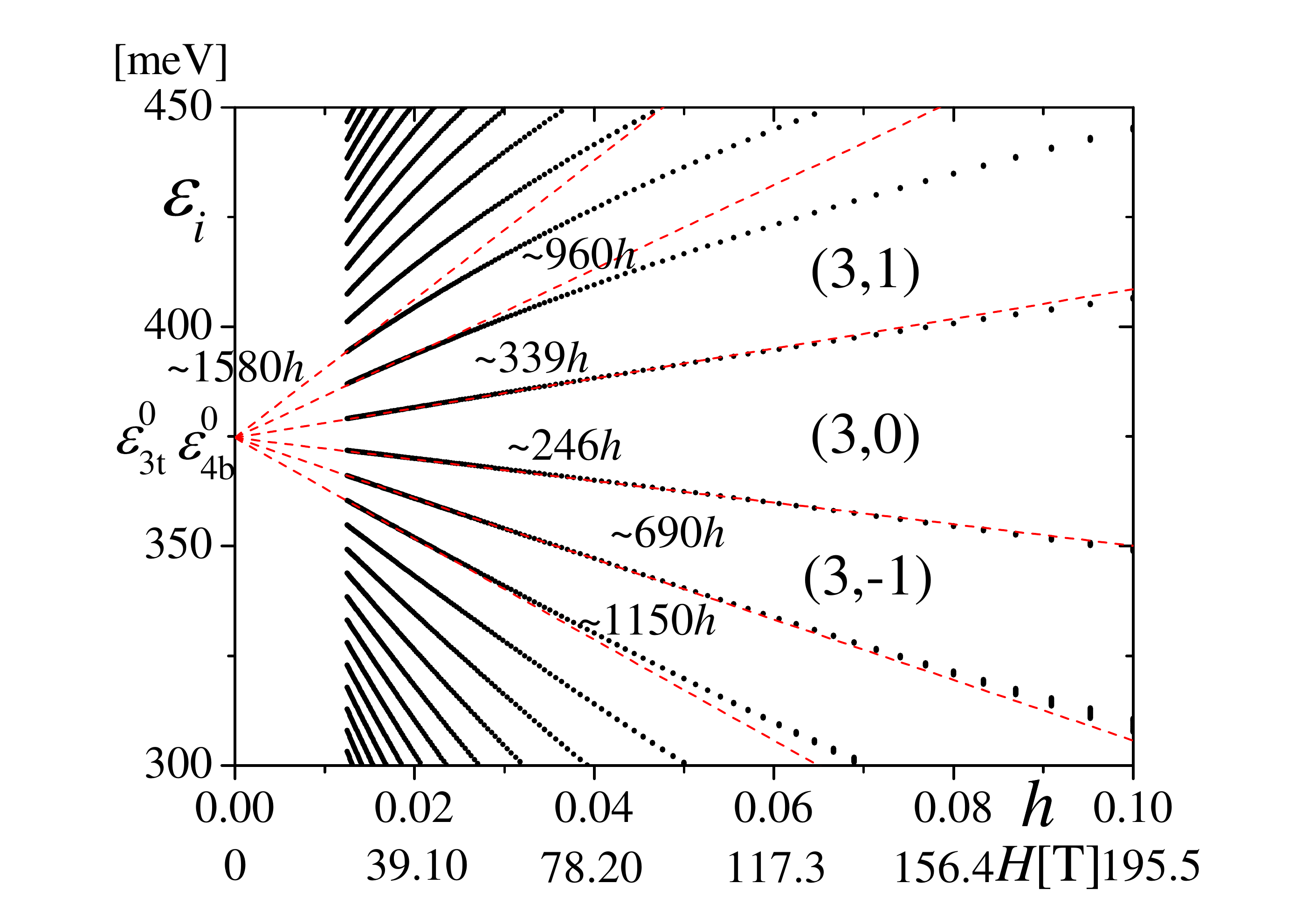}\vspace{0.0cm}
\begin{flushleft} \hspace{0.5cm}(b) \end{flushleft}\vspace{-0.0cm}
\includegraphics[width=0.38\textwidth]{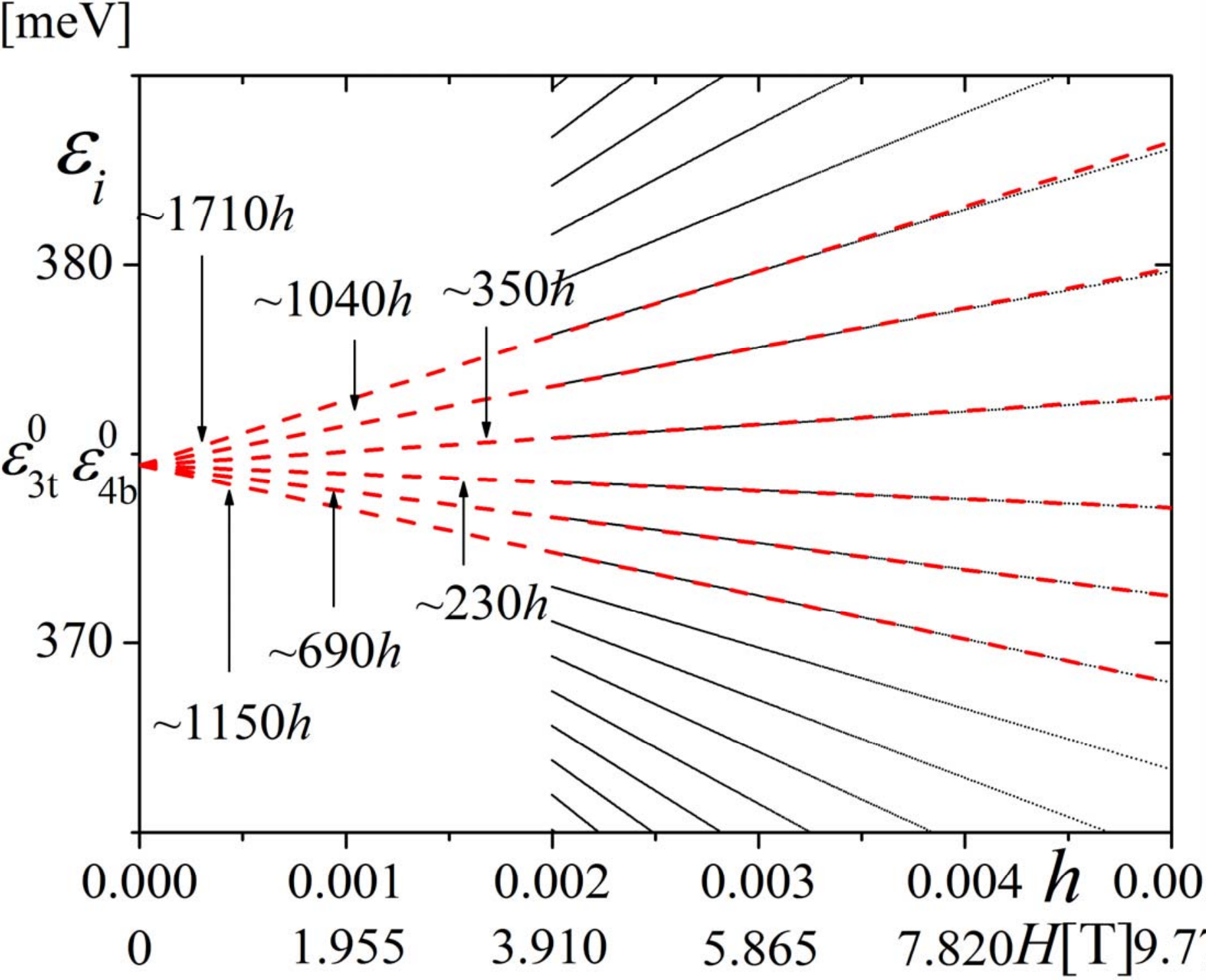}\vspace{-0.0cm}
\caption{(Color online) 
(a) Energy as a function of $h$ at $V=86.5$ meV, where $2\Delta\simeq 123.2$ meV, $\varepsilon^0_{\rm F}=\varepsilon_{\rm 3t}^0=\varepsilon_{\rm 4b}^0\simeq 374.7$ meV. We take $h=1/q$ with $10 \leq q \leq 80$ and $h=2/q$ with $q=2m+1$ and $20 \leq m \leq 59$, where the wave number $(k_x, k_y) = (n_x \pi/(30a),0)$ with $0 \leq n_x \leq 30$. 
 (b) An enlarged figure of (a). 
We take $h=1/q$ with $200 \leq q \leq 500$, where $(k_x, k_y) = (n_x \pi/(18a),0)$ with $0 \leq n_x \leq 18$. 
}
\label{Figure12}
\end{figure}

We also study the case of $V=86.5$~meV, when the top of the third band $\varepsilon_{\rm 3t}$  and the bottom of the fourth band $\varepsilon_{\rm 4b}$
are the same and 
the electron and hole pockets disappear at 3/4-filling,
as shown in Fig.~\ref{Figure4}. We plot the energy as a function of $h$ in Fig.~\ref{Figure12}. The band widths are very narrow. 
Since the ratio of the slopes of the Landau levels is close to $1: 3: 5: \cdots$, 
the bands are recognized to Landau levels for free electrons and holes. 

If holes or electrons are doped, the chemical potential is above or below the dotted blue line in Fig.\ref{Figure7}(b) ($V=12.38$ meV) or the dotted orange line in Fig. \ref{Figure10}(b) ($V=37.14$ meV). The Hall conductance is quantized when the chemical potential is in the energy gap, but it is not quantized when the chemical potential is within the broadened band. For the reasonable value of anion potential ($V=12.38$ meV), the energy band is broadened. Therefore, the Hall conductance is quantized only in some regions of the magnetic field, if electrons or hales are doped, and it is not quantized in other regions of the magnetic field.



\section{magnetization and de Haas van Alphen oscillations}

The oscillatory part of the magnetization with the fixed chemical potential $(\mu$) at the temperature ($T$) is given by the LK 
formula\cite{LK,shoenberg,nakano,alex2001,champel,CM,KH,Igor2004PRL,Igor2011,Sharapov}. In the generalized LK formula in two-dimensional metals
the magnetization oscillates periodically as a function $1/H$ with the period
\begin{equation}
 f = \frac{c \hbar A}{2 \pi e},
\end{equation}
where $A$ is the area of the Fermi surface at $H=0$. 
The generalized LK formula at $T=0$ for 
the two-dimensional metals is given by
\begin{equation}
M^{\rm LK}=-\frac{e}{2\pi^2 c\hbar}
 \frac{A}{\frac{\partial A}{\partial \mu}}
\sum_{l=1}^{\infty}\frac{1}{l}\sin\left[2\pi l\left(\frac{f}{H}-\gamma\right)
\right].\label{LK_0}
\end{equation}
Note that the oscillation part of the magnetization in LK formula is zero at
\begin{equation}
 H=H_n
\end{equation}
and we obtain 
\begin{equation}
 \frac{1}{H_n} = \frac{1}{f} (n+\gamma) 
\end{equation}
with $n=0,1,2,\cdots$. Namely, $M^{\rm LK}=0$ appears periodically as a function of $1/H$ with the frequency, $f$. 
The amplitude of the oscillation at $T=0$ is independent of $H$ in the LK formula. In the LK formula the broadening of the Landau levels in the tight-binding model is not taken into account. 

In this section we study dHvA oscillation in (TMTSF)$_2$NO$_3$ by taking the 
effects of the magnetic field in the tight-binding model. 
The energy $\varepsilon_{i, \mathbf{k}}$ in the magnetic field is obtained as the eigenvalues of $4q \times 4q$ matrix $\tilde{\varepsilon}_{\mathbf{k}}$
given in Eq.~(\ref{j5}), where $i=1 \sim 4q$. 
The thermodynamic potential $\Omega$ per sites at $T$ is obtained as
\begin{equation}
\Omega=-\frac{k_BT}{4qN_k} \sum_{i=1}^{4q}\sum_{{\bf k}} \log\left
\{\exp\left(\frac{\mu-\varepsilon_{i,{\bf k}}}{k_BT} \right) +1
\right\},
\end{equation}
where $k_B$ is the Boltzmann constant and 
$N_k$ is the number of $\mathbf{k}$ points taken in the magnetic Brillouin zone.
At $T=0$, $\Omega$ becomes the total energy with fixed $\mu$, 
\begin{equation}
E_{\mu}=\frac{1}{4qN_k}\sum_{\varepsilon_{i,{\bf k}}\leq\mu}  (\varepsilon_{i,{\bf k}}-\mu).
\end{equation} 
The magnetization is obtained in grand canonical ensemble by 
\begin{equation}
M_{\mu} = -\frac{\partial \Omega}{\partial h}. 
\end{equation}

On the other hand, 
when the electron number is fixed (in (TMTSF)$_2$X, electrons are $\nu$-filling 
with $\nu=3/4$), 
the magnetic-field dependence of the chemical potential 
is not negligible in the two-dimensional systems in 
general\cite{shoenberg,nakano,alex2001,champel,KH}, whereas it can be neglected in the three-dimensional metals. 
In this study, the Helmholtz free energy and the magnetization are calculated in the canonical ensemble with the fixed electron number. In that case, 
the chemical potential, $\mu$, should be obtained by the equation,
\begin{equation}
\nu =\frac{1}{4qN_k} \sum_{i=1}^{4q} \sum_{{\bf k}}
\frac{1}{\exp\left(\frac{\varepsilon_{i,{\bf k}}-\mu}{k_BT} \right) +1}.
\end{equation}
The Helmholtz free energy ($F$) per sites at $T$ is given by
\begin{equation}	
F=-\frac{k_BT}{4qN_k} \sum_{i=1}^{4q}\sum_{{\bf k}} \log\left
\{\exp\left(\frac{\mu-\varepsilon_{i,{\bf k}}}{k_BT} \right) +1
\right\} + \mu \nu.
\end{equation}
At $T=0$ it becomes
\begin{equation}
E_{\nu}=\frac{1}{4qN_k}\sum_{\varepsilon_{i,{\bf k}}\leq\mu}  \varepsilon_{i,{\bf k}},
\end{equation}
The magnetization for fixed electron filling $\nu$ is given by 
\begin{equation}
M_{\nu} = -\frac{\partial F}{\partial h}.
\end{equation}
We obtain the magnetization by the numerical differentiation. 

\begin{figure}[bt]
\begin{flushleft} \hspace{0.5cm}(a) \end{flushleft}\vspace{-0.5cm}
\includegraphics[width=0.48\textwidth]{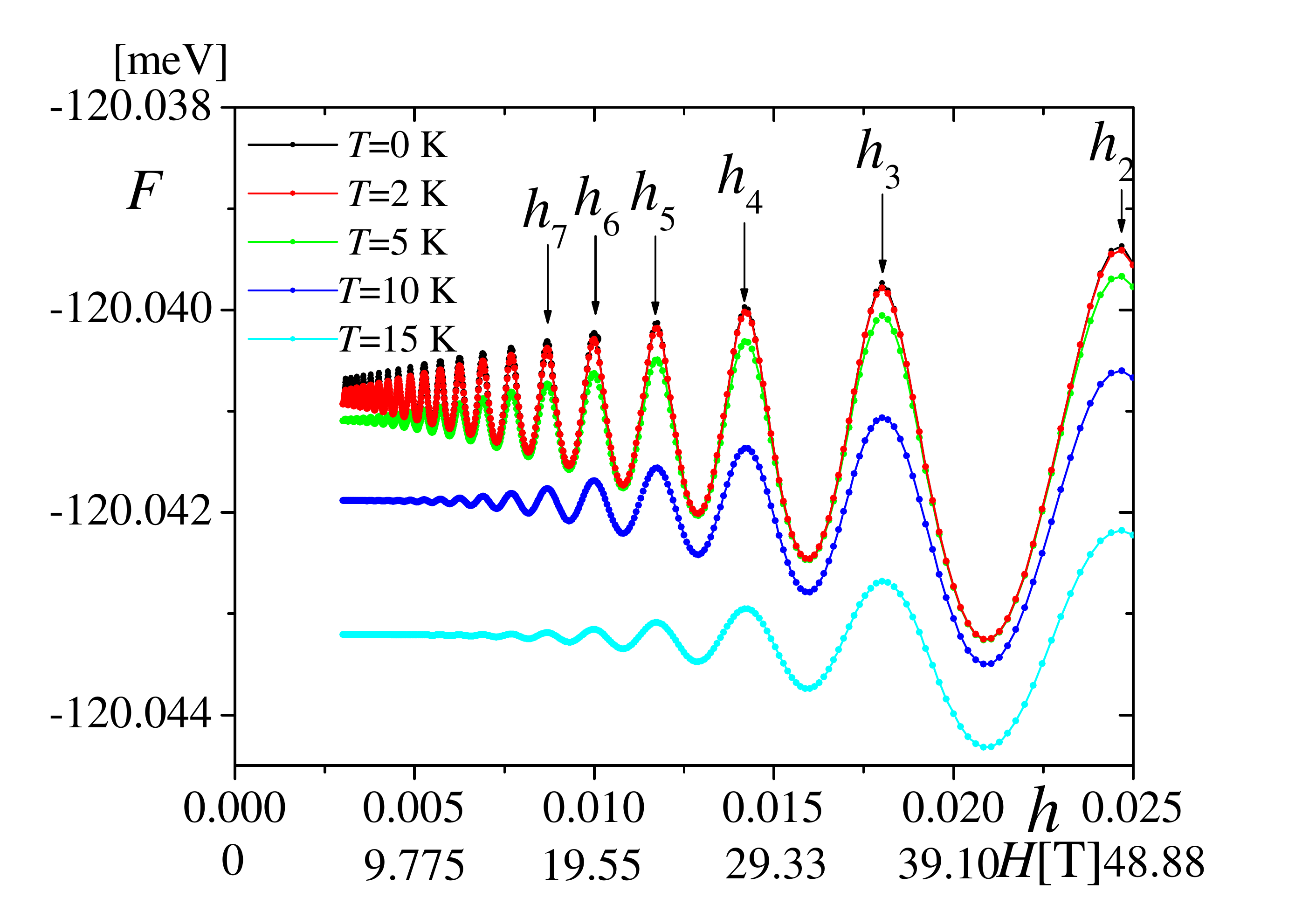}\vspace{-0.5cm}
\begin{flushleft} \hspace{0.5cm}(b) \end{flushleft}\vspace{-0.5cm}
\includegraphics[width=0.48\textwidth]{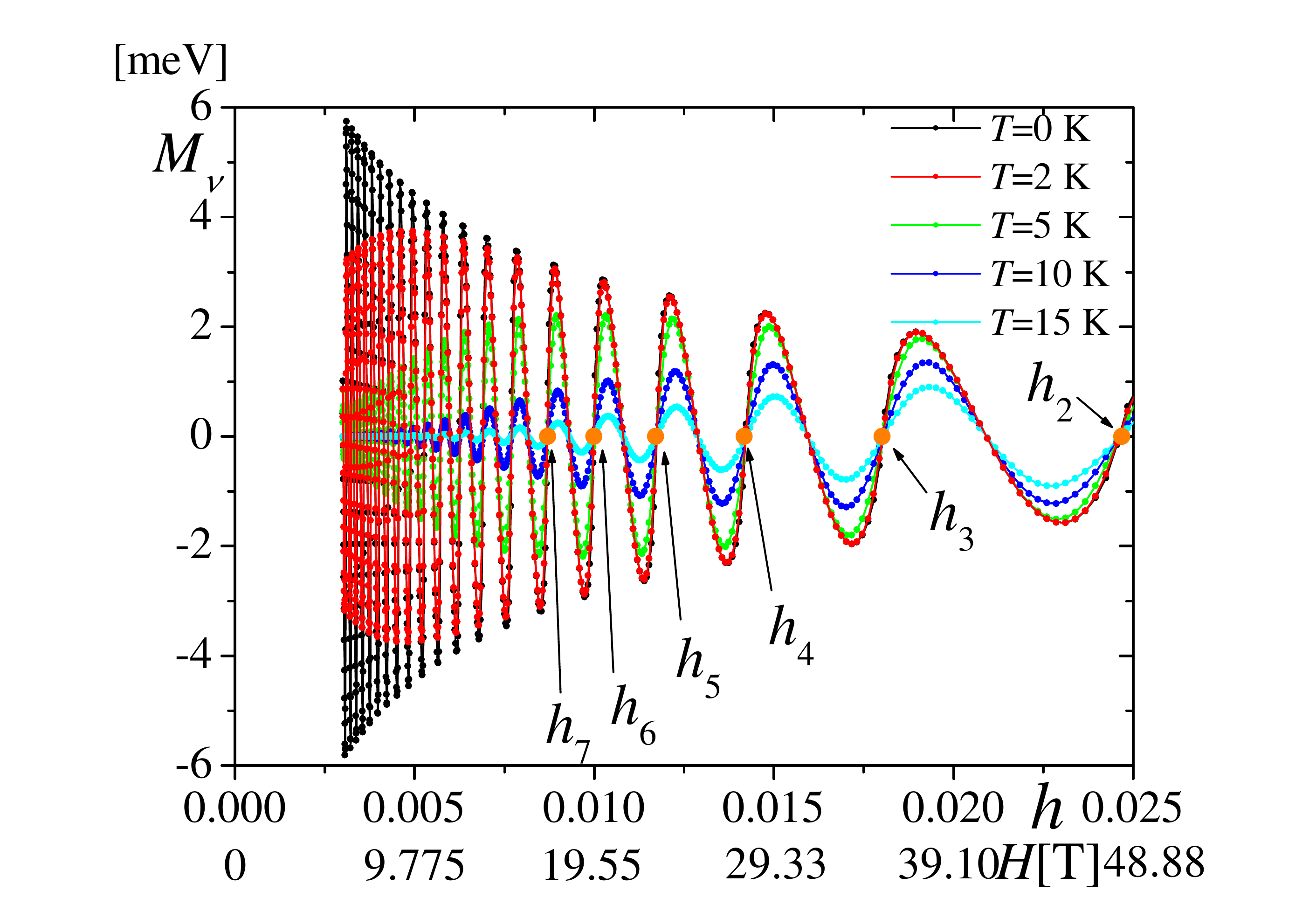}\vspace{-0.5cm}
\begin{flushleft} \hspace{0.5cm}(c) \end{flushleft}\vspace{-0.5cm}
\includegraphics[width=0.48\textwidth]{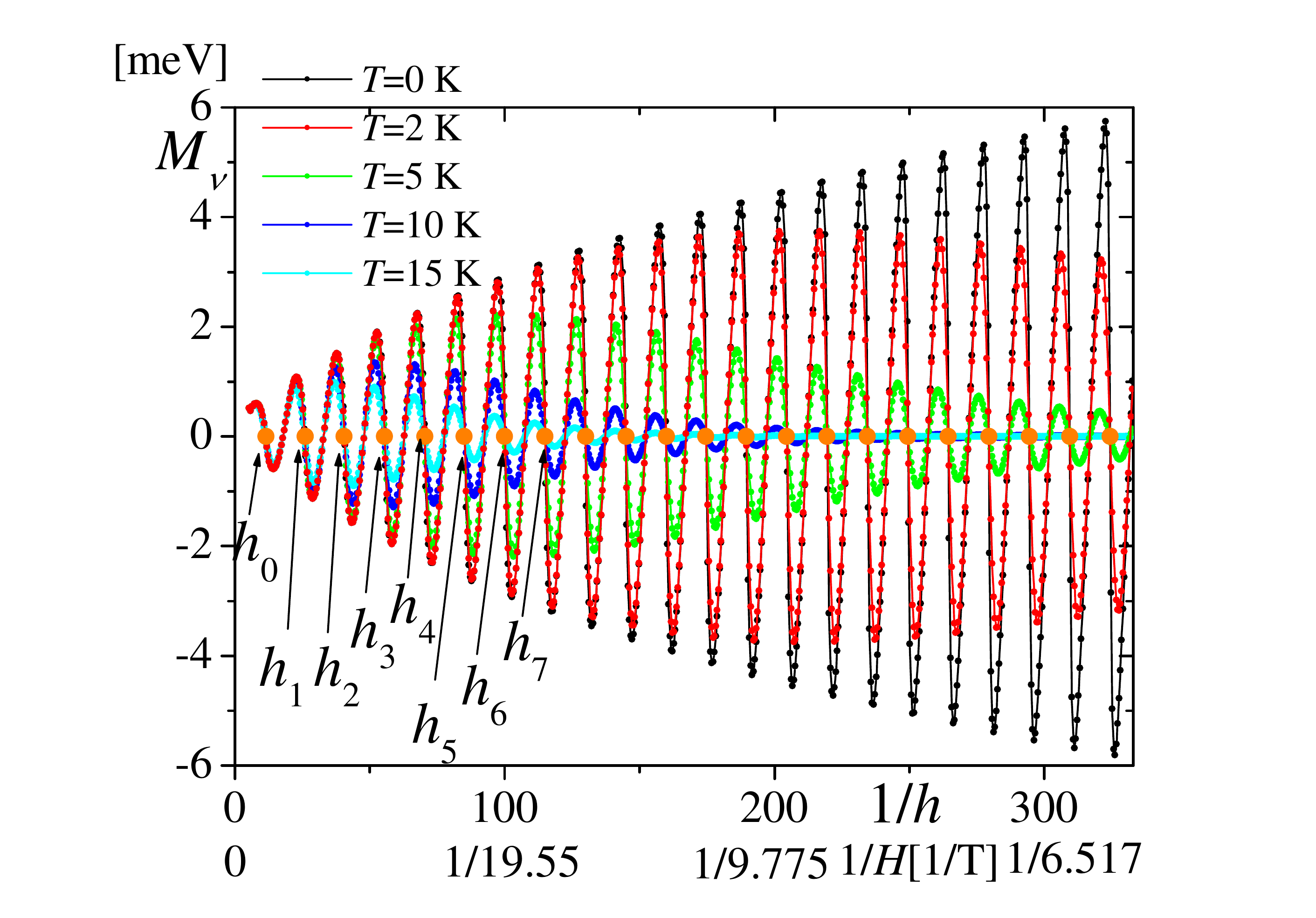}\vspace{-0.0cm}
\caption{(Color online) 
Free energy (a) and magnetizations as a function of $h$ (b) and as a function of $1/h$ (c) at $V=12.38$ meV for $T=0, 5, 10 $
 and 15 K. We take $h=1/q$ with $5 \leq q \leq 333$ and $h=2/q$ 
with $q=2m+1$ and $5 \leq m \leq 332$, where 
$(k_x, k_y) = (n_x \pi/(6a),0)$ with $0 \leq n_x \leq 6$ for $q>333$, 
$(k_x, k_y) = (n_x \pi/(13a),0)$ with $0 \leq n_x \leq 13$ for $200<q\leq 333$, $(k_x, k_y) = (n_x \pi/(30a),0)$ with $0 \leq n_x \leq 30$ for $80<q\leq 200$ and 
$(k_x, k_y) = (n_x \pi/(61a),0)$ with $0 \leq n_x \leq 61$ for $q\leq 80$. 
}
\label{Figure13}
\end{figure}

\begin{figure}[bt]
\begin{flushleft} \hspace{0.5cm}(a) \end{flushleft}\vspace{-0.5cm}
\includegraphics[width=0.48\textwidth]{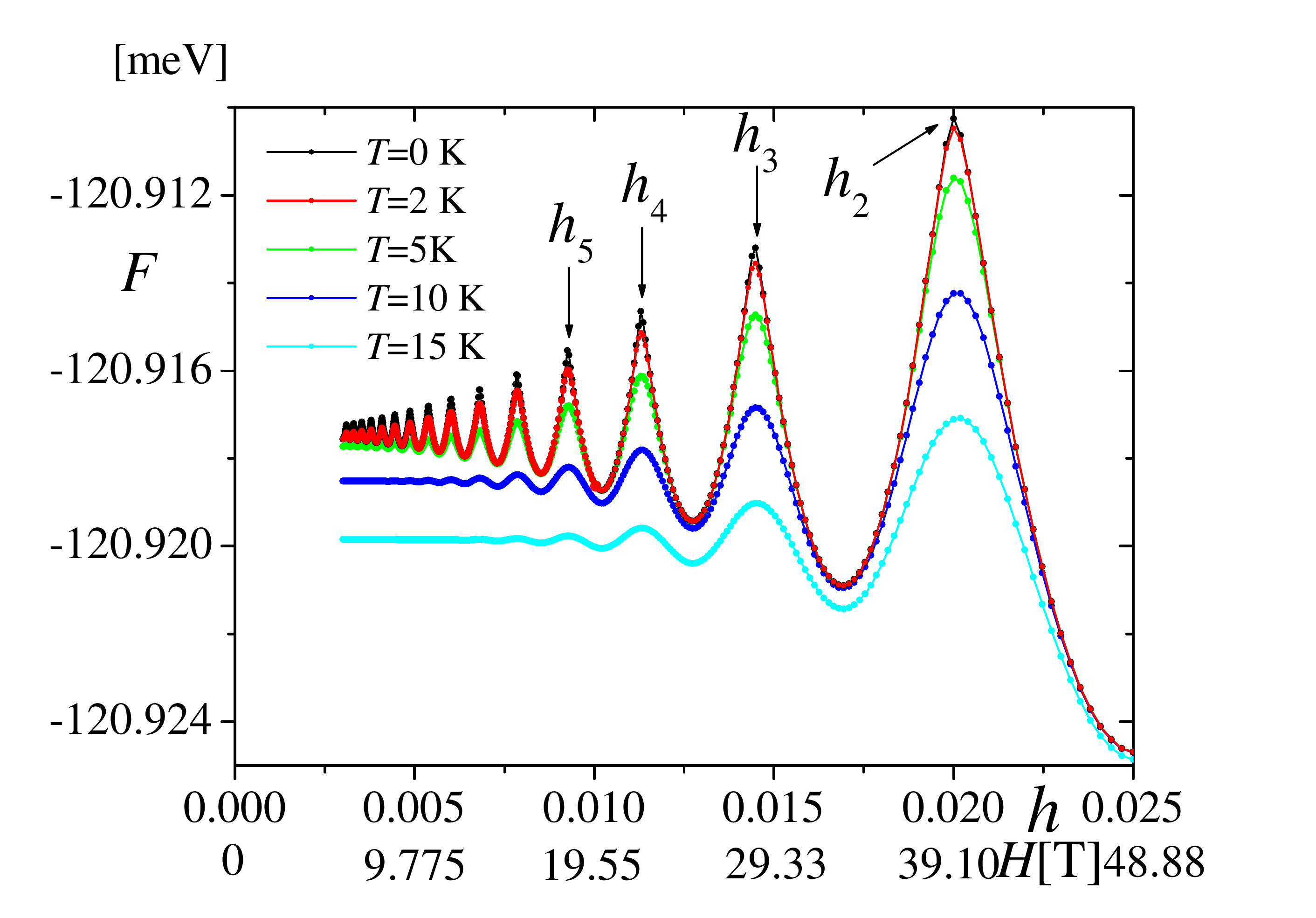}\vspace{-0.5cm}
\begin{flushleft} \hspace{0.5cm}(b) \end{flushleft}\vspace{-0.5cm}
\includegraphics[width=0.48\textwidth]{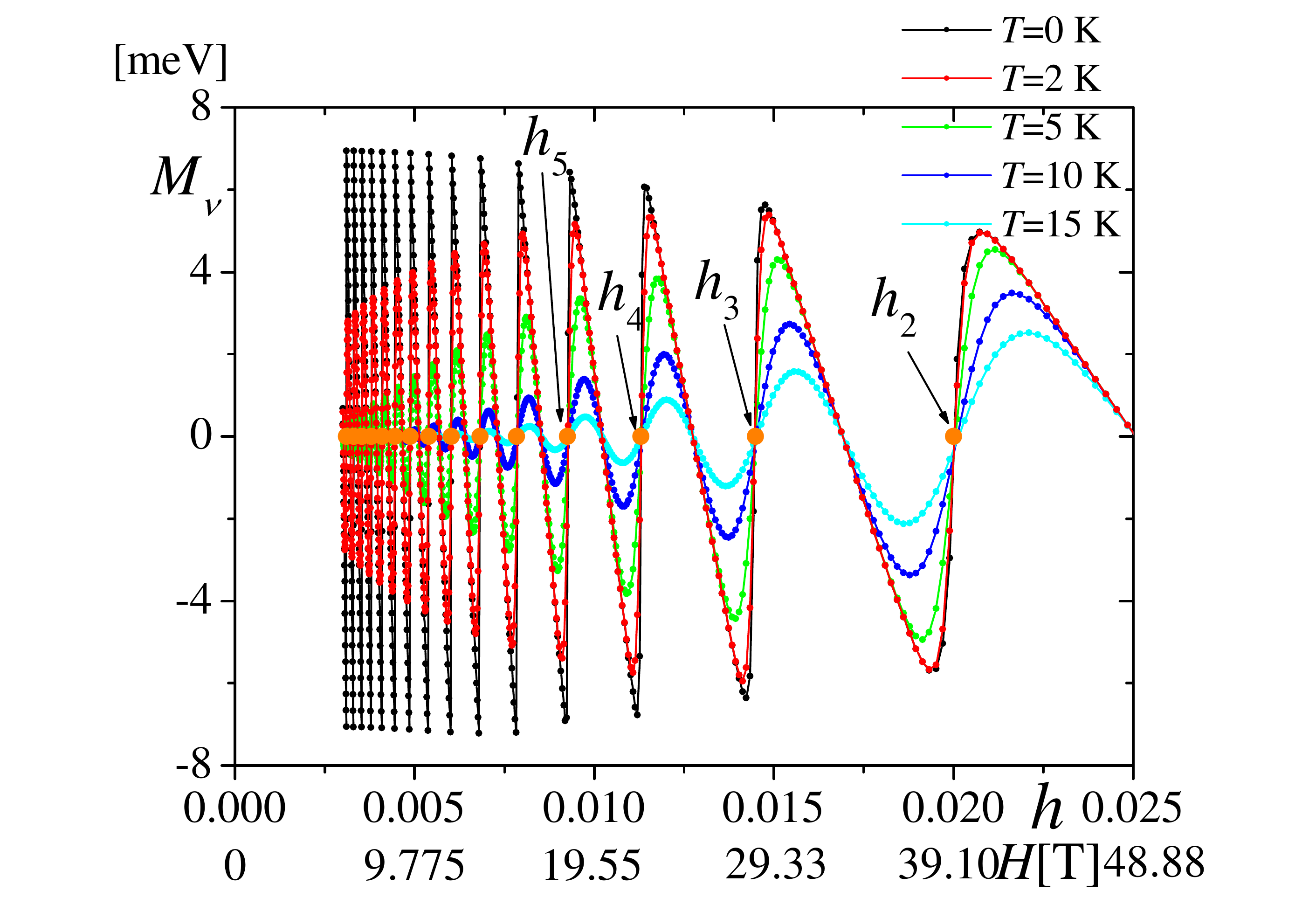}\vspace{-0.5cm}
\begin{flushleft} \hspace{0.5cm}(c) \end{flushleft}\vspace{-0.5cm}
\includegraphics[width=0.48\textwidth]{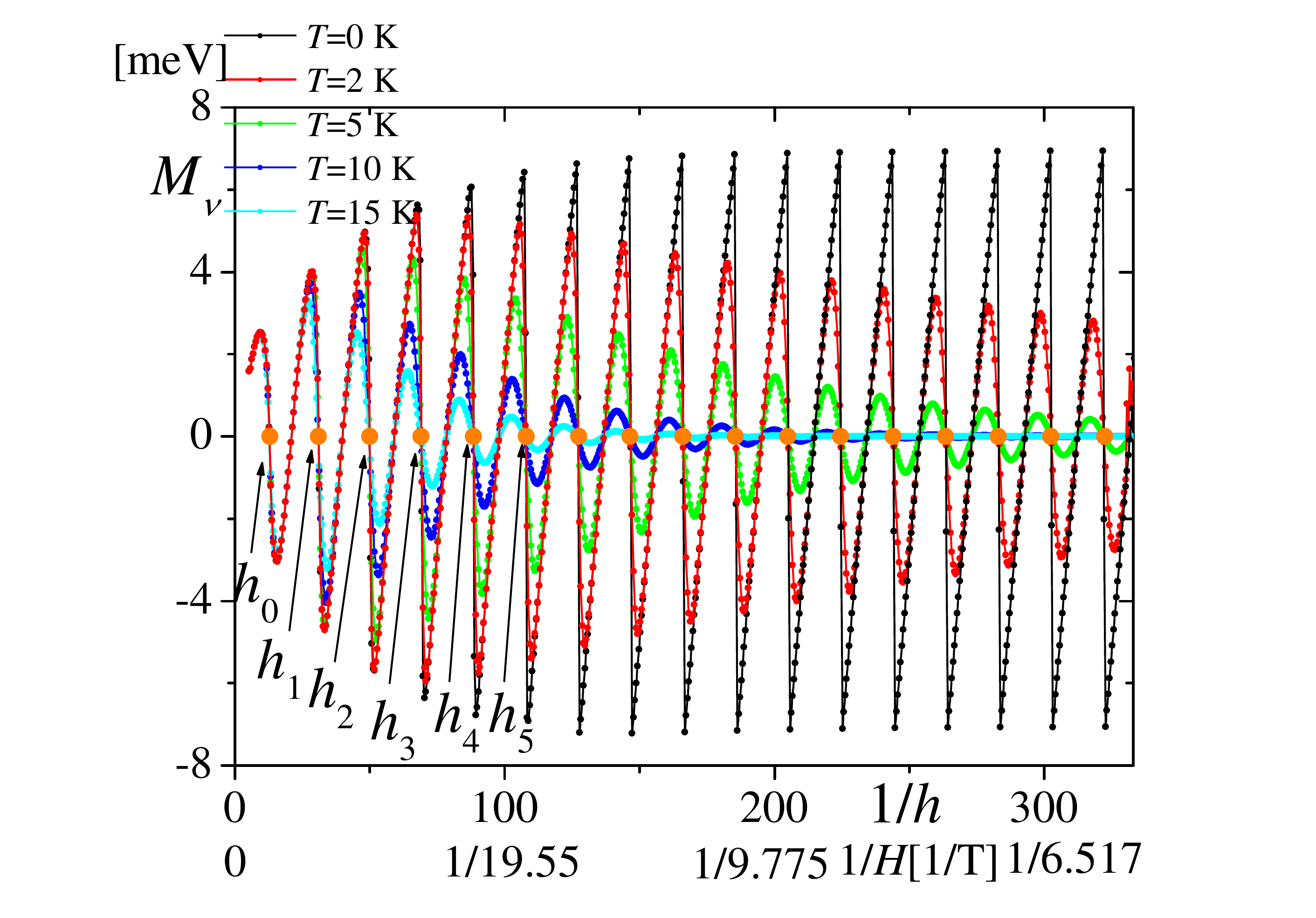}\vspace{-0.0cm}
\caption{(Color online) 
Free energy (a) and magnetizations as a function of $h$ (b) and as a function of $1/h$ (c) at $V=37.14$ meV for $T=0, 5, 10 $ and 15 K. We take the same values for $q$ and ($k_x, k_y$) as those of Fig. \ref{Figure13}. 
}
\label{Figure14}
\end{figure}

As seen in Figs.~\ref{Figure7} (b) and \ref{Figure10} (b), 
the chemical potentials (dotted black and orange lines) for $\nu=3/4$ are in the gap labeled by $(3,0)$ and almost independent of $h$. 
Therefore, in both cases of $V=12.38$ meV and 37.14 meV, 
$M_{\mu}$ and $M_{\nu}$ are expected to be almost the same.
Indeed we obtained the negligible difference between $M_\mu$ and $M_\nu$ in the numerical calculation.
In Figs.~\ref{Figure13} (a) and (b) and Figs. \ref{Figure14} (a) and (b), we plot $F$ and $M_{\nu}$ as a function $h$ with $V=12.38$~meV and $37.14$~meV, respectively. 
The periodical oscillations as a function of $1/h$ are seen in 
Figs.~\ref{Figure13} (c) and \ref{Figure14} (c), respectively. 
These oscillations are thought to correspond to the dHvA oscillation for electrons and holes in semi-classical theory. 
Free energy, $F$, has local maximum values at $h=h_n$, which are shown as blue circles in Figs.~\ref{Figure7} and \ref{Figure10}. At $h=h_n$ gaps labeled  
by $(3,0)$ is closed or almost closed. The free energy may be lowered by opening the finite gap at the Fermi energy. Therefore it is reasonable that the free energy is a local maximum at $h=h_n$. 
As a result, magnetization is zero at $h=h_n$. Since $1/h_n$ is fitted by the straight line (see Figs.~\ref{Figure8} and ~\ref{Figure11}) as proportional to 
$(n+\gamma)$, the magnetization oscillates periodically as a function $1/h$, (dHvA oscillation). These frequencies (0.0675 and 0.0670) in Fig.~\ref{Figure8} are almost same as the areas of the electron and hole pockets in Fig. \ref{Figure2}(b) per the area of the Brillouin zone. 
It is expected that, in the dHvA experiment of (TMTSF)$_2$NO$_3$, $\gamma=0.76$ in Eq. (\ref{17}) ($\gamma=0.68$ in Eq. (\ref{18})) are estimated at 17 $\leq H \leq 170$ T (6.025 T $\leq H \leq 15.04$ T). 
From the experiment of SdH oscillation\cite{ulmet}, $\gamma$ is estimated to be 0. 
The SdH experiment was performed in the SDW state. The amplitude of the SDW order parameter may depend on the magnetic field. Therefore, it is not easy to compare the experiment with our result calculated in the metallic state without SDW order.

For larger $V$ ($V=37.14$~meV), the amplitude of the magnetization oscillation is almost constant for $1/h \gtrsim 100$ 
(i.e. $H \lesssim 19.55$~T) at $T=0$ as shown in Figs.~\ref{Figure14}~(b) and (c). The almost constant field dependence of the amplitude and the saw tooth shape of
$M_\nu$ (Fig.~\ref{Figure14}~(c)) are the same as those 
of $M^{\rm LK}$ (Eq. (\ref{LK_0})) 
for the fixed chemical potential case in two-dimensional metals. 
For the realistic value of $V$ ($V=12.38$~meV), the saw tooth shape is similar. 
However, the amplitude of magnetization oscillation is an increase function of $1/h$ for
$1/h \lesssim 333$ ($ H \gtrsim 5.865$~T) at $T=0$ 
as shown in Figs.~\ref{Figure13}~(b) and (c). 
The $h$-dependence of the amplitude of magnetization oscillation is caused by the broadening of Landau levels.




\section{Conclusions}

We use the spinless tight-binding model on a two-dimensional rectangular lattice for (TMTFS)$_2$NO$_3$ with realistic band parameters and
potentials due to the effect of the anion ordering. 
The effects of a uniform magnetic field $\sim$ 6 Tesla are treated as the phase factors for the electron hoppings. 
We think this quantum mechanical treatment of the uniform magnetic field provides us the more appropriate results than the semi-classical theory\cite{fortin,fortin_2009}, in which the Landau quantization for the semi-classical closed orbits of electrons and holes is assumed by the magnetic breakdown phenomenon with a phenomenological parameter. 

For a reasonable value of anion potential ($V=12.38$~meV), energy bands in the magnetic field are broadened (Fig.~\ref{Figure7}), which is caused by the tight-binding nature of electrons. There should be much smaller gaps in the broadened Landau levels in principle, but it is very small and may not be seen in experiments at finite temperature. If electrons or holes are doped, the region of the non-quantized Hall effect is wider as the magnetic field increases due to the broadening of Landau levels. 
This broadening causes an interesting phenomenon that the amplitude of de Haas van Alphen oscillation at $T=0$ decreases as the magnetic field increases. 
This is different from the LK formula although the chemical potential is almost constant in this calculation.

For the larger value of anion potential ($V=37.14$~meV), the energy bands in the magnetic field are narrow and are seen as 
a slightly broadened Landau levels (Fig.~\ref{Figure10}), which is similar to energy obtained from the semi-classical theory\cite{fortin}. In this case the amplitude of de Haas van Alphen oscillation at $T=0$ is almost independent of magnetic field at low field, as in the semi-classical LK formula\cite{shoenberg,LK,nakano,alex2001,champel,CM,KH,Igor2004PRL,Igor2011,Sharapov}. 
The energy gaps at $3/4$-filling are closed or almost closed periodically at the inverse magnetic field, which was seen in both cases of $V=12.38$~meV and $V=37.14$ meV.

We would like to emphasize the difference between the quantum mechanical theory and semi-classical theory\cite{fortin} for (TMTSF)$_2$NO$_3$, which has electron and hole pockets at $h=0$. 
Unlike the cases in the semi-classical theory, 
we have shown that the Landau levels are sufficiently broadened near the Fermi energy and the energy gaps are closed or almost closed periodically as a function of the inverse magnetic field. 
Since we have neglected the hoppings between the
conducting plane, we have not discussed the effects of the direction of the magnetic field. We have not studied the transport properties in this paper, either. Therefore, the angular-dependent magnetoresistance have to be studied quantum mechanically in future.


It is possible to observe the results shown about quantum Hall conductance and dHvA oscillation 
in (TMTSF)$_2$NO$_3$ without SDW (for example, at 
$T_{\mathrm{SDW}}<T <T_{\mathrm{AO}}$, where SDW state does not exist). 
The wider region of the non-quantized Hall effect upon increasing the magnetic field will be observed under doping when the broadening of Landau levels is larger than thermal broadening. 
The Hall conductance\cite{basletic} and magnetization\cite{Naughton1997} have been observed experimentally in (TMTSF)$_2$NO$_3$ in the SDW state, but not in the metallic state. If the SDW state is suppressed by pressure, which affects the tight-binding parameters slightly but changes the nesting of the Fermi surface drastically, the magnetic field dependence of the amplitude of dHvA oscillation will be observed at low temperature.

%


\section*{Acknowledgement}
One of the authors (KK) thanks Noriaki Matsunaga for useful discussions and information of experiments.

\appendix
\section{energy at $H=0$}
\label{appendixA}
We use the spinless two-dimensional tight-binding model on 
a rectangular lattice in the unit cell with the four sites (A, B, A$^{\prime}$, B$^{\prime}$), 
where TMTSF molecules correspond to  sites. The effect of the anion ordering 
is represented by the on-site potential
along $x$-axis, ($V, V, -V, -V$), as shown in Figs. \ref{Figure1} (b) and (c). 
We show the Fermi surface in Figs. \ref{Figure2} (a) and (b) for $V=0$ and $V=12.38$ meV, respectively. 

The Bravais lattices for a rectangular lattice 
are given by 
\begin{eqnarray}
{\bf v}_1=(4a, 0)  \label{eqn:1.1}
\end{eqnarray}
and 
\begin{eqnarray}
{\bf v}_2=(0, b),  \label{eqn:1.2}
\end{eqnarray}
where $a$ and $b$ are the lattice spacings of TMTSF molecules. 
In this model, 
the Hamiltonian is given by 
\begin{align}
 \hat{\cal H}_0&=\sum_{i,j}t_{ij} c^{\dag}_i c^{}_j +  \sum_{i}V_i c_i^{\dagger} c_i \nonumber \\
  &=\displaystyle \sum_{{\bf r}_j}\biggl[ 
t_{\rm S1} 
\bigl( a^\dag_{{\bf r}_j} b_{{\bf r}_j}  + a^{\prime \dag}_{{\bf r}_j}b^{\prime}_{{\bf r}_j}
      + b^\dag_{{\bf r}_j} a_{{\bf r}_j}  + b^{\prime \dag}_{{\bf r}_j}{a^{\prime}}_{{\bf r}_j} \bigr)
\nonumber \\
&+t_{\rm S2} 
\bigl( b^\dag_{{\bf r}_j}a^{\prime}_{{\bf r}_j}     + b^{\prime \dag}_{{\bf r}_j}a_{{\bf r}_j+{\bf v}_1}
      + a^{\prime \dag}_{{\bf r}_j}{b}_{{\bf r}_j}   + a^\dag_{{\bf r}_j+{\bf v}_1}b^{\prime}_{{\bf r}_j} \bigr)
 \nonumber \\
&+t_{\rm I1} 
\bigl( a^\dag_{{\bf r}_j+{\bf v}_2}b_{{\bf r}_j}    + a^{\prime\dag}_{{\bf r}_j+{\bf v}_2}b^{\prime}_{{\bf r}_j}
      + b^\dag_{{\bf r}_j} a_{{\bf r}_j+{\bf v}_2}   + b^{\prime \dag}_{{\bf r}_j}a^{\prime}_{{\bf r}_j+{\bf v}_2} \bigr)
\nonumber \\
&+t_{\rm I2} 
\bigl(   b^\dag_{{\bf r}_j+{\bf v}_2}a^{\prime}_{{\bf r}_j}    + a^{\prime\dag}_{{\bf r}_j} b_{{\bf r}_j+{\bf v}_2}   \nonumber \\
 &\hspace{0.5cm}   
      + b^{\prime\dag}_{{\bf r}_j+{\bf v}_2}a_{{\bf r}_j+{\bf v}_1}   + a^{\dag}_{{\bf r}_j+{\bf v}_1}  b^{\prime}_{{\bf r}_j+{\bf v}_2} \bigr)
\nonumber \\
 &+t_{\rm I3}
\bigl( a^\dag_{{\bf r}_j} a_{{\bf r}_j +{\bf v}_2}  + a^{\prime \dag}_{{\bf r}_j}a^{\prime}_{{\bf r}_j+{\bf v}_2} 
        +b^\dag_{{\bf r}_j} b_{{\bf r}_j+{\bf v}_2} + b^{\prime \dag}_{{\bf r}_j}b^{\prime}_{{\bf r}_j+{\bf v}_2}
\nonumber \\
&\hspace{0.5cm}
                     + a^\dag_{{\bf r}_j+{\bf v}_2} a_{{\bf r}_j}  + a^{\prime\dag}_{{\bf r}_j+{\bf v}_2}a^{\prime}_{{\bf r}_j} 
                     +b^\dag_{{\bf r}_j+{\bf v}_2}b_{{\bf r}_j}  + b^{\prime\dag}_{{\bf r}_j+{\bf v}_2}b^{\prime}_{{\bf r}_j} \bigr)
\nonumber \\
&+t_{\rm I4}
\bigl( a^\dag_{{\bf r}_j+{\bf v}_2}a^{\prime}_{{\bf r}_j}      +b^\dag_{{\bf r}_j+{\bf v}_2}b^{\prime}_{{\bf r}_j} \nonumber \\
&\hspace{0.5cm}
+ a^{\prime\dag}_{{\bf r}_j+{\bf v}_2}a_{{\bf r}_j+{\bf v}_1} + b^{\prime\dag}_{{\bf r}_j+{\bf v}_2}b_{{\bf r}_j+{\bf v}_1} \nonumber \\
&\hspace{0.5cm}
         + a^{\prime\dag}_{{\bf r}_j}a_{{\bf r}_j+{\bf v}_2} + b^{\prime\dag}_{{\bf r}_j} b_{{\bf r}_j+{\bf v}_2} \nonumber \\
&\hspace{0.5cm}
         + a^{\dag}_{{\bf r}_j+{\bf v}_1}a^{\prime}_{{\bf r}_j+{\bf v}_2}+ b^{\dag}_{{\bf r}_j+{\bf v}_1}b^{\prime}_{{\bf r}_j+{\bf v}_2} \bigr)
\nonumber \\
&+V
\bigl( a^\dag_{{\bf r}_j}a_{{\bf r}_j}+b^\dag_{{\bf r}_j}b_{{\bf r}_j}
-a^{\prime\dag}_{{\bf r}_j}a^{\prime}_{{\bf r}_j}-b^{\prime\dag}_{{\bf r}_j}b^{\prime}_{{\bf r}_j} \bigr)
\biggr],  
\label{eqn:01}
\end{align}
where $a^\dag_{{\bf r}_j}$, $b^\dag_{{\bf r}_j}$, 
$a^{\prime\dag}_{{\bf r}_j}$ and $b^{\prime\dag}_{{\bf r}_j}$ ($a_{{\bf r}_j}$,  
$b_{{\bf r}_j}$, $a^{\prime}_{{\bf r}_j}$ and  
$b^{\prime}_{{\bf r}_j}$) are creation (annihilation) operators for 
A, B, A$^{\prime}$ and B$^{\prime}$  sites in $j$-th unit cell, respectively. 
By using the following Fourier transform,
\begin{eqnarray}
a_{{\bf r}_j} 
&=&\displaystyle \sum_{\bf k} e^{i\mathbf{k}\cdot{\bf r}_j}a_{\bf k},\label{eqn:02}\\
b_{{\bf r}_j}&=&\displaystyle \sum_{\bf k} e^{i\mathbf{k}
\cdot({\bf r}_j+ \frac{1}{4} {\bf v}_1)}b_{\bf k}, \label{eqn:03}\\
a^{\prime}_{{\bf r}_j} 
&=&\displaystyle \sum_{\bf k} e^{i\mathbf{k}\cdot({\bf r}_j+ \frac{1}{2}{\bf v}_{1})}
a^{\prime}_{\bf k},\label{eqn:04}\\
b^{\prime}_{{\bf r}_j}&=&\displaystyle \sum_{\bf k} e^{i\mathbf{k}
\cdot({\bf r}_j+\frac{3}{4}{\bf v}_{1})}b^{\prime}_{\bf k}, \label{eqn:05}
\end{eqnarray}
we obtain the Hamiltonian in the momentum space  as 
\begin{eqnarray}
{\cal {\hat H}}_{0}=\sum_{{\bf k}}
C^{\dagger}_{{\bf k}}\varepsilon_{\bf k}C_{{\bf k}}, 
\label{ham_0}
\end{eqnarray}
where 
\begin{eqnarray}
{C}^{\dagger}_{{\bf k}}=(a^\dag_{{\bf k}}, b^{\dag}_{{\bf k}},a^{\prime\dag}_{{\bf k}}, b^{\prime\dag}_{{\bf k}})
\end{eqnarray}
and 
\begin{eqnarray}
{C}_{{\bf k}}=\begin{pmatrix}
a^{}_{{\bf k}}\\
b^{}_{{\bf k}}\\
a^{\prime\dag}_{{\bf k}}\\
b^{\prime\dag}_{{\bf k}}
\end{pmatrix}.
\end{eqnarray}
In this equation, $\varepsilon_{\bf k}$ is a 4$\times 4$ matrix as follows; 
\begin{eqnarray}
\varepsilon_{\bf k} &=& 
\left(
\begin{array}{cccc}
{\epsilon}_{\mathbf{k} AA} & {\epsilon}_{\mathbf{k} AB} & {\epsilon}_{\mathbf{k} AA^{\prime}} & {\epsilon}_{\mathbf{k} AB^{\prime}} \\
{\epsilon}_{\mathbf{k} BA} & {\epsilon}_{\mathbf{k} BB} &{\epsilon}_{\mathbf{k} BA^{\prime}} & {\epsilon}_{\mathbf{k} BB^{\prime}} \\
\epsilon_{\mathbf{k} A^{\prime}A} & \epsilon_{\mathbf{k} A^{\prime}B} &\epsilon_{\mathbf{k} A^{\prime}A^{\prime}} & \epsilon_{\mathbf{k} A^{\prime}B^{\prime}} \\
\epsilon_{\mathbf{k} B^{\prime}A}&{\epsilon}_{\mathbf{k} B^{\prime}B}&{\epsilon}_{\mathbf{k} B^{\prime} A^{\prime}}&{\epsilon}_{\mathbf{k} B^{\prime}B^{\prime}} 
\end{array}
\right) \label{J3}
\end{eqnarray}
with 
\begin{align}
{\epsilon}_{\mathbf{k} AA} &= {\epsilon}_{\mathbf{k} BB}=2t_{\rm I3}\cos(bk_y)+V,\\
{\epsilon}_{\mathbf{k} A^{\prime}A^{\prime}} &= {\epsilon}_{\mathbf{k} B^{\prime}B^{\prime}}=2t_{\rm I3}\cos(bk_y)-V,\\
{\epsilon}_{\mathbf{k} AB} &= {\epsilon}_{\mathbf{k} A^{\prime}B^{\prime}}=t_{\rm S1}e^{iak_x}+t_{\rm I1}e^{i(ak_x-bk_y)}, \\
{\epsilon}_{\mathbf{k} BA^{\prime}} &= {\epsilon}_{\mathbf{k} B^{\prime}A}=
t_{\rm S2}e^{iak_x}+t_{\rm I2}e^{i(ak_x-bk_y)}, \\
{\epsilon}_{\mathbf{k} BA} &= {\epsilon}_{\mathbf{k} B^{\prime}A^{\prime}}={\epsilon}^*_{\mathbf{k} AB},\\
{\epsilon}_{\mathbf{k} AB^{\prime}} &= {\epsilon}_{\mathbf{k} A^{\prime}B}={\epsilon}^*_{\mathbf{k} BA^{\prime}}, \\
{\epsilon}_{\mathbf{k} AA^{\prime}} &= {\epsilon}_{\mathbf{k}BB^{\prime}}
=2t_{\rm I4}\cos(2ak_x-bk_y), \\
{\epsilon}_{\mathbf{k} A^{\prime}A} &= {\epsilon}_{\mathbf{k}B^{\prime}B}={\epsilon}_{\mathbf{k}AA^{\prime}}.
\label{ham_matrix}
\end{align}

When $V=0$, the Hamiltonian matrix of ${\cal {\hat H}}_0$ can be reduced to the 2$\times$2 
as 
\begin{eqnarray}
\varepsilon_{\bf k}^0 &=& 
\left(
\begin{array}{cc}
{\epsilon}_{\mathbf{k} AA}^{(0)} & {\epsilon}_{\mathbf{k} AB}^{(0)}  \\
{\epsilon}_{\mathbf{k} BA}^{(0)} & {\epsilon}_{\mathbf{k} BB}^{(0)}
\end{array}
\right) \label{J3_v0}
\end{eqnarray}
with 
\begin{align}
{\epsilon}_{\mathbf{k} AA}^{(0)} 
  &= {\epsilon}_{\mathbf{k} BB}^{(0)} \nonumber \\
  &= 2t_{\rm I3}\cos(bk_y)+2t_{\rm I4}\cos(2ak_x-bk_y),\\
{\epsilon}_{\mathbf{k} AB}^{(0)}
  &= t_{\rm S1} e^{iak_x}+t_{\rm S2}e^{-iak_x}\nonumber \\
  & \ + t_{\rm I1} e^{i(ak_x-bk_y)} + t_{\rm I2}e^{-i(ak_x-bk_y)}, \\
{\epsilon}_{\mathbf{k} BA}^{(0)}
 &= ({\epsilon}^{(0)}_{\mathbf{k}AB})^*.
\label{ham_matrix_v0}
\end{align}


\section{energy in the magnetic field}
\label{refappB}
The Hamiltonian in a spinless two-dimensional tight-binding model in the magnetic field 
becomes 
\begin{eqnarray}
{\cal {\hat H}}=\sum_{i,j}t_{ij} c^{\dag}_i c^{}_je^{i2\pi\phi_{ij}} + \sum_i V_i c^{\dagger}_i c_i,
\label{ham_H}
\end{eqnarray}
where  $c_i$  is $a_{\mathrm{r}_i}$, $b_{\mathrm{r}_i}$, $a'_{\mathrm{r}_i}$,  or $b'_{\mathrm{r}_i}$, and  the phase factor ($\phi_{ij}$) is given by  
\begin{eqnarray}
\phi_{ij}=\frac{e}{ch}\int_{\mathbf{r}_i}^{\mathbf{r}_j}{\bf A}\cdot d{\bf l},
\end{eqnarray}

In this study,
the magnetic field is applied perpendicular to the $x-y$ plane and we take
 the Landau gauge ${\bf A}=(0,Hx,0)$. 
The flux thorough the unit cell ($4ab$) is
\begin{eqnarray}
\Phi=4abH.
\end{eqnarray}
The phase factors are given as 
\begin{align}
\phi_{{\rm I1}AB}^{(n)}                          &= -\phi_{{\rm I1}BA}^{(n)}                          =\frac{\Phi}{\phi_0} (n+\frac{1}{8} ), \\
\phi_{{\rm I1}A^{\prime}B^{\prime}}^{(n)} &=-\phi_{{\rm I1}B^{\prime}A^{\prime}}^{(n)} = \frac{\Phi}{\phi_0}(n+\frac{5}{8}), \\
\phi_{{\rm I2}BA^{\prime}}^{(n)}              &= -\phi_{{\rm I2}A^{\prime}B}^{(n)}            =\frac{\Phi}{\phi_0}(n+\frac{3}{8}), \\
\phi_{{\rm I2}B^{\prime}A}^{\prime (n-1,n)}            &= \frac{\Phi}{\phi_0}(n-\frac{1}{8}),\\
\phi_{{\rm I2}AB^{\prime}}^{\prime (n+1,n)}            &= - \frac{\Phi}{\phi_0}(n+\frac{7}{8}),\\
\phi_{{\rm I3}AA}^{(n)}                          &= \frac{\Phi}{\phi_0}n, \\
\phi_{{\rm I3}BB}^{(n)}                          &= \frac{\Phi}{\phi_0}(n+\frac{1}{4}), \\
\phi_{{\rm I3}A^{\prime}A'}^{(n)}             &= \frac{\Phi}{\phi_0}(n+\frac{1}{2}), \\
\phi_{{\rm I3}B^{\prime}B'}^{(n)}             &= \frac{\Phi}{\phi_0}(n+\frac{3}{4}), \\
\phi_{{\rm I4}AA^{\prime}}^{(n)}              &= - \phi_{{\rm I4}A^{\prime}A}^{(n)}  = \frac{\Phi}{\phi_0}(n+\frac{1}{4}), \\
\phi_{{\rm I4}BB^{\prime}}^{(n)}              &= - \phi_{{\rm I4}B^{\prime}B}^{(n)}   =\frac{\Phi}{\phi_0}(n+\frac{1}{2}),\\
\phi_{{\rm I4}A^{\prime}A}^{\prime(n-1,n)}              &=  \frac{\Phi}{\phi_0}(n-\frac{1}{4}), \\
\phi_{{\rm I4}B^{\prime}B}^{\prime(n-1,n)}              &= \frac{\Phi}{\phi_0}n,
\end{align}
and
the phase factor is zero for the transfer integrals of $t_{\mathrm{a1}}$, $t_{\mathrm{a2}}$ and $t_{\mathrm{a3}}$. 
The phase factor $\phi_{\mu \alpha \beta}^{(n)}$ ($\mu= \mathrm{b1}, \mathrm{b2}, \mathrm{b3}$ or $\mathrm{b4}$, $\alpha$ and $\beta$ are
1, 2, 3, or 4) is the phase factor for the hopping $\mu$ from the site $\beta$ to the site $\alpha$ both in the $n$th unit cell 
($4 n a \leq x_i < 4 (n+1) a$ for both $\alpha$ and $\beta$).  When $\alpha \neq \beta$, the direction of the hopping is uniquely determined, and when $\alpha=\beta$ we take the hopping to the $y$-direction. 
The phase factor $\phi_{\mu \alpha \beta}^{\prime (m,n)}$ ($m= n-1$ or $m=n+1$)
 is for the hopping $\mu$ from the  $\beta$ site in the $n$th unit cell to the $\alpha$ site in the  $m$th unit cell 
($4 n a \leq x_i < 4(n+1) a$ for the $\beta$ site and $4 m a \leq x_i <4(m+1)a$ for the $\alpha$ site).

When the magnetic field is commensurate with the lattice
period, i.e.,
\begin{eqnarray}
\frac{\Phi}{\phi_0}=\frac{p}{q}, 
\end{eqnarray}
where 
$p$ and $q$ are integers, the magnetic unit cell is $4 q a \times b$. 
The Hamiltonian is  written as
\begin{equation}
\hat{\mathcal{H}} = \sum_{\mathbf{k}} \tilde{C}^{\dagger}_{\mathbf{k}} \tilde{\varepsilon}_{\mathbf{k}} \tilde{C}_{\mathbf{k}},
\end{equation}
where the summation over ${\bf k}$ is taken in the magnetic Brillouin zone,
\begin{eqnarray}
-\frac{\pi}{4qa}&\leq &k_x<\frac{\pi}{4qa}, \\
-\frac{\pi}{b}&\leq &k_y<\frac{\pi}{b},
\end{eqnarray}
$\tilde{C}^{\dagger}_{\mathbf{k}}$ and $\tilde{C}_{\mathbf{k}}$
have $4q$ components of creation and annihilation operators,
\begin{equation}
{\tilde C}^{\dagger}_{{\bf k}}=(a^{(0)\dag}_{{\bf k}}, b^{(0)\dag}_{{\bf k}}, a^{\prime(0)\dag}_{{\bf k}}, b^{\prime(0)\dag}_{{\bf k}},
\cdots, a^{\prime(q-1)\dag}_{{\bf k}}, b^{\prime(q-1)\dag}_{{\bf k}}),
\end{equation}
and
\begin{equation}
{\tilde C}_{{\bf k}}=\begin{pmatrix}
a^{(0)}_{{\bf k}}\\b^{(0)}_{{\bf k}}\\
a^{\prime(0)}_{{\bf k}}\\b^{\prime(0)}_{{\bf k}}\\
\vdots \\
a^{\prime (q-1)}_{{\bf k}}\\b^{\prime(q-1)}_{{\bf k}}
\end{pmatrix}.
\end{equation}
The $4q \times 4q$ matrix $\tilde{\varepsilon}_{\mathbf{k}}$ is expressed with $4 \times 4$ matrices $D_{\mathbf{k}}^{(n)}$ and
$F_{\mathbf{k}}^{(n)}$ as
\begin{align}
&\tilde{\varepsilon}_{\mathbf{k}} \nonumber \\
=& \left( \begin{array}{cccccc}
D^{(0)}_{\mathbf{k}}  & F^{(1)}_{\mathbf{k}} & 0 & \cdots & 0 &  F^{(0)\dagger}_{\mathbf{k}} \\
  F^{(1) \dagger }_{\mathbf{k}} & D^{(1)}_{\mathbf{k}}  & F^{(2)}_{\mathbf{k}} & \ddots & \ddots & 0 \\
  0  &   F^{(2) \dagger}_{\mathbf{k}}   &  D^{(2)}_{\mathbf{k}} &    F^{(3)}_{\mathbf{k}}    & \ddots & \vdots \\
  \vdots & \ddots & \ddots & \ddots & \ddots & 0 \\
  0 & \ddots & \ddots &  F^{(q-2) \dagger}_{\mathbf{k}}  & D^{(q-2)}_{\mathbf{k}} &  F^{(q-1)}_{\mathbf{k}} \\
   F^{(0)}_{\mathbf{k}} & 0 & \dots & 0 &  F^{(q-1) \dagger}_{\mathbf{k}} & D^{(q-1)}_{\mathbf{k}}\\
   \end{array} \right),
\label{j5} 
\end{align}
where
\begin{equation}
D_{\mathbf{k}}^{(n)} = \left( \begin{array}{cccc}
\epsilon_{\mathbf{k} AA}^{(n)} & \epsilon_{\mathbf{k} AB}^{(n)} & \epsilon_{\mathbf{k} AA'}^{(n)} & 0 \\
\epsilon_{\mathbf{k} BA}^{(n)} & \epsilon_{\mathbf{k} BB}^{(n)} &   \epsilon_{\mathbf{k} BA'}^{(n)} & \epsilon_{\mathbf{k} BB'}^{(n)} \\
\epsilon_{\mathbf{k} A'A}^{(n)} & \epsilon_{\mathbf{k} A'B}^{(n)} &   \epsilon_{\mathbf{k} A'A'}^{(n)} & \epsilon_{\mathbf{k} A'B'}^{(n)} \\
0                         & \epsilon_{\mathbf{k}B'B}^{(n)} &   \epsilon_{\mathbf{k}B'A'}^{(n)} & \epsilon_{\mathbf{k}B'B'}^{(n)} \\
 \end{array} \right),
\end{equation}
\begin{align}
\epsilon^{(n)}_{\mathbf{k} AA}&=2t_{\rm I3}\cos\bigg[bk_y +2 \pi \phi_{\mathrm{I3}AA}^{(n)} \bigg]+V,\\
\epsilon^{(n)}_{\mathbf{k} BB}&=2t_{\rm I3}\cos\bigg[bk_y +2 \pi \phi_{\mathrm{I3}BB}^{(n)} \bigg]+V, \\
\epsilon^{(n)}_{\mathbf{k} A^{\prime}A^{\prime}}&=2t_{\rm I3}\cos\bigg[bk_y + 2 \pi \phi_{\mathrm{I3}A'A'}^{(n)}\bigg]-V,\\
\epsilon^{(n)}_{\mathbf{k} B^{\prime}B^{\prime}}&=2t_{\rm I3}\cos\bigg[bk_y+2 \pi \phi_{\mathrm{I3}B'B'}^{(n)}\bigg]-V ,
\end{align}
\begin{align}
\epsilon^{(n)}_{\mathbf{k} AB}&=\epsilon^{(n)*}_{\mathbf{k} BA} \nonumber \\
  &=t_{\rm S1}e^{iak_x}\nonumber \\
  &+t_{\rm I1} \exp\bigg[ i \left(a k_x - bk_y - 2\pi \phi_{\mathrm{I1}AB}^{(n)} \right)\bigg], \\
\epsilon^{(n)}_{\mathbf{k} BA^{\prime}}&=\epsilon^{(n)*}_{\mathbf{k} A^{\prime}B} \nonumber \\
&= t_{\rm S2}e^{iak_x}\nonumber \\
&+ t_{\rm I2} \exp\bigg[ i \left( a k_x - bk_y - 2\pi  \phi_{\mathrm{I2}BA'}^{(n)} \right)\bigg], \\
\epsilon^{(n)}_{\mathbf{k} A^{\prime}B^{\prime}}&=\epsilon^{(n)*}_{\mathbf{k} B^{\prime}A^{\prime}} \nonumber \\
 &= t_{\rm S1}e^{iak_x}\nonumber \\
&+t_{\rm I1} \exp\bigg[ i \left( ak_x - bk_y - 2\pi  \phi_{\mathrm{I1}A'B'}^{(n)} \right) \bigg], 
\end{align}
\begin{align}
\epsilon^{(n)}_{\mathbf{k} AA^{\prime}}&=\epsilon^{(n)*}_{\mathbf{k} A^{\prime}A} \nonumber \\
  &= t_{\rm I4}\exp\bigg[ i \left(2ak_x-bk_y  - 2\pi \phi_{\mathrm{I4}AA'}^{(n)} \right) \bigg],\\
\epsilon^{(n)}_{\mathbf{k} BB^{\prime}}&=\epsilon^{(n)*}_{\mathbf{k} B^{\prime}B} \nonumber \\
  &= t_{\rm I4}\exp\bigg[i\left(2ak_x-bk_y  - 2\pi \phi_{\mathrm{I4}BB'}^{(n)} \right)  \bigg], 
\end{align}
\begin{equation}
F_{\mathbf{k}}^{(n)} = \left( \begin{array}{cccc}
   0                      &   0                         &   0                         & 0 \\
   0                        &   0                         &     0                         &    0                         \\
\epsilon_{\mathbf{k} A'A}^{\prime(n)}  &   0                         &      0                         &    0                        \\
\epsilon_{\mathbf{k} B'A}^{\prime(n)}   & \epsilon_{\mathbf{k} B'B}^{\prime(n)} &    0 & 0\\
 \end{array} \right),
\end{equation}
\begin{align}
\epsilon^{\prime (n)}_{\mathbf{k} A^{\prime}A}&=t_{\rm I4}\exp\bigg[i \left( 2ak_x - bk_y  - 2\pi \phi_{\mathrm{I4}A'A}^{\prime (n-1,n)} \right) \bigg],\\
\epsilon^{\prime (n)}_{\mathbf{k} B^{\prime}B}&=t_{\rm I4}\exp\bigg[i \left( 2ak_x - bk_y  - 2\pi \phi_{\mathrm{I4}B^{\prime}B}^{\prime (n-1,n)} \right) \bigg],\\ 
\epsilon^{\prime (n)}_{\mathbf{k} B^{\prime}A}&=t_{\rm S2} \exp\left[iak_x\right] \nonumber \\
&+t_{\rm I2} \exp\bigg[ i \left( ak_x - bk_y - 2\pi \phi_{\mathrm{I2}B'A}^{\prime (n-1,n)} \right) \bigg].
\end{align}
The matrix of Eq. (\ref{j5}) can be numerically diagonalized.

\end{document}